\documentclass[onecolumn,draftcls,dvips,letter]{IEEEtran}

\usepackage{amsmath,amssymb,epsfig,color}
\usepackage{graphicx,verbatim}
\usepackage[section]{placeins}
\usepackage{afterpage}

\newtheorem{definition}{Definition}
\newtheorem{theorem}{Theorem}

\newtheorem{proposition}{Proposition}
\newtheorem{corollary}{Corollary}

\newcommand{\eps}{\epsilon}
\newcommand{\styp}{A^{*(n)}_{\eps}}
\newcommand{\stypd}{A^{*(n)}_{\delta}}
\newcommand{\stypdp}{A^{*(n)}_{\delta'}}

\newcommand{\typ}{A_{\epsilon}^{(n)}}
\newcommand{\typm}{A_{\delta}^{(m)}}
\newcommand{\ttyp}{\tilde{A}_{\epsilon, \delta}^{(n)}}
\newcommand{\btyp}{\bar{A}_{\epsilon, \delta}^{(n)}}
\newcommand{\hP}{\hat{P}}
\newcommand{\hY}{\hat{Y}}
\newcommand{\hhY}{\hat{\hat{Y}}}
\newcommand{\hy}{\hat{y}}
\newcommand{\hhy}{\hat{\hat{y}}}
\newcommand{\hw}{\hat{w}}

\newcommand{\mN}{\mathcal{N}}

\newcommand{\mD}{\mathcal{D}}
\newcommand{\mS}{\mathcal{S}}
\newcommand{\mU}{\mathcal{U}}
\newcommand{\mV}{\mathcal{V}}
\newcommand{\mL}{\mathcal{L}}

\newcommand{\mW}{\mathcal{W}}
\newcommand{\mX}{\mathcal{X}}
\newcommand{\mY}{\mathcal{Y}}
\newcommand{\mZ}{\mathcal{Z}}
\newcommand{\mM}{\mathcal{M}}

\newcommand{\mhY}{\hat{\mathcal{Y}}}

\newcommand{\tmL}{\tilde{\mathcal{L}}}
\newcommand{\xvec}{\mathbf{x}}
\newcommand{\yvec}{\mathbf{y}}
\newcommand{\Xvec}{\mathbf{X}}
\newcommand{\Yvec}{\mathbf{Y}}
\newcommand{\hYvec}{\hat{\mathbf{Y}}}
\newcommand{\tYvec}{\tilde{\mathbf{Y}}}
\newcommand{\tZvec}{\tilde{\mathbf{Z}}}
\newcommand{\tTvec}{\tilde{\mathbf{T}}}
\newcommand{\tXvec}{\tilde{\mathbf{X}}}
\newcommand{\uvec}{{\bf u}}
\newcommand{\vvec}{{\bf v}}

\newcommand{\wvec}{{\bf w}}
\newcommand{\avec}{{\bf a}}
\newcommand{\hyvec}{\hat{\mathbf{y}}}
\newcommand{\hhyvec}{\hat{\hat{\mathbf{y}}}}
\newcommand{\Pe}{P_{e}^{(n)}}

\newcommand{\tend}{\hfill$\blacksquare$}

\newcommand{\Rgood}{R_{x1}}
\newcommand{\Rbad}{R_{x2}}

\newcommand{\Bt}{\mbox{Bin}_N(\theta)}

\newcommand{\Bj}{\mbox{Bin}_{L_i'}(j)}

\newcommand{\sigR}{\sigma_1^2}
\newcommand{\sigW}{\sigma_W^2}
\newcommand{\sigD}{\sigma^2}
\newcommand{\sigQ}{\sigma_Q^2}

\newcommand{\nQ}{N_Q}
\newcommand{\negdista}{\!\!\!\!\!\!\!\!\!\!}
\newcommand{\ners}{\mbox{\scriptsize no erase}}
\newcommand{\ers}{\mbox{\scriptsize erase}}

\title{On the Role of Estimate-and-Forward with Time-Sharing in Cooperative Communication
\thanks{The authors are with the School of Electrical and Computer
Engineering, Cornell University, Ithaca, NY. URL: {\tt
http://cn.ece.cornell.edu/}.
Work supported by the National Science Foundation, under awards
CCR-0238271 (CAREER), CCR-0330059, and ANR-0325556.}}
\author{Ron Dabora \hspace{2cm} Sergio D.\ Servetto}

\begin{document}
\maketitle
\begin{picture}(0,0)
\put(0,70){\tt\small Submitted to the IEEE Transactions on
Information Theory, October 2006.}
\end{picture}
\begin{abstract}
    \it\noindent
    In this work we focus on the general relay channel.
    We investigate the application of estimate-and-forward (EAF) to different scenarios. Specifically,
    we consider assignments of the auxiliary random variables that always satisfy the feasibility constraints.
    We first consider the multiple relay channel and obtain an achievable rate without decoding at the relays. We demonstrate
    the benefits of this result via an explicit discrete memoryless multiple relay scenario
    where multi-relay EAF is superior to multi-relay decode-and-forward (DAF).
    We then consider the Gaussian relay channel with coded modulation, where we show that a three-level quantization outperforms the
    Gaussian quantization commonly used to evaluate the achievable rates in this scenario. Finally we consider
    the cooperative general broadcast scenario with a multi-step conference. We apply
    estimate-and-forward to obtain a general multi-step achievable rate region. We then give an
    explicit assignment of the auxiliary random variables, and use this result to
    obtain an explicit expression for the single common message broadcast scenario with a two-step conference.
\end{abstract}

\section{Introduction}
\begin{figure}[h]
    \centering
    \scalebox{0.6}{\includegraphics{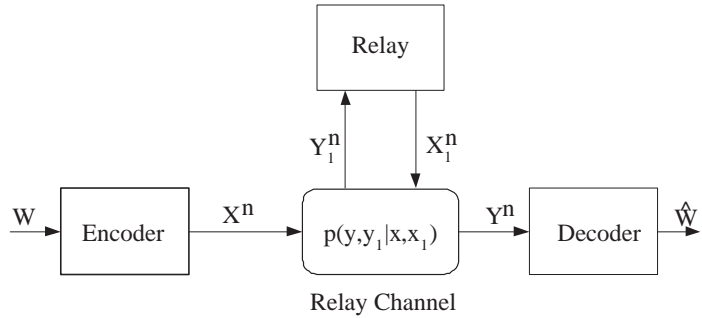}}
    \caption{The relay channel. The encoder sends a message $W$ to the decoder.}
    \label{fig:relay_setup}
\end{figure}

The relay channel was introduced by van der Meulen in 1971
\cite{Meulen:71}. In this setup, a single transmitter with channel input $X^n$ communicates with a single receiver with channel
output $Y^n$, where the superscript $n$ denotes the length of a vector. In addition, an external transceiver, called a relay,
listens to the channel and is able to output signals to the channel. We denote the relay output with $Y_1^n$ and its input with $X_1^n$.
This setup is depicted in figure \ref{fig:relay_setup}.

\subsection{Relaying Strategies}
\label{sec:relay_strategies}
In \cite{CoverG:79} Cover \& El-Gamal introduced two relaying
strategies commonly referred to as decode-and-forward (DAF) and
estimate-and-forward (EAF). In DAF the relay decodes the message
sent from the transmitter and then, at the next time interval,
transmits a codeword based on the decoded message. The rate
achievable with DAF is given in \cite[theorem 1]{CoverG:79}:
\begin{theorem}
    \label{thm:CEG_DAF}
    \it (achievability of \cite[theorem 1]{CoverG:79}) For the general relay channel any rate $R$ satisfying
    \begin{equation}
    \label{eqn:CEG_DAF}
        R \le \min \left\{I(X,X_1;Y), I(X;Y_1|X_1)\right\}
    \end{equation}
    for some joint distribution $p(x,x_1,y,y_1) = p(x,x_1)p(y,y_1|x,x_1)$, is achievable.
\end{theorem}
We note that for DAF to be effective, the rate to the relay has to be greater than the point-to-point rate
i.e.
\begin{equation}
    \label{eqn:DAF_condition}
    I(X;Y_1|X_1) > I(X;Y|X_1),
\end{equation}
otherwise higher rates could be obtained without using the relay at all.
For relay channels where DAF is not useful or not optimal, \cite{CoverG:79} proposed the EAF strategy. In this strategy,
the relay sends an estimate of its channel input to the destination, without decoding the source message at all.
The achievable rate with EAF is given in
\cite[theorem~6]{CoverG:79}:
\begin{theorem}
    \label{thm:CEG_EAF}
    \it (\cite[theorem 6]{CoverG:79}) For the general relay channel any rate $R$ satisfying
    \begin{eqnarray}
    \label{eqn:EAF_rate}
        R  &\le & I(X;Y,\hY_1|X_1),\\
    \label{eqn:EAF_feasible}
     \mbox{subject to }   I(X_1;Y) & \ge & I(Y_1;\hY_1|X_1,Y),
    \end{eqnarray}
    for some joint distribution $p(x,x_1,y,y_1,\hy_1) = p(x)p(x_1)p(y,y_1|x,x_1)p(\hy_1|y_1,x_1)$, where
    $||\mhY_1|| < \infty$, is achievable.
\end{theorem}

Of course, one can combine the DAF and EAF schemes by performing partial decoding at the relay, thus obtaining
higher rates as in \cite[theorem 7]{CoverG:79}.

\subsection{Related Work}
In recent years, the research in relaying has mainly focused on multiple-level
relaying and the MIMO relay channel. In the context of multiple-level relaying based on DAF, several DAF variations
were considered.
 In \cite{GuptaKumar:2003} Cover \& El-Gamal's block Markov encoding/succesive decoding DAF method was applied to the
multiple-relay case. Later work \cite{XieKumar:2004}, \cite{XieKumar:2005} and
\cite{Kramer:2003} applied the so-called regular encoding/sliding-window decoding and
the regular encoding/backward decoding techniques to the multiple-relay scenario.
In \cite{Madsen:2005} the DAF strategy was applied to the MIMO relay channel.
The EAF strategy was also applied to the multiple-relay scenario.
The work in \cite{Kramer:2005}, for example, considered the EAF strategy for multiple relay scenarios and the Gaussian relay
channel, in addition to considering the DAF strategy.
Also \cite{Gastpar:2002} considered the EAF strategy in the multiple-relay setup.
% and in \cite{SchienGallager:2000}
%communication over two parallel relay channels to a destination, without a direct link
%between the source and the destination, was considered .
Another approach applied recently to the relay channel is that
of iterative decoding. In \cite{ElGamalH:2006} the three-node network in the half-duplex regime was considered.
In the relay case, \cite{ElGamalH:2006} uses a feedback scheme where the receiver first uses EAF to
send information to the relay and then the relay decodes and uses DAF at the next time interval to help
the receiver decode its message. Combinations of EAF and DAF were also considered in \cite{Goldsmith:2006}, where
conferencing schemes over orthogonal relay-receiver channels were analyzed and compared.
Both \cite{ElGamalH:2006} and \cite{Goldsmith:2006} focus on the Gaussian case.
% In \cite{Mine:06} we applied simultaneous decoding to the EAF method which resulted in an
% increased feasible region for this strategy compared to \cite[theorem 6]{CoverG:79}.
% Another work that should be noted in that context is \cite{Motani:2005} where simultaneous decoding
% is used to improve upon Cover and El-Gamal's combined DAF/EAF result of \cite[theorem 7]{CoverG:79}.
% However, when specialized to the EAF setup, the result of \cite{Motani:2005} converges to
% \cite[theorem 6]{CoverG:79}.

An extension of the relay scenario to a hybrid broadcast/relay system was
introduced in \cite{DraperFK:03}
in which the authors applied a combination of EAF and DAF strategies to the independent broadcast
channel with a single common message, and then extended this strategy to the multi-step conference.
In \cite{RonSer:2005} we used both a single-step and a two-step conference
with orthogonal conferencing channels in the discrete memoryless framework.
A thorough investigation of the broadcast-relay channel was done in \cite{LiagV:2005}, where the authors
applied the DAF strategy to the case where only one user is helping the other user, and also presented an upper bound for
this case. Then, the fully cooperative scenario was analyzed. The authors applied both the
DAF and the EAF methods to that case.

\subsection{The Gaussian Relay Channel with Coded Modulation}
One important instance of the relay channel we consider in this work is the Gaussian relay channel with
coded modulation. This scenario is important in evaluating the rates achievable with practical communication
systems, where components in the receive chain, such as equalization for example, require
a uniformly distributed finite constellation for optimal operation.
In Gaussian relay channel scenarios, most often three types for relaying techniques are encountered:
\begin{itemize}
    \item The first technique is decode-and-forward. This technique achieves capacity for the physically degraded
        Gaussian relay channel (see \cite[section IV]{CoverG:79}), and also for more general relay
        channels under certain conditions (see \cite{Goldsmith:2006}).

    \item The second technique is estimate-and-forward, where the auxiliary variable $\hY_1$ is assigned a Gaussian
    distribution. For example, in \cite[section IV]{ElGamal:06} a Gaussian auxiliary random variable (RV) is used in conjunction with
    time-sharing at the transmitter, and in \cite{HostMadsen:05} the ergodic capacity for full duplex
    transmission with Gaussian EAF is obtained.

    \item The third technique is linear relaying, where the relay transmits a weighted sum of
    all its previously received inputs \cite[section V]{ElGamal:06}. An important subclass of this
    family of relaying functions is when the relay transmits a scaled version of its input. This method is called
    amplify-and-forward \cite{Laneman:2000}, and was later combined with DAF to produce the
    decode-amplify-and-forward method of \cite{Bao:2005}.
\end{itemize}
% In this paper we also consider the relay channel with coded modulation.
% The coded modulation relay scenario is important when evaluating the rates that can be obtained by practical
%systems,
Several recent papers consider the Gaussian relay channel with coded modulation.
In \cite{Kramer:Asi05} the author considered variations of DAF for different practical systems. In
\cite{Laneman:2000} DAF and amplify-and-forward were considered for coherent orthogonal BPSK signalling, and in
\cite{Stankovic:05} a practical construction that implements a half-duplex EAF coding scheme was proposed.

As indicated by several authors (see \cite{ElGamal:06}) it is not obvious if a Gaussian relay function is
indeed optimal. In this paper we show that for the case of coded modulation, there are scenarios where
non-Gaussian assignments of the auxiliary RV result in a higher rate than the commonly applied Gaussian assignment.

\subsection{Main Contributions}

In the following we summarize the main contributions of this work:
\begin{itemize}
    \item We give an intuitive insight into the relay channel in terms of information flow on a graph,
    and show how to obtain \cite[theorem 6]{CoverG:79} from flow considerations. Using flow considerations
    we also obtain the rate of the EAF strategy when the receiver uses joint-decoding.
    A similar expression can be obtained by specializing the result of
    \cite{Motani:06} to the case where the relay does not perform partial decoding.
    We then show that joint-decoding does not increase the maximum rate of the EAF strategy, and
    find the time-sharing assignment that obtains the joint-decoding rate from the general EAF expression. We also
    present another time-sharing assignment that always exceeds the joint-decoding rate.

    \item We introduce an  achievable rate expression for the multiple relay scenario based on EAF, that is also practically computabe.
    As discussed in section \ref{sec:relay_strategies}, in
    the ``noisy relay" case EAF outperforms DAF. However, for the multiple relay scenario there is no explicit, computationally practical  expression
    based on EAF that can be compared with the DAF-based result presented in \cite{XieKumar:2005}, so that the
    best strategy can be selected. As indicated in \cite[remark 22, remark 23]{Kramer:2005}, applying general EAF to
    a network with an arbitrary number of relays
    is computationally impractical due to the large number of constraints that characterize the feasible region.
    Therefore, it is interesting to explore a computationally simple assignment that allows to derive a
    result that extends to an arbitrary number of relays. We also provide an explicit numerical example
    to demonstrate that indeed there are cases where multi-relay EAF outperforms the multi-relay DAF.

    \item We consider the optimization of the EAF auxiliary random variable for the
    Gaussian relay channel with an orthogonal relay. We consider the coded modulation scenario, and
    show that there are three regions: high SNR on the source-relay link, where DAF is the best strategy,
    low SNR on the source-relay link in which the common
    EAF with Gaussian assignment is best, and an intermediate region where EAF with hard-decision
    per symbol is optimal. For this intermediate SNR region we consider two kinds of hard-decisions: deterministic and
    probabilistic, and show that each one of them can be superior, depending on the channel conditions.

    \item Lastly, we consider the cooperative broadcast scenario with a multi-step conference. We present a
    general rate region, extending the Marton rate region of \cite{Marton:79} to the case where the
    receivers hold a $K$-cycle conference prior to decoding the messages. We then specialize this result
    to the single common message case and obtain explicit expressions (without auxiliary RVs)
    for the two-step conference.
    %  that demonstrate that indeed it exceeds the two-step conference.
 %   Contrary to the two-step conference, the three-step scheme achieves
 %   the full cooperation bound when the conference capacities are less than those given by the Slepian-Wolf
 %   theorem \cite[theorem 14.4.1]{cover-thomas:it-book}.

\end{itemize}

%In the third part of this paper we demonstrate our new strategy in the cooperative broadcast channel with a
%single common message scenario. For this setup
%we present an explicit three-step cooperation scheme that does not require
%auxiliary random variables. This new cooperation scheme yields a rate
%increase over the non-cooperative rate for any given cooperation capacity. In addition, this scheme achieves
%the full cooperation bound when the conference capacities are less than those given by the Slepian-Wolf
%theorem \cite[theorem 14.4.1]{cover-thomas:it-book}.

The rest of this paper is organized as follows:
%in section \ref{sec:defs} we define the mathematical framework and also
%present an intuitive formulation of the relay channel using information flow on a graph.
in section \ref{sec:timeshare_single} we discuss the single relay case. We consider the EAF strategy with
time-sharing (TS) and relate it to the EAF rate expression for joint-decoding at the destination receiver.
In section \ref{sec:MultipleRelays} we present an achievable region for the multiple-relay channel, and
in section \ref{sec:Gauss_relay} we examine the Gaussian relay channel with coded modulation.
In section \ref{sec:application_multi_step} we investigate the general cooperative broadcast scenario, and
obtain an explicit rate expression by applying TS-EAF to the general multi-step conference.
Finally, section \ref{sec:conclu} presents concluding remarks.

%%%%%%%%%%%%%%%%%%%%%%%%%%%%%%%%%%%%%%%%%%%%%%%%%%%%%%%%%%%%%%%%%%%%%%%%%%%%%%%%%%%%%%%%%%%%%%%%%%%%%%%%%%%%%%%%%
%%%%%%%%%%%%%%%%%%%%%%%%%%%%%%%%%%%%%%%%%%%%%%%%%%%%%%%%%%%%%%%%%%%%%%%%%%%%%%%%%%%%%%%%%%%%%%%%%%%%%%%%%%%%%%%%%
%%%%%%%%%%%%%%%%%%%%%%%%%%%%%%%%%%%%%%%%%%%%%%%%%%%%%%%%%%%%%%%%%%%%%%%%%%%%%%%%%%%%%%%%%%%%%%%%%%%%%%%%%%%%%%%%%
%%%%%%%%%%%%%%%%%%%%%%%%%%%%%%%%%%%%%%%%%%%%%%%%%%%%%%%%%%%%%%%%%%%%%%%%%%%%%%%%%%%%%%%%%%%%%%%%%%%%%%%%%%%%%%%%%
%%%%%%%%%%%%%%%%%%%%%%%%%%%%%%%%%%%%%%%%%%%%%%%%%%%%%%%%%%%%%%%%%%%%%%%%%%%%%%%%%%%%%%%%%%%%%%%%%%%%%%%%%%%%%%%%%
%%%%%%%%%%%%%%%%%%%%%%%%%%%%%%%%%%%%%%%%%%%%%%%%%%%%%%%%%%%%%%%%%%%%%%%%%%%%%%%%%%%%%%%%%%%%%%%%%%%%%%%%%%%%%%%%%

\section{Time-Sharing for the Single-Relay Case}
\label{sec:timeshare_single}

\subsection{Definitions}
\label{sec:defs}
First, a word about notation:
we denote discrete random variables with capital letters e.g. $X$, $Y$, and their realizations with lower case letters
$x$, $y$. A random variable $X$ takes values in a set $\mX$. We use $||\mX||$ to denote the cardinality
of a finite discrete set $\mX$, and $p_X(x)$ denotes the probability distribution function (p.d.f.) of $X$ on $\mX$. For brevity we may omit the subscript $X$ when it is obvious from
the context. We denote vectors with boldface letters, e.g. $\xvec$, $\yvec$; the $i$'th element of a vector $\xvec$ is
denoted by $x_i$ and we use $\xvec_i^j$ where $i<j$ to denote $(x_i, x_{i+1},...,x_{j-1},x_j)$.
We use $\styp(X)$ to denote the set of $\eps$-strongly typical sequences w.r.t. distribution
$p_X(x)$ on $\mX$, as defined in \cite[ch. 5.1]{YeungBook} and $\typ(X)$ to denote the $\eps$-weakly typical set
as defined in \cite[ch. 3]{cover-thomas:it-book}.

We also have the following definitions:
\begin{definition}
    \label{def:relay_channel}
    The {\em discrete relay channel} is defined by two discrete input alphabets $\mX$ and $\mX_1$, two
    discrete output alphabets $\mY$ and $\mY_1$ and a probability density function $p(y,y_1|x,x_1)$ giving the
    probability distribution on $\mY \times \mY_1$ for each $(x,x_1) \in \mX \times \mX_1$.
    The relay channel is called {\em memoryless} if the probability of a block of $n$ transmissions is given by
    $p(\yvec,\yvec_1|\xvec,\xvec_1) = \prod_{i=1}^n p\left(y_i, y_{1,i}|x_i,x_{1,i}\right)$.
\end{definition}
In this paper we consider only the memoryless relay channel.
\begin{definition}
    \label{def:code}
    A {\em $(2^{nR},n)$ code} for the relay channel consists of a source message set
    $\mW = \left\{1,2,...,2^{nR}\right\}$, a mapping function $f$ at the encoder,
    \[
        f: \mW \mapsto \mX^n,
    \]
    a set of $n$ relay functions
    \[
        x_{1,i} = t_i\left(y_{1,1},y_{1,2},...,y_{1,i-1} \right),
    \]
    where the $i$'th relay function $t_i$ maps the first $i-1$ channel outputs at the relay into a transmitted
    relay symbol at time $i$. Lastly we have a decoder
    \[
        g: \mY^n \mapsto \mW.
    \]
\end{definition}
\begin{definition}
    \label{def:Perr}
    The {\em average probability of error} for a code of length $n$ for the relay channel is defined as
    \[
        \Pe = \Pr(g(Y^n) \ne W),
    \]
    where $W$ is selected uniformly over $\mW$.
\end{definition}
\begin{definition}
    A rate $R$ is called {\em achievable} if there exists a sequence of $(2^{nR},n)$ codes with
    $\Pe \rightarrow 0$ as $n \rightarrow \infty$.
\end{definition}

\subsection{The Single Relay EAF with Time-Sharing}
\label{sec:ts-single-subsec}
Consider the following assignment of the auxiliary random variable
of theorem \ref{thm:CEG_EAF}:
\begin{equation}
    \label{eqn:time-sharing-mapping}
    p(\hy_1|y_1,x_1) = \left\{
            \begin{array}{cl}
                q &, \hy_1 = y_1\\
                1-q & ,\hy_1 = \Omega \notin \mY_1.
            \end{array}
        \right.
\end{equation}
Under this assignment, the feasibility condition of
\eqref{eqn:EAF_feasible} becomes
\begin{eqnarray*}
    I(X_1;Y) & \ge & I(Y_1;\hY_1|X_1,Y) \\
             & = &   H(Y_1|X_1,Y) - H(Y_1|X_1,Y,\hY_1) \\
             & = &   H(Y_1|X_1,Y) - (1-q)H(Y_1|X_1,Y) - q H(Y_1|X_1,Y,Y_1)\\
             & = &   q H(Y_1|X_1,Y),
\end{eqnarray*}
and the rate expression \eqref{eqn:EAF_rate} becomes
\begin{eqnarray*}
    R & \le & I(X;Y,\hY_1|X_1)\\
        & = & I(X;Y|X_1) + I(X; \hY_1|X_1,Y)\\
        & = & I(X;Y|X_1) + H(X| X_1,Y) - H(X|X_1,Y,\hY_1)\\
        & = & I(X;Y|X_1) + H(X| X_1,Y) - (1-q) H(X|X_1,Y) - q H(X|X_1,Y,Y_1)\\
        & = & I(X;Y|X_1) + q I(X;Y_1|X_1,Y).
\end{eqnarray*}
Clearly, maximizing the rate implies maximizing $q$ subject to the
constraint $q\in [0,1]$. This gives the following corollary to theorem \ref{thm:CEG_EAF}:
\begin{corollary}
    \label{corr:single_relay_TAF}
    \it For the general relay channel any rate $R$ satisfying
    \begin{equation}
    \label{eqn:main_corr}
        R \le I(X;Y|X_1) + \left[ \frac{I(X_1;Y)}{H(Y_1|X_1,Y)} \right]^* I(X;Y_1|X_1,Y),
    \end{equation}
    for the joint distribution $p(x,x_1,y,y_1) = p(x) p(x_1) p(y,y_1|x,x_1)$, with $[x]^* \triangleq \min(x,1)$,
    is achievable.
\end{corollary}

Now, consider the following distribution chain:
\begin{equation}
\label{eqn:extended_prob_chain}
    p(x,x_1,y,y_1,\hy_1,\hhy_1) = p(x)p(x_1)p(y,y_1|x,x_1) p(\hy_1|x_1,y_1) p(\hhy_1 | \hy_1).
\end{equation}
We note that this extended chain can be put into the standard form by letting $p(\hhy_1|x_1,y_1) = \sum_{\mhY_1}p(\hy_1,\hhy_1|x_1,y_1) =
\sum_{\mhY_1}p(\hy_1|x_1,y_1)p(\hhy_1|\hy_1) $.
After compression of $Y_1$ into $\hY_1$, there is a second compression operation, compressing $\hY_1$ into $\hhY_1$. The output
of the second compression is used to facilitate cooperation between the relay and the destination. Therefore, the
receiver decodes the message based on $\hhyvec_1$ and $\yvec$, repeating exactly the same step as in the standard relay decoding, with
$\hhyvec$ replacing $\hyvec$. Then, the expressions of theorem \ref{thm:CEG_EAF} become
\begin{eqnarray}
    \label{eqn:EAF_rate_extended}
        R  &\le & I(X;Y,\hhY_1|X_1),\\
    \label{eqn:EAF_feasible_extended}
     \mbox{subject to }   I(X_1;Y) & \ge & I(Y_1;\hhY_1|X_1,Y).
\end{eqnarray}
Now, applying TS to $\hhY_1$ with
\begin{equation}
    \label{eqn:assignment_hhy}
        p(\hhy_1|\hy_1) = \left\{
            \begin{array}{cl}
                q &,\hhy_1 = \hy_1\\
                1-q & ,\hhy_1 = \Delta \notin \mhY_1
            \end{array}
        \right.,
\end{equation}
the expressions in \eqref{eqn:EAF_rate_extended} and \eqref{eqn:EAF_feasible_extended} become
\begin{eqnarray}
    R & \le & I(X;Y|X_1) + I(X;\hhY_1|X_1,Y)\nonumber\\
     & = & I(X;Y|X_1) + H(X|X_1,Y) - H(X|\hhY_1,X_1,Y)\nonumber\\
     & = & I(X;Y|X_1) + q(H(X|X_1,Y)  -  H(X|\hY_1,X_1,Y)) \nonumber\\
     \label{eqn:rate_CEG_extended_chain}
     & = & I(X;Y|X_1) + q I(X;\hY_1|X_1,Y), \\
    I(X_1;Y) & \ge & I(Y_1;\hhY_1|X_1,Y)\nonumber\\
        & = & H(Y_1|X_1,Y) - H(Y_1|\hhY_1,X_1,Y)\nonumber\\
        & = & H(Y_1|X_1,Y) - (1-q)H(Y_1|X_1,Y) - q H(Y_1|\hY_1,X_1,Y)\nonumber\\
    \label{eqn:feasibility_CEG_extended_chain}
        & = & q I(Y_1;\hY_1|X_1,Y).
\end{eqnarray}
Combining this with the constraint $q \in [0,1]$ we obtain the following corollary to theorem \ref{thm:CEG_EAF}:
    \begin{proposition}
    \label{prop:TAF}
    \it
        For the general relay channel, any rate $R$ satisfying
        \[
            R \le I(X;Y|X_1) + \left[ \frac{I(X_1;Y)}{I(Y_1;\hY_1|X_1,Y)}\right]^* I(X;\hY_1|X_1,Y),
        \]
        for some joint distribution $p(x,x_1,y,y_1,\hy_1) = p(x)p(x_1)p(y,y_1|x,x_1)p(\hy_1|x_1,y_1)$,
        is achievable.
    \end{proposition}
\smallskip
This proposition generalizes on corollary \ref{corr:single_relay_TAF} by performing a general Wyner-Ziv (WZ) compression combined with
TS (which is a specific type of WZ compression), intended to guarantee feasibility of the first compression step.
In section \ref{sec:Gauss_relay} we apply a similar idea to the EAF relaying in the Gaussian relay channel scenario with coded modulation.
Before we discuss the relationship between joint-decoding and time-sharing we present an intuitive way to view the EAF strategy.

\subsection{An Intuitive View of Estimate-and-Forward}
\label{sec:intuitive_explanation}
Consider the rate bound and the feasible region of theorem \ref{thm:CEG_EAF}
given in equations \eqref{eqn:EAF_rate} and \eqref{eqn:EAF_feasible}.
We note that the following intuitive explanation does not constitute a proof but it does provide an insight into the
relay achievability results. We emphasize that the achievable rates stated in this section can also be proved rigorously.
In the following we provide an intuitive insight into these expressions in terms of a flow on a graph.

In constructing the intuitive information flow representation for the relay channel, we first need to specify
the underlaying assumptions and the operations performed at the source, the relay and the destination receiver:
\begin{itemize}
    \item The source and the relay generate their codebooks independently.

    \item The relay compresses its channel output $\yvec_1$ into $\hyvec_1$, which represents the information
    conveyed to the destination receiver to assist in decoding the source message.

    \item Based on the above two restrictions we have the following Markov chain:
    $p(x)p(x_1)x(y,y_1|x,x_1)p(\hy_1|x_1,y_1)$.

    \item The relay input signal $\xvec_1$ is based only on the compressed $\hyvec_1$.

    \item The destination uses $\xvec_1$, $\hyvec_1$ and $\yvec$ to decode the source message $\xvec$.
\end{itemize}
We also use the following representation for transmission, reception and compression:
\begin{itemize}
    \item  We represent an information source
        as a source whose output flow is equal to its information rate.

    \item We represent the compression
        operation as a flow sink whose flow consumption is equal to the mutual information between the
        original and the compressed sequences.

    \item The destination is represented as a flow sink.

    \item  As in a standard flow on a graph, the flows are additive, following the
        chain rule of mutual information.
\end{itemize}

\begin{figure}[ht]
    \centering
    \scalebox{0.6}{\includegraphics{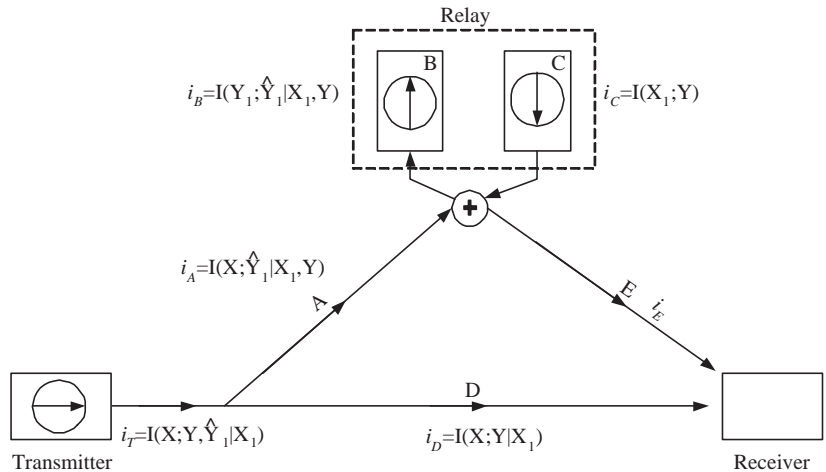}}
    \caption{The information flow budget for the general relay channel with compression at the relay.}
    \label{fig:Relay_flow}
\end{figure}

Now consider the following flow diagram of figure \ref{fig:Relay_flow}.
As can be observed from the figure, the source has an output flow of
\[
    i_T = I(X;Y,\hY_1,X_1) = I(X;Y,\hY_1|X_1).
\]
This follows from the fact that the destination uses $\xvec_1,\hyvec_1$ and $\yvec$ to decode $\xvec$ and the fact that
$X$ and $X_1$ are independent. This total
flow reaches the receiver through two branches, the direct branch (D) which carries a flow of $i_D = I(X;Y|X_1)$ and
the relay branch (ABCE). Now, the quantities in the relay branch are calculated given $X_1$ and $Y$ to represent only the
rate increase over the direct path.
The relay branch has four parts: an edge
(A) which carries a flow of $I(X;\hY_1|X_1,Y)$, a sink (B) with consumption $I(Y_1;\hY_1|X_1,Y)$,
a relay source (C) with an output flow of $I(X_1;Y)$ and an edge (E) from the relay to the destination.
Here, the relay transmission to the destination (C) is done at a fixed rate $I(X_1;Y)$, independent of the type
of compression $p(\hy_1|x_1)$ used at the relay, since we always transmit from the relay to the destination
at the maximum possible rate in order to obtain the best performance.
The rate loss due to compression is represented by $I(\hY_1;Y_1|X_1,Y)$, since we consider only the excess
rates over the direct one.

Now, from the laws of flow addition and conservation, the overall flow from the source to the destination through
the relay branch is $i_E = i_A + i_B + i_C$. To assist the direct link (D) we need
the flow on (ABCE) to be positive. In theorem \ref{thm:CEG_EAF} the scheme considers only the last two elements,
$i_B + i_C$, and verifies that their net flow is positive, namely
\begin{equation}
\label{eqn:intuive_CEG}
    -I(Y_1;\hY_1|X_1,Y) + I(X_1;Y) > 0.
\end{equation}
This condition guarantees a net positive flow on (ABCE) since always $i_A \ge 0$.
Now, the flow to the destination can be obtained as the minimum
\begin{equation}
\label{eqn:intuit_CEG_Rate}
    R \le \min \left\{ i_D + i_E, i_T\right\},
\end{equation}
where, the second term in the minimum is obtained from the transmitter, since
trivially the information rate at the receiver cannot exceed $i_T$. We note that because $i_B + i_C \ge 0$, the minimum in \eqref{eqn:intuit_CEG_Rate}
is $i_T$. Therefore, the resulting achievable rate is
\[
    R \le I(X;Y,\hY_1|X_1),
\]
which combined with \eqref{eqn:intuive_CEG} gives the result of \cite[theorem 6]{CoverG:79}.

However, the condition in \eqref{eqn:intuive_CEG} is not tight since
even when $i_B + i_C < 0$ the  flow on (ABCE) is still non-negative if the entire sum $i_A + i_B + i_C$ is
non-negative, i.e.
\begin{equation}
\label{eqn:tighter_condition}
    I(X;\hY_1|X_1,Y) - I(\hY_1;Y_1|X_1,Y) + I(X_1;Y) \ge 0.
\end{equation}
Then, the achievable rate to the destination is bounded by
\begin{equation}
\label{eqn:intuit_jt_rate}
    R \le i_D + i_E = I(X;Y|X_1) + I(X_1;Y) - I(\hY_1;Y_1|X,X_1,Y).
\end{equation}
Indeed, when the flow through the relay branch (ABCE) is zero we obtain the
non-cooperative rate $I(X;Y|X_1)$.
 Plugging the expression \eqref{eqn:intuit_jt_rate} into \eqref{eqn:intuit_CEG_Rate} yields the following achievable rate:
\begin{eqnarray*}
    R  & \le & \min\left\{i_D + i_E , i_T\right\}\\
       &  =  & \min\left\{ I(X;Y|X_1) + I(X_1;Y) - I(\hY_1;Y_1|X,X_1,Y), I(X;Y,\hY_1|X_1)\right\}\\
       &  =  & I(X;Y|X_1) + \min\left\{  I(X_1;Y) - I(\hY_1;Y_1|X,X_1,Y), I(X;\hY_1|X_1,Y)\right\}.
%       &  =  & \min\left\{ I(X,X_1;Y)  - I(\hY_1;Y_1|X,X_1,Y), I(X;Y,\hY_1|X_1)\right\}.
\end{eqnarray*}
Combining this with \eqref{eqn:tighter_condition}, (informally) proves the following proposition:
\begin{proposition}
    \label{prop:jt-rate}
    \it
    For the general relay channel, any rate $R$ satisfying
    \begin{eqnarray*}
        R  & \le & I(X;Y|X_1) + \min\left\{  I(X_1;Y) - I(\hY_1;Y_1|X,X_1,Y), I(X;\hY_1|X_1,Y)\right\},\\
        \mbox{subject to } I(X_1;Y) & \ge & I(\hY_1;Y_1|X,X_1,Y) =  I(\hY_1;Y_1|X_1,Y) - I(X;\hY_1|X_1,Y),
    \end{eqnarray*}
    for some joint distribution $p(x,x_1,y,y_1,\hy_1) = p(x)p(x_1)p(y,y_1|x,x_1)p(\hy_1|x_1,y_1)$, is achievable.
\end{proposition}
\bigskip
The proof of proposition \ref{prop:jt-rate} can be made formal using joint-decoding at the destination receiver,
but as in the next subsection we show that this expression is a special case of \cite[theorem 6]{CoverG:79} obtained by time-sharing, we omit the
details of the proof here.

\subsection{Joint-Decoding and Time-Sharing}
In the original work of \cite[theorem 6]{CoverG:79}, the decoding
procedure at the destination receiver for decoding the message
$w_{i-1}$ at time $i$ is composed of three steps (the notations
below are identical to \cite[theorem 6]{CoverG:79}. The reader is referred to the proof of \cite[theorem 6]{CoverG:79}
to recall the definitions of the sets and variables used in the following description):
\begin{enumerate}
    \item Decode the relay index $s_i$ using $\yvec(i)$, the received signal at time $i$.
    \item Decode the relay message $z_{i-1}$, using $s_i$, the received
        signal $\yvec(i-1)$ and the previously decoded $s_{i-1}$.
    \item Decode the source message $w_{i-1}$ using $\yvec(i-1)$,
    $z_{i-1}$ and $s_{i-1}$.
\end{enumerate}

Evidently, when decoding the relay message $z_{i-1}$ at the second step, the receiver does not make use of
the statistical dependence
between $\hyvec_{1}(i-1)$, the relay sequence at time $i-1$, and $\xvec(w_{i-1})$, the transmitted source codeword at time $i-1$.
The way to use this dependence is to jointly decode $z_{i-1}$ and $w_{i-1}$ after decoding $s_i$ and $s_{i-1}$. The joint-decoding procedure
then has the following steps:
    \begin{enumerate}
        \item From $\yvec(i)$, the received signal at time $i$, the receiver decodes $s_i$ by looking for a unique
        $s \in \mS$, the set of indices used to select $\xvec_1$, such that $\big(\xvec_1(s), \yvec(i)\big) \in \styp$.
        As in \cite[theorem 6]{CoverG:79},
        the correct $s_i$ can be decoded with an arbitrarily small probability of error by taking $n$ large
        enough as long as
        \begin{equation}
        \label{eqn:R0_conds}
            R_0 \le I(X_1;Y),
        \end{equation}
        where $||\mS|| = 2^{nR_0}$.

        \item The receiver now knows the set $S_{s_i}$ into which $z_{i-1}$ (the relay message at time $i-1$) belongs.
        Additionally, from decoding at time $i-1$
        the receiver knows $s_{i-1}$, used to generate $z_{i-1}$.

        \item The receiver generates the set
        $\mL(i-1) = \left\{ w \in \mW: \big(\xvec(w), \yvec(i-1), \xvec_1(s_{i-1})\big) \in \styp\right\}$.

        \item The receiver now looks for a unique $w \in \mL(i-1)$ such that
            $\big( \xvec(w), \yvec(i-1), \hyvec_1(z|s_{i-1}), \xvec_1(s_{i-1})\big) \in \styp$ for some
            $z \in S_{s_i}$.
            If such a unique $w$ exists then it is the decoded $\hw_{i-1}$,
            otherwise the receiver declares an error.
    \end{enumerate}
We do not give here a formal proof for the resulting rate expression, but as indicated in section
\ref{sec:intuitive_explanation}, the rate expression resulting from this decoding procedure is given by
proposition \ref{prop:jt-rate}.

Let us now compare the the rates obtained with joint-decoding (proposition \ref{prop:jt-rate}) with the rates obtained with the sequential
decoding of \cite[thoerem 6]{CoverG:79}:
to that end we consider the joint-decoding result of proposition \ref{prop:jt-rate} with the extended probability chain of
\eqref{eqn:extended_prob_chain}:
\[
%    \label{eqn:extended_chain}
    p(x,x_1,y,y_1,\hy_1,\hhy_1) = p(x)p(x_1)p(y,y_1|x,x_1) p(\hy_1|x_1,y_1) p(\hhy_1 | \hy_1),
\]
where $\hhY_1$ represents the information relayed to the destination.
%Applying exactly the same steps as in the proof of proposition \ref{prop:jt-rate} we obtain the expression:
%\begin{eqnarray}
%    \label{eqn:rate_2_hats}
%        R & \le & I(X;Y|X_1)  + \min\left\{I(X_1;Y) - I(\hhY_1;Y_1|X,X_1,Y),  I(X;\hhY_1|X_1,Y)\right\}\\
%    \label{eqn:constr_2_hats}
%        \mbox{subject to }I(X_1;Y) & \ge & I(\hhY_1;Y_1|X,X_1,Y) = I(\hhY_1;Y_1|X_1,Y) - I(X;\hhY_1|X_1,Y).
%\end{eqnarray}
%Now consider the expressions in \eqref{eqn:rate_2_hats} and \eqref{eqn:constr_2_hats}. Setting $p(\hhy_1 | \hy_1)$ the same as in
%\eqref{eqn:assignment_hhy}
%subject to $q \in [0,1]$, we obtain that \eqref{eqn:rate_2_hats} and \eqref{eqn:constr_2_hats} become
Expanding the expressions of proposition \ref{prop:jt-rate} using the assignment \eqref{eqn:assignment_hhy}, similarly to proposition
\ref{prop:TAF}, we obtain the expressions:
\begin{eqnarray}
    \label{eqn:rate_2_hats_q}
        R & \le & I(X;Y|X_1)  + \min\left\{I(X_1;Y) - q I(\hY_1;Y_1|X,X_1,Y),  q I(X;\hY_1|X_1,Y)\right\}\\
    \label{eqn:constr_2_hats_q}
        \mbox{subject to }I(X_1;Y) & \ge & q I(\hY_1;Y_1|X,X_1,Y) = q\left(I(\hY_1;Y_1|X_1,Y) - I(X;\hY_1|X_1,Y)\right).
\end{eqnarray}

We can now make the following observations:
\begin{enumerate}
    \item Setting $q = 1$ we obtain proposition \ref{prop:jt-rate}. Additionally, if
        $I(X_1;Y) > I(\hY_1;Y_1|X_1,Y)$ then both proposition \ref{prop:jt-rate} and \cite[theorem 6]{CoverG:79} give
        identical expressions.
    \item When $q=1$ and
        \begin{equation}
        \label{eqn:cond_joint}
            I(\hY_1;Y_1|X_1,Y) - I(X;\hY_1|X_1,Y) < I(X_1;Y) < I(\hY_1;Y_1|X_1,Y),
        \end{equation}
        then {\em for the same} mapping $p(\hy_1|x_1, y_1)$ we obtain that proposition \ref{prop:jt-rate} provides rate but
        \cite[theorem 6]{CoverG:79} does not. The rate expression under these conditions is
        \begin{equation}
        \label{eqn:rate_q_is_one}
            R  \le  I(X;Y|X_1)  + I(X_1;Y) - I(\hY_1;Y_1|X,X_1,Y).
        \end{equation}

    \item
        Now, fix the probability chain $p(x)p(x_1)p(y,y_1|x,x_1)p(\hy_1|x_1,y_1)$ and  examine the expressions
        \eqref{eqn:rate_2_hats_q} and \eqref{eqn:constr_2_hats_q} when \eqref{eqn:cond_joint} holds:
        when $q < 1$, then \eqref{eqn:cond_joint}
        guarantees that condition \eqref{eqn:constr_2_hats_q} is still satisfied.
        If $q$ is close enough to $1$ such that we also have
        $I(X_1;Y) \le q I(\hY_1;Y_1|X_1,Y)$, the rate from \eqref{eqn:rate_2_hats_q}, i.e.,
        \[
            R \le  I(X;Y|X_1)  + I(X_1;Y) - qI(\hY_1;Y_1|X,X_1,Y),
        \]
        is now greater than \eqref{eqn:rate_q_is_one}. In this case can  keep decreasing $q$ until
        \begin{equation}
        \label{eqn:optim_q}
            I(X_1;Y) - qI(\hY_1;Y_1|X,X_1,Y) = qI(X;\hY_1|X_1,Y)
        \end{equation}
        at which point the rate becomes
        \begin{equation}
        \label{eqn:rate-jt-optim}
            R \le I(X;Y|X_1)  + q I(X;\hY_1|X_1,Y).
        \end{equation}
        This rate can be obtained from \cite[theorem 6]{CoverG:79} by applying the extended probability chain of \eqref{eqn:extended_prob_chain},
        as long as $I(X_1;Y) \ge q I(\hY_1,Y_1|X_1,Y)$.
\end{enumerate}
We therefore conclude that all the rates that joint decoding allows can also be obtained  or exceeded by the original EAF with an
appropriate time sharing\footnote{This argument is due to Shlomo Shamai and Gerhard Kramer.}.

    Note that equality in \eqref{eqn:optim_q} implies
    \[
        q_{opt} = \min\left\{1,\frac{I(X_1;Y)}{I(\hY_1;Y_1|X,X_1,Y)+ I(X;\hY_1|X_1,Y)}\right\}
            =\min\left\{1, \frac{I(X_1;Y)}{I(\hY_1;Y_1|X_1,Y)}\right\}
            ,
    \]
    hence $q_{opt}$ is the maximum $q$ that makes the mapping $p(\hy_1|x_1,y_1)$ feasible for \cite[theorem 6]{CoverG:79}.
    Plugging $q_{opt}$ into \eqref{eqn:rate-jt-optim}, we obtain the rate expression of proposition \ref{prop:TAF}.

    Finally, consider again the region where joint decoding is useful \eqref{eqn:cond_joint}:
    \begin{eqnarray*}
        I(\hY_1;Y_1|X,X_1,Y) & \le I(X_1;Y) \le &   I(\hY_1;Y_1|X_1,Y)\\
    \Rightarrow   0 & \le I(X_1;Y) - I(\hY_1;Y_1|X,X_1,Y) \le &   I(\hY_1;Y_1|X_1,Y) - I(\hY_1;Y_1|X,X_1,Y)\\
   \Rightarrow   0 & \le I(X_1;Y) - I(\hY_1;Y_1|X,X_1,Y) \le &    I(X_1;\hY_1|X_1,Y)\\
        \Rightarrow   0 & \le \frac{I(X_1;Y) - I(\hY_1;Y_1|X,X_1,Y)}{I(X;\hY_1|X_1,Y)} \le &   1.
    \end{eqnarray*}
    If $I(X;\hY_1|X_1,Y) > 0$, then using time-sharing on $\hY_1$ with
    \begin{equation}
    \label{eqn:assign_q_joint}
        q = \frac{I(X_1;Y)-I(\hY_1;Y_1|X,X_1,Y)}{I(X;\hY_1|X_1,Y)}
    \end{equation}
    into equations \eqref{eqn:rate_CEG_extended_chain} and \eqref{eqn:feasibility_CEG_extended_chain} yields:
    \[
        I(X;Y|X_1) + q I(X;\hY_1|X_1,Y) = I(X;Y|X_1) + I(X_1;Y)-I(\hY_1;Y_1|X,X_1,Y),
    \]
    as long as $I(X_1;Y) \ge qI(\hY_1;Y_1|X_1,Y)$, or equivalently
    \begin{equation}
    \label{eqn:cond_TS_CEG6}
        q \le \frac{I(X_1;Y)}{I(\hY_1;Y_1|X_1,Y)}.
    \end{equation}
    Plugging assignment \eqref{eqn:assign_q_joint} into \eqref{eqn:cond_TS_CEG6} we obtain:
    \begin{eqnarray*}
        \frac{I(X_1;Y)-I(\hY_1;Y_1|X,X_1,Y)}{I(X;\hY_1|X_1,Y)} & \le & \frac{I(X_1;Y)}{I(\hY_1;Y_1|X_1,Y)}\\
        \Rightarrow \quad \left(I(X_1;Y)-I(\hY_1;Y_1|X,X_1,Y)\right)I(\hY_1;Y_1|X_1,Y)
            & \le & I(X_1;Y)I(X;\hY_1|X_1,Y)\\
        \Rightarrow \quad I(X_1;Y)I(\hY_1;Y_1|X_1,Y)- I(X_1;Y)I(X ;\hY_1|X_1,Y)
            & \le & I(\hY_1;Y_1|X ,X_1,Y)I(\hY_1;Y_1|X_1,Y)\\
        \Rightarrow \quad I(X_1;Y)I(\hY_1;Y_1|X ,X_1,Y) & \le & I(\hY_1;Y_1|X ,X_1,Y)I(\hY_1;Y_1|X_1,Y)\\
        \Rightarrow \quad I(X_1;Y) & \le & I(\hY_1;Y_1|X_1,Y),
    \end{eqnarray*}
    as long as $I(\hY_1;Y_1|X ,X_1,Y) > 0$,
    which is the region where joint-decoding is supposed to be useful.
    Hence the joint-decoding rate of proposition \ref{prop:jt-rate} can be obtained by time sharing
    on the \cite[theorem 6]{CoverG:79} expression. Therefore, joint-decoding does not improve on the
    rate of \cite[theorem 6]{CoverG:79}. In fact the rate of proposition \ref{prop:TAF} is always at least as large as
    that of proposition \ref{prop:jt-rate}.

%%%%%%%%%%%%%%%%%%%%%%%%%%%%%%%%%%%%%%%%%%%%%%%%%%%%%%%%%%%%%%%%%%%%%%%%%%%%%%%%%%%%%%%%%%%%%%%%%%%%%%%%%%%%%%%%%%%
%%%%%%%%%%%%%%%%%%%%%%%%%%%%%%%%%%%%%%%%%%%%%%%%%%%%%%%%%%%%%%%%%%%%%%%%%%%%%%%%%%%%%%%%%%%%%%%%%%%%%%%%%%%%%%%%%%%
%%%%%%%%%%%%%%%%%%%%%%%%%%%%%%%%%%%%%%%%%%%%%%%%%%%%%%%%%%%%%%%%%%%%%%%%%%%%%%%%%%%%%%%%%%%%%%%%%%%%%%%%%%%%%%%%%%%
%%%%%%%%%%%%%%%%%%%%%%%%%%%%%%%%%%%%%%%%%%%%%%%%%%%%%%%%%%%%%%%%%%%%%%%%%%%%%%%%%%%%%%%%%%%%%%%%%%%%%%%%%%%%%%%%%%%
%%%%%%%%%%%%%%%%%%%%%%%%%%%%%%%%%%%%%%%%%%%%%%%%%%%%%%%%%%%%%%%%%%%%%%%%%%%%%%%%%%%%%%%%%%%%%%%%%%%%%%%%%%%%%%%%%%%

\section{An Achievable Rate for the Relay Channel with Multiple Relays}
\label{sec:MultipleRelays}
When the source-relay
% The multiple-relay channel was studied by in \cite{GuptaKumar:2003}, \cite{XieKumar:2004}, \cite{XieKumar:2005}
% and \cite{Kramer:2003}. These results are based on decoding at the relays (according to an hierarchy)
% prior to generating the relay codeword. However, when the
channel is very noisy then,
as discussed in the introduction, it may be better not to use the relay at all than to
employ the decode-and-forward strategy. Alternatively, when decode-and-forward is not useful, one
could employ estimate-and-forward. One result for multiple relays based on EAF can be found in
\cite{Gastpar:2002} which considered the two-relay case. In \cite[theorem 3]{Kramer:2005} the EAF
strategy, with partial decoding was applied to the multiple-relay case, and in \cite[theorem 4]{Kramer:2005} a mixed
EAF and DAF strategy was applied.
However, as stated in \cite[remark 22, remark 23]{Kramer:2005} applying the general estimate-and-forward to
a network with an arbitrary number of relays
is computationally impractical due to the large number of constraints that
characterize the feasible region (for two relays
we need to satisfy $9$ constraints). Moreover, the rate computation is prohibitive since
it would imply solving a non-convex optimization problem. In conclusion, an alternative achievable rate
to that based on decode-and-forward, which can also be evaluated with a reasonable effort, has not been presented to date.
In this section we derive an explicit achievable rate based on estimate-and-forward.
The strategy we use is to pick the auxiliary random variable
such that the feasibility constraints are satisfied. This is not a trivial choice since setting the
auxiliary random variable in theorem \ref{thm:CEG_EAF} to be the relay channel output (i.e. $\hY_1 = Y_1$) does
not remove this constraint, and we therefore need to incorporate time-sharing as discussed in the following.

\subsection{A General Achievable Rate}
\label{sec:achieve_general}
%The rate derived in the previous section is based on separate decoding of the pair $(\xvec_i, \hyvec_i)$ for
%the relay $i$. However, we can improve on this rate if, when decoding the information for relay $i$, we
%use all the information obtained from decoding for previous relays. This results in the following theorem:
We extend the idea of section \ref{sec:ts-single-subsec} to the relay channel with $N$ relays. This channel consists of
a source with channel input $X$, $N$ relays where for relay $i$, $X_i$ denotes the channel input and $Y_i$ denotes the channel output,
and a destination with channel output $Y$. This channel is denoted by
$\left( \mX \times_{i=1}^N \mX_i,p(y,y_1,...,y_N|x,x_1,...,x_N), \mY \times_{i=1}^N \mY_i\right)$.
Let $\Xvec = \left(X_1,X_2,...,X_N\right)$ and $\Yvec = \left(Y_1,Y_2,...,Y_N\right)$. We now have the
following theorem:
\begin{theorem}
    \label{thm:achieve_N_result_2}
    \it
    For the general multiple-relay channel with $N$ relays,
        $\Big( \mX \times_{i=1}^N \mX_i,p(y,y_1,...,y_N|x,x_1,...,x_N),$ ${\mY \times_{i=1}^N \mY_i}\Big)$, any rate $R$ satisfying
        \[
            R \le I(X;Y|\Xvec) + \sum_{\theta = 1}^{2^N-1} P(\Bt)I(X;\Yvec_{\Bt}|\Xvec,Y),
        \]
        where $\Bt$ is an $N$-element vector that contains $'1'$ in the locations where the $N$-bit binary representation
        of the integer $\theta$ contains $'1'$,
        $P(\Bt) = \prod_{i:\Bt_i = 0} (1-q_i) \prod_{i:\Bt_i = 1} q_i$, $\Bt_i$ is the $i$'th bit in the
        $N$-bit binary representation of $\theta$,
        $\Yvec_{\Bt} = \left(Y_{i_1}, Y_{i_2},...,Y_{i_M} \right)$, where $i_1$, $i_2$, ..., $i_M$ are
        the locations of the $'1'$ in $\Bt$, and
        \begin{equation}
            \label{eqn:q_i_assgn_full_thm}
  %     q_i = \left[\frac{\sum_{j = 1}^{L_i} P_l(\Bj)I(X_i;Y|\tXvec_i(\tZvec_i),\tYvec_{l,{\Bj}}(\tZvec_i)) }
           q_i = \left[\frac{I(X_i;Y|\tZvec_i) }
                        {H(Y_i|\Xvec,Y) -\sum_{j = 1}^{2^{L'_i}-1} P_{l'}(\Bj)I(Y_i;\tYvec_{l',\Bj}(\tTvec_i)|\Xvec,Y)}
            \right]^*,
        \end{equation}
                for the joint distribution
        $p(x,x_1,x_2,...,x_N,y,y_1,y_2,...,y_N)=p(x)p(x_1)...p(x_N)p(y,y_1,...,y_N|x,x_1,...,x_N)$ is achievable.
        In \eqref{eqn:q_i_assgn_full_thm} $\tZvec_i$ is the vector containing all the variables
        $X_j$ decoded prior to decoding $X_i$, $\tTvec_i$ is a vector that contains all the variables $\hY_p$ decoded
        prior to decoding $\hY_i$, and $\tYvec_{l',\Bj}(\tTvec_i)$ contains all the $Y_{l_r'}$, such that
         $\hY_{l_r'} \in \tTvec_i$, and $r$ is a location of  $\; '1'$ in the $L_i'$-bit binary representation of $j$.
         $L_i'$ if the number of elements in $\tTvec_i$. Note that if $\hY_p \in \tTvec_i$ then we must have
         $X_p \in \tZvec_i$.

\end{theorem}

        To facilitate the understanding of the expressions in theorem \ref{thm:achieve_N_result_2}, we first look at a simplified case
        where the destination decodes each relay message independently of the messages of the other relays.
%Therefore, each $s_{i,k}$ is decoded using only $\yvec(k)$
%and each $\mL_i(k-1)$ is generated based only on $\xvec_1(s_{1,k-1}),\xvec_2(s_{2,k-1}),...,\xvec_N(s_{N,k-1})$ and $\yvec(k-1)$.
%This is the simplest implementation of the multi-relay EAF strategy.
This can be obtained from theorem
\ref{thm:achieve_N_result_2} by setting $\tZvec_i = \varnothing$ and $\tTvec_i = \varnothing$, $i = 1,2,...,N$. The result is summarized in the
following corollary:
\begin{corollary}
    \label{corr:achieve_N_result_1}
    \it
    For the general multiple-relay channel
        $\left( \mX \times_{i=1}^N \mX_i,p(y,y_1,...,y_N|x,x_1,...,x_N), \mY \times_{i=1}^N \mY_i\right)$, any rate $R$ satisfying
        \begin{equation}
        \label{eqn:rate_expression_multi_relay}
            R \le I(X;Y|\Xvec) + \sum_{\theta = 1}^{2^N-1} P(\Bt)I(X;\Yvec_{\Bt}|\Xvec,Y),
        \end{equation}
        is achievable,
%        where $\Bt$ is an $N$-element vector that contains $'1'$ in the location where the $N$-bit binary representation
%        of the integer $\theta$ contains $'1'$,
%        $P(\Bt) = \prod_{i:\Bt_i = 0} (1-q_i) \prod_{i:\Bt_i = 1} q_i$,
%        $\Yvec_{\Bt} = \left(Y_{i_1}, Y_{i_2},...,Y_{i_M} \right)$, where $i_1$, $i_2$, ..., $i_M$ are
%        the locations of the $'1'$ in $\Bt$, and
        where
        \begin{equation}
            \label{eqn:q_i_assgn_simple}
            q_i = \left[ \frac{I(X_i;Y)}{H(Y_i|\Xvec,Y)} \right]^*,
        \end{equation}
        for the joint distribution
        $p(x,x_1,x_2,...,x_N,y,y_1,y_2,...,y_N)=p(x)p(x_1)...p(x_N)p(y,y_1,...,y_N|x,x_1,...,x_N)$.
\end{corollary}
\bigskip

    In the multi-relay strategy we employ in this section
    each relay transmits its channel output $Y_i$ with probability $q_i$, independent of the other relays.
    Therefore, when considering a group of $N$ relays,
    the probability that any subgroup of relays will transmit their channel outputs simultaneously is simply the product of all transmission
    probabilities $q_i$ at each relay in the group, multiplied by the product of erasure probabilities $(1-q_i)$ for each relay in the complement
    group. Now, considering the rate expression of \eqref{eqn:rate_expression_multi_relay} we observe
    that the rate is obtained by taking all possible groupings of relays. For each grouping the resulting rate is the
    rate obtained when using all the channel outputs of all the relays in that group to assist in decoding. This is indicated by the
    term $\Yvec_{\Bt}$. This rate has to be weighted by the probability of such an overlap occurring, which is given by $P(\Bt)$.
    We then sum over all such groupings to obtain the achievable rate. The parameter $q_i$ for each relay, which is determined by
    \eqref{eqn:q_i_assgn_simple}, can be interpreted by considering the terms in the denominator and numerator: the denominator
    $H(Y_i|\Xvec,Y)$ is the (exponent of the) size of uncertainty at the destination receiver about relay $i$'s output $Y_i$. The numerator is
    the (exponent of the) size of the information set that can be transmitted from relay $i$ to the destination receiver. Therefore, the fraction
    $\frac{I(X_i;Y)}{H(Y_i|\Xvec,Y)}$
     can be interpreted as the maximal fraction of the uncertainty at the destination about relay $i$'s channel output $Y_i$,
     that can be compensated by the relay transmission. Of course, this faction has to be upper bounded by one. In the more general setup
     of theorem \ref{thm:achieve_N_result_2}, the decoding of the relay information from relay $i$ is done by using the information
     from the relays which were decoded before relay $i$ to assist in decoding. This results in the conditioning at the numerator and
     the negative terms in the denominator, both contribute to increasing the value of $q_i$.

\subsection{Proof of Theorem \ref{thm:achieve_N_result_2}}
    \subsubsection{Overview of Coding Strategy}
    The transmitter generates its codebook independent of the relays. Next, each relay generates
    its own codebook independent of the other relays following the construction of \cite[theorem 6]{CoverG:79}, with the mapping
    $p(\hy_i|x_i,y_i)$ at each relay set to the time-sharing mapping of \eqref{eqn:time-sharing-mapping} with parameter
    $q_i$. The destination receiver first needs to decode all the relay codewords $\left\{X_i^n\right\}_{i=1}^N$ and use this information to decode the relay messages
    $\left\{\hY_i^n\right\}_{i=1}^N$. To this end, the relay decides on a decoding order for the $X_i^n$ sequences and
    a decoding order for the $\hY_i^n$ sequences. These decoding orders determine the maximum value of $q_i$ that can be selected for each relay,
    thereby allowing us to determine the auxiliary variables' mappings and obtain an explicit rate expression. Finally, the receiver uses all the
    decoded $\left\{X_i^n\right\}_{i=1}^N$ and $\left\{\hY_i^n\right\}_{i=1}^N$ sequences, together with its channel input to decode the
    source message.

    \bigskip
    We now give the details of the construction:
    fix the distributions $p(x)$, $p(x_1)$, $p(x_2)$,...,$p(x_N)$, and
    \begin{equation}
    \label{eqn:aux_assign_simple}
        p(\hy_i|x_i,y_i) = \left\{
            \begin{array}{cl}
                q_i &, \hy_i = y_i\\
                1-q_i & ,\hy_i = \Omega \notin \mY_i
            \end{array}
        \right.,
    \end{equation}
    $i = 1,2,...,N$. Let $\mW = \left\{1,2,...,2^{nR}\right\}$ be the source message set.
    \subsubsection{Code Construction at the Transmitter and the Relays}
    \begin{itemize}
        \item  Code construction and transmission at the transmitter are the same as in \cite[theorem 6]{CoverG:79}.

        \item Code construction at the relays is done by repeating the relay code construction
            of \cite[theorem 6]{CoverG:79}
            for each relay, where relay $i$ uses the distributions $p(\hy_i|x_i,y_i)$ and
            $p(x_i)$. We denote the relay message, the transmitted message and the partition
            set at relay $i$ at time $k$ with $z_{i,k}$, $s_{i,k}$ and $S^{(i)}_{s_{i,k}}$ respectively. The message set
            for $s_i$ is denoted $\mW_i$, where $||\mW_i|| = 2^{n R_i}$. The message set for $z_i$ is denoted
            $\mW_i'$, $||\mW_i'|| = 2^{n R_i'}$. The relay codewords at relay $i$ are denoted $\hyvec_i(z_i|s_i)$, and
            the transmitted codewords at relay $i$ are denoted $\xvec_i(s_i)$, $s_i \in \mW_i$, $z_i \in \mW_i'$.
    \end{itemize}

    \subsubsection{Decoding and Encoding at the Relays}$ $

    Consider relay $i$ at time $k-1$:
    \begin{itemize}
        \item From the relay transmission at time $k-1$, the relay knows $s_{i,k-1}$. Now the relay looks for a message
            $z_i \in \mW_i'$, such that
            \[
                \big(\hyvec_i(z_i|s_{i,k-1}), \yvec_i(k-1), \xvec_i(s_{i,k-1}) \big) \in \styp(\hY_i, Y_i,X_i).
            \]
            Following the argument in \cite[theorem 6]{CoverG:79}, for $n$ large enough there is such a message $z_i$ with
            a probability that is arbitrarily close to $1$, as long as
            \begin{equation}
            \label{eqn:relay_rate_constr_1}
                R_i' > I(\hY_i;Y_i|X_i) + \eps = q_iH(Y_i|X_i) + \eps.
            \end{equation}
        Denote this message with $z_{i,k-1}$.
        \item Let $s_{i,k}$ be the index of the partition of $\mW_i'$ into which $z_{i,k-1}$ belongs, i.e.,
            $z_{i,k-1} \in S^{(i)}_{s_{i,k}}$.

        \item At time $k$ relay $i$ transmits $\xvec_i(s_{i,k})$.
    \end{itemize}

    \subsubsection{Decoding at the Destination}
%    Therefore, at each relay we have the rate constraint \ref{eqn:relay_rate_constr_1}, i.e.
%    \[
%        R'_i > q_i H(Y_i|X_i) + \eps
%    \]
    \begin{itemize}
    \item Consider the decoding of $w_{k-1}$ at time $k$, for a
    fixed decoding order: let $\tZvec_i$ contain all the $X_j$'s whose $s_{j,k}$'s are decoded prior
    to decoding $s_{i,k}$. Therefore, decoding $s_{i,k}$ is done by looking for a unique message $s_i \in \mW_i$
    such that
    \begin{eqnarray*}
       &  &  \big(\xvec_i(s_{i}),\xvec_{m_1}(s_{m_1,k}), \xvec_{m_2}(s_{m_2,k}),...,\xvec_{m_{M_i}}(s_{m_{M_i},k}),\yvec(k)
       \big) \in \styp(X_i,\tZvec_i,Y),
%       &  &  \qquad\qquad \hyvec_{l_1}(z_{l_1,k-1}|s_{l_1,k-1}),\hyvec_{l_2}(z_{l_2,k-1}|s_{l_2,k-1}),...,
%            \hyvec_{l_{L_i}}(z_{l_{L_i},k-1}|s_{l_{L_i},k-1})\big) \in \styp(X_i,Y,\tZvec_i),
    \end{eqnarray*}
    where $m_1$, $m_2$,...,$m_{M_i}$ enumerate all the $X_j$'s in
    $\tZvec_i = \left(X_{m_1},X_{m_2},...X_{m_{M_i}} \right)$.
%    and $l_1$, $l_2$,...,$l_{L_i}$ enumerate all the $\hY_l$'s in $\tZvec_i$. Of course, if $\hY_l$ is in
%    $\tZvec_i$ then also $X_l$ must be there (i.e. we use only legal orderings).
    Assuming correct decoding at the previous steps, then by the point-to-point channel achievability proof
    we obtain that the probability of error for decoding
    $s_{i,k}$ can be made arbitrarily small by taking $n$ large enough as long as
    \begin{equation}
        R_i < I(X_i;Y,\tZvec_i) - \eps = I(X_i;Y|\tZvec_i) - \eps.
    \end{equation}
    Let $\tTvec_i$ contain all the $\hY_{l'}$'s whose $z_{l',k-1}$'s are decoded prior to decoding $z_{i,k-1}$. Note that all
    the $\left\{s_{i,k-1}\right\}_{i=1}^N$ were already decoded at the previous time interval when $w_{k-2}$ was decoded.

        \item The destination generates the set
            \begin{eqnarray}
            &  & \mL_i(k-1) = \bigg\{ z_i \in \mW'_i : \big(\yvec(k-1), \hyvec_i(z_i|s_{i,k-1}),
                \hyvec_{l'_1}(z_{l'_1,k-1}|s_{l'_1,k-1}), ...,\hyvec_{l'_{L'_i}}(z_{l'_{L'_i},k-1}|s_{l'_{L'_i},k-1}),\nonumber\\
            \label{eqn:set_for_multi_relay_EAF}
            &  & \qquad \qquad \qquad \qquad\qquad\qquad
%                    \xvec_{m'_1}(s_{m'_1,k-1}),...,\xvec_{m'_{M'_i}}(s_{m'_{M'_i},k-1}) \big) \in \styp(\hY_i,Y,\tTvec_i)\bigg\},
                    \xvec_{1}(s_{1,k-1}),\xvec_{2}(s_{2,k-1}),...,\xvec_{N}(s_{N,k-1}) \big) \in \styp(Y,\hY_i,\tTvec_i,\Xvec)\bigg\},
            \end{eqnarray}
            where
%            $m'_1$, $m'_2$,...,$m'_{M'_i}$ enumerate all the $X_j$'s in $\tTvec_i$ and
            $l'_1$, $l'_2$,...,$l'_{L_i}$ enumerate all the $\hY_{l'}$'s in $\tTvec_i$.
            The average size of $\mL_i(k-1)$ can be bounded using the standard technique of
            \cite[equation (36)]{CoverG:79} and the fact that when $z_i \ne z_{i,k-1}$, then the corresponding
            $\hyvec_i(z_i|s_{i,k-1})$ is independent of all the variables in \eqref{eqn:set_for_multi_relay_EAF}
            except $\xvec_i(s_{i,k-1})$. The resulting bound is
            \[
                 E\left\{||\mL_i(k-1)||\right\} \le 1 + 2^{n(R_i' - I(\hY_i;Y,\Xvec_{-i},\tTvec_i|X_i) + 3\eps)},
            \]
            where $\Xvec_{-i}$ is an $N-1$ element vector that contains all the elements of $\Xvec$ except $X_i$.
%    \begin{figure}[ht]
%         \epsfxsize=0.26\textwidth \leavevmode\centering\epsffile{Markov_chain.eps}
%        \caption{The Markov relationship between the random variables used for decoding $z_i$, for the case
%        of $L_i' = 2$. Edges in the figure represent Markov relationship.}
%        \label{fig:Markov-relation}
%    \end{figure}

        \item Now, the destination looks for a unique $z_{i} \in \mL_i(k-1) \bigcap S^{(i)}_{s_{i,k}}$.
            Therefore, making the probability of error arbitrarily small by taking $n$ large enough
            can be done as long as
            \begin{equation}
            \label{eqn:Ri'_upper_bound}
                R'_i < I(\hY_i;Y,\Xvec_{-i},\tTvec_i|X_i) + I(X_i;Y|\tZvec_i) -  4\eps.
            \end{equation}
    \end{itemize}
     We note that using the assignment \eqref{eqn:aux_assign_simple} we can write
         \begin{eqnarray*}
             I(\hY_i;Y,\Xvec_{-i},\tTvec_i|X_i) & =  & H(Y,\Xvec_{-i},\tTvec_i|X_i) - H(Y,\Xvec_{-i},\tTvec_i|X_i,\hY_i)\\
                    & = & H(Y,\Xvec_{-i},\tTvec_i|X_i)  - (1-q_i)H(Y,\Xvec_{-i},\tTvec_i|X_i) - q_i H(Y,\Xvec_{-i},\tTvec_i|X_i,Y_i)\\
                    & = & q_iH(Y,\Xvec_{-i},\tTvec_i|X_i)  - q_i H(Y,\Xvec_{-i},\tTvec_i|X_i,Y_i)\\
                    & = & q_i I(Y_i;Y,\Xvec_{-i},\tTvec_i|X_i)\\
                    & = & q_i\left(H(Y_i|X_i) - H(Y_i|Y,\Xvec_{-i},X_i, \hY_{l_1'},\tTvec_{i,2}^{L_i'})  \right)\\
                    & = & q_i\Big(q_{l_1'}H(Y_i|X_i) + (1-q_{l_1'}) H(Y_i|X_i)\\
                    &   & \qquad \qquad  - q_{l_1'}H(Y_i|Y,\Xvec_{-i},X_i, Y_{l_1'},\tTvec_{i,2}^{L_i'})   - (1-q_{l_1'})H(Y_i|Y,\Xvec_{-i},X_i,\tTvec_{i,2}^{L_i'})\Big)\\
                    & = & q_i\Big(q_{l_1'}I(Y_i;Y,\Xvec_{-i}, Y_{l_1'},\tTvec_{i,2}^{L_i'}|X_i)
                             + (1-q_{l_1'}) I(Y_i;Y,\Xvec_{-i}, \tTvec_{i,2}^{L_i'}|X_i)\Big)\\
                    & ... &\\
                    & = & q_i \sum_{j = 0}^{2^{L'_i}-1} P_{l'}(\Bj)I(Y_i;Y,\Xvec_{-i} ,\tYvec_{l',\Bj}(\tTvec_i)|X_i),
         \end{eqnarray*}
    where $P_{l'}(\Bj) = \prod_{r:\Bj_{r} = 1} q_{l'_r} \times \prod_{r:\Bj_{r} = 0}(1- q_{l'_r})$,
    $\Bj_{r}$ is the $r$-th bit of the $L_i'$-bit binary representation of $j$, and
    $\tYvec_{l',\Bj}(\tTvec_i) = \left(Y_{l_{n_1}'}, Y_{l_{n_2}'},...,Y_{l_{n_M}'}\right)$,
    $n_1, n_2,...,n_M$ are the locations of '1' in the $L_i'$-bit binary representation of $j$, and
    $l_{n_1}', l_{n_2}',...,l_{n_M}'$ are the indices of the $\hY_i$'s in locations $n_1, n_2,...,n_M$ in $\tTvec_i$.
    For example, if $L_i' = 3$ and $j = 3$ then $\mbox{Bin}_3(3) = (1,0,1)$ and $M = 2$,
    $n_1 = 1, n_2 = 3$. Letting $\tTvec_i = \left(\hY_3,\hY_1,\hY_2\right)$
    then $l_1' = 3, l_2' = 1$ and $l_3' = 2$, and
                \begin{eqnarray*}
                    P_{l'}(\mbox{Bin}_3(3)) & = & q_{l_1'}(1-q_{l_2'})q_{l_3'},\\
                    \tYvec_{l',{\mbox{Bin}_3(3)}}(\tTvec_i)) & = & (Y_{l_1'}, Y_{l_3'}) = (Y_3,Y_2).
                \end{eqnarray*}

    \subsubsection{Combining the Bounds on $R'_i$}
    Applying the above scheme requires that $R'_i$ satisfies \eqref{eqn:relay_rate_constr_1} and
    \eqref{eqn:Ri'_upper_bound}:
    \begin{eqnarray*}
        q_i H(Y_i|X_i) + \eps < R'_i & < & q_i \sum_{j = 0}^{2^{L'_i}-1} P_{l'}(\Bj)I(Y_i;Y,\Xvec_{-i} ,\tYvec_{l',\Bj}(\tTvec_i)|X_i)
            + I(X_i;Y|\tZvec_i) - 4\eps,
%            &  & \qquad + \sum_{j = 0}^{L_i} P_l(\Bj)I(X_i;Y|\tXvec_i(\tZvec_i),\tYvec_{l,{\Bj}}(\tZvec_i)) - 4\eps
    \end{eqnarray*}
    which is satisfied if
    \begin{eqnarray*}
%        q_i \le \frac{\sum_{j = 1}^{L_i} P_l(\Bj)I(X_i;Y|\tXvec_i(\tZvec_i),\tYvec_{l,{\Bj}}(\tZvec_i)) - 5\eps}
%                        {H(Y_i|X_i) -\sum_{j = 1}^{L'_i} P_{l'}(\Bj)I(Y_i;Y|\tXvec_i(\tTvec_i),\tYvec_{l',\Bj}(\tTvec_i))}
        q_i & < & \frac{I(X_i;Y|\tZvec_i) - 5\eps}
                        {H(Y_i|X_i) -\sum_{j = 0}^{2^{L'_i}-1} P_{l'}(\Bj)I(Y_i;Y,\Xvec_{-i} ,\tYvec_{l',\Bj}(\tTvec_i)|X_i)}\\
        & = & \frac{I(X_i;Y|\tZvec_i) - 5\eps}
                        {H(Y_i|X_i) - I(Y_i;Y,\Xvec_{-i}|X_i) -\sum_{j = 1}^{2^{L'_i}-1} P_{l'}(\Bj)I(Y_i;\tYvec_{l',\Bj}(\tTvec_i)|\Xvec,Y)}\\
        & = & \frac{I(X_i;Y|\tZvec_i) - 5\eps}
                        {H(Y_i|\Xvec,Y) -\sum_{j = 1}^{2^{L'_i}-1} P_{l'}(\Bj)I(Y_i;\tYvec_{l',\Bj}(\tTvec_i)|\Xvec,Y)}.
    \end{eqnarray*}
    Combining with the constraint $0 \le q_i \le 1$ gives the condition in \eqref{eqn:q_i_assgn_full_thm}.

    Finally, %having set all the $q_i$'s, $i=1,2,...,N$
    the achievable rate is obtained as
    follows: using the decoded $\left\{\hyvec_i(z_{i,k-1}|s_{i,k-1})\right\}_{i=1}^N$ (assuming
    correct decoding of all $\left\{z_{i,k-1} \right\}_{i=1}^N$) the receiver decodes the source
    message $w_{k-1}$ by looking for a message $w \in \mW$ such that
    \begin{eqnarray*}
       &  &\Big(\xvec(w), \hyvec_1(z_{1,k-1}|s_{1,k-1}), \hyvec_2(z_{2,k-1}|s_{2,k-1}),...,
            , \hyvec_N(z_{N,k-1}|s_{N,k-1}),\\
       &  &\qquad \qquad       \xvec_1(s_{1,k-1}\big),\xvec_2(s_{2,k-1}\big),...,\xvec_N(s_{N,k-1}),\yvec(k-1)\Big)
        \in \styp(X,\hYvec,\Xvec,Y),
    \end{eqnarray*}
    where $\hYvec = \left(\hY_1, \hY_2,...,\hY_N\right)$.
    This results in an achievable rate of
    \[
        R \le I(X;Y,\hYvec,\Xvec) = I(X;Y,\hYvec|\Xvec).
    \]
    Plugging in the assignments of all the $\hY_i$'s, we get the following explicit rate expression:
    \begin{eqnarray*}
            I(X;Y,\hYvec|\Xvec) & = & I(X;Y|\Xvec) + I(X;\hYvec|\Xvec,Y)\\
            & = & I(X;Y|\Xvec) + H(X|\Xvec,Y)  -  H(X|\Xvec,Y,\hYvec)\\
            & = & I(X;Y|\Xvec) + H(X|\Xvec,Y)  -  (1-q_1) H(X|\Xvec,Y,\hYvec_2^N) - q_1 H(X|\Xvec,Y,\hYvec_2^N,Y_1)\\
            & = & I(X;Y|\Xvec) +  (1-q_1) I(X;\hYvec_2^N|\Xvec,Y) + q_1 I(X;\hYvec_2^N,Y_1|\Xvec,Y)\\
            & ... &\\
            & = & I(X;Y|\Xvec) + \sum_{\theta = 1}^{2^N-1} P(\Bt)I(X;\Yvec_{\Bt}|\Xvec,Y).
        \end{eqnarray*}
\tend

\subsection{Discussion}
To demonstrate the usefulness of the explicit EAF-based achievable rate of theorem \ref{thm:achieve_N_result_2} we
compare it with the DAF-based method of
\cite[theorem 3.1]{XieKumar:2005} for the two-relay case.
For this scenario there are five possible DAF setups, and the maximum of the five resulting rates is taken as the
DAF-based rate:
\begin{eqnarray*}
    R^{DAF} & = & \sup_{p(x,x_1,x_2)} \max \left\{R_1, R_2, R_{12}, R_{21}, R_G \right\}\\
    R_1 & = & \max_{x_2\in \mX_2} \min\left\{I(X; Y_1|X_1,x_2), I(X; Y|X_1, x_2) + I(X_1; Y|x_2) \right\}\\
    R_2 & = & \max_{x_1\in \mX_1} \min\left\{I(X; Y_2|X_2,x_1), I(X; Y|X_2, x_1) + I(X_2; Y|x_1) \right\}\\
    R_{12} & = & \min\left\{I(X; Y_1|X_1, X_2), I(X; Y_2|X_1, X_2) + I(X_1; Y_2|X_2), I(X; Y|X_1, X_2)+ I(X_1; Y|X_2) + I(X_2; Y)  \right\}\\
    R_{21} & = & \min\left\{I(X; Y_2|X_1, X_2), I(X; Y_1|X_1, X_2) + I(X_2; Y_1|X_1), I(X; Y|X_1, X_2)+ I(X_2; Y|X_1) + I(X_1; Y)  \right\}\\
    R_G & = & \min \left\{I(X; Y_1|X_1, X_2), I(X; Y_2|X_1, X_2), I(X,X_1,X_2; Y) \right\},
\end{eqnarray*}
where $R_1$ is the rate obtained when only relay 1 is active, $R_2$ is the rate obtained when only relay 2 is active,
$R_{12}$ is the rate obtained when relay 1 decodes first and relay 2 decodes second and $R_{21}$ is
the rate obtained when this order is reversed.
 $R_G$ is the rate obtained when both relays form one group\footnote{In fact, since we take the supremum over all p.d.f.'s
 $p(x,x_1,x_2)$ we do not need to explicitly include $R_1$ and $R_2$ in the maximization, but
 it is included here to provide a complete presentation.}.
Now, as in the single-relay case, DAF is limited by the worst source-relay link. Therefore, if
\begin{equation}
    \label{eqn:DAF_inequality}
     R^{PTP} >
        \max_{p(x|x_1,x_2), (x_1,x_2) \in \mX_1 \times \mX_2} \big\{I(X;Y_1|x_1,x_2), I(X;Y_2|x_1,x_2)\big\},
\end{equation}
where $  R^{PTP} = \max_{p(x|x_1,x_2), (x_1,x_2) \in \mX_1 \times \mX_2} I(X; Y|x_1,x_2)$
is the point-to-point rate,
then it is better not to use \cite[theorem 3.1]{XieKumar:2005} at all, but rather set the relays to transmit
the symbol pair $(x_1,x_2) \in \mX_1 \times \mX_2$ such that the point-to-point rate is maximized.
However, the rate obtained using corollary \ref{corr:achieve_N_result_1} for the two-relay case is given by
\begin{eqnarray*}
    R^{TS-EAF} & \le &  \sup_{p(x)p(x_1)p(x_2)} I(X; Y|X_1,X_2) + q_1(1-q_2) I(X;Y_1|X_1,X_2,Y)  \\
    &  & \phantom{xxxxxxxxxxxxxxx}  +(1-q_1)q_2I(X;Y_2|X_1,X_2,Y) + q_1q_2I(X;Y_1,Y_2|X_1,X_2,Y),
\end{eqnarray*}
where $q_1$ and $q_2$ are positive and determined according to \eqref{eqn:q_i_assgn_simple}.
This expression can, in general be greater than
%$\max_{p(x|x_1,x_2), (x_1,x_2) \in \mX_1 \times \mX_2} I(X; Y|x_1,x_2)$
$R^{PTP}$
even when
\eqref{eqn:DAF_inequality} holds, for channels where the relay to destination links are very good.
 Hence, this explicit achievable expression provides an easy way to improve upon the
DAF-based achievable rates when the source-to-relay links are very noisy.

To demonstrate this, consider the channel given in table \ref{table:channel_for_example} over binary RVs
$X$, $X_1$, $X_2$, $Y$, $Y_1$ and $Y_2$. The channel
\begin{table}
       \caption{$p(y,y_1,y_2|x,x_1,x_2)$ for the EAF example.}
        \label{table:channel_for_example}
       \begin{tabular}[h!]{|c||c|c|c|c|c|c|c|c|}
               \hline
               $(x,x_1,x_2)$& \multicolumn{7}{c}{$p(y,y_1,y_2|x,x_1,x_2)$} &\\
               \cline{2-9}
               & 000 & 001 & 010 & 011 & 100  & 101 & 110 & 111\\
              \hline \hline
             000 &  8.047314e-2  &  1.948360e-1 &   2.041506e-1  &  4.523933e-2 &
                 2.423322e-1  &  7.057734e-3 &   1.310053e-1  &  9.490483e-2\\

             001 &  8.601616e-1  &  6.643713e-2  &  1.662897e-2  &  1.937227e-2 &
                 1.859104e-2  &  1.741020e-2 &   8.833169e-4  &  5.154431e-4 \\

             010 &  3.131504e-1  &  1.821840e-1  &  5.618147e-2  &  1.522841e-1 &
                 5.290856e-2  &  1.555570e-1 &   3.214581e-2  &  5.558854e-2  \\

             011 &  5.183921e-3  &  3.704625e-1  &  1.641795e-2  &  2.208356e-1 &
                 1.660775e-3  &  2.355928e-1  &  9.590170e-4  &  1.488874e-1 \\

             100 &  8.116746e-3  &  8.139504e-3  &  9.387860e-2  &  1.736515e-2 &
                 1.039350e-1  &  7.308714e-3  &  7.612555e-1  &  7.612563e-7\\

             101 &  4.824126e-2  &  1.196128e-1  &  1.705739e-1  &  7.127199e-2 &
                 4.631349e-2  &  1.955324e-1  &  1.928693e-1  &  1.555848e-1\\

             110 &  9.367321e-2  &  1.248830e-1  &  1.873302e-1  &  6.161358e-2 &
                 5.827773e-2  &  1.906660e-1  &  1.589616e-1  &  1.245946e-1\\

             111 &  9.141272e-7  &  9.141263e-1  &  7.618061e-3  &  3.435473e-2 &
                 7.974830e-4  &  4.117531e-2  &  9.302643e-4  &  9.969457e-4\\
           \hline
       \end{tabular}
\end{table}
distribution was constructed under the independence constraint
\[
    p(y,y_1,y_2|x,x_1,x_2) = p(y_1|x,x_1,x_2) p(y_2|x,x_1,x_2) p(y|x,x_1,x_2,y_1,y_2),
\]
i.e. given the channel inputs, the two relay outputs are independent.
This channel is characterized by noisy source-relay links, while
the link from relay $1$ to the destination has low noise. Therefore, DAF is inferior to the point-to-point
transmission but EAF is able to exceed this rate, by giving up a small amount of rate on the direct link (compared
to the point-to-point rate) and gaining more rate through the relays. The numerical evaluation of the
rates for this channel produces\footnote{The resulting rates were obtained by optimizing for the rates with
random initial input distributions. The optimization was repeated $50$ times for each rate and the maximum resulting rate
was recorded. The m-files used for this evaluation are available at {\tt http://cn.ece.cornell.edu}.}
\begin{eqnarray*}
R^{PTP}  & = & 0.2860323,\\
R^{DAF} & = & 0.2408629,\\
R^{TS-EAF} & = & 0.2924798,
\end{eqnarray*}
where the optimal distributions that achieve these rates are summarized in tables \ref{table:opt_DAF_dist} and
\ref{table:opt_EAF_dist}.
\begin{table}
\centering
\begin{minipage}{5cm}
\centering
    \caption{Optimal distribution for DAF}
    \label{table:opt_DAF_dist}
    \vspace{-0.2cm}
       \begin{tabular}[!h]{|c||c|}
               \hline
               $(x,x_1,x_2)$ & $p(x,x_1,x_2)$ \\
                \hline \hline
                000 & 5.698189907239905e-009\\
                001 & 5.259061814752764e-017\\
                010 & 4.301809992760095e-009\\
                011 & 4.424193267301109e-001\\
                100 & 6.792096128437060e-009\\
                101 & 4.740938235494830e-017\\
                110 & 3.207903771562940e-009\\
                111 & 5.575806532698892e-001\\
               \hline
       \end{tabular}
\end{minipage}
\phantom{xxxxxxxxx}
\begin{minipage}{5cm}
    \centering
    \caption{Optimal distribution for EAF}
    \label{table:opt_EAF_dist}
        \vspace{-0.2cm}
       \begin{tabular}[!h]{|c|}
               \hline
            $\Pr(X = 0)  = 4.3752093552645e-001$\\
%            \hline
            $\Pr(X_1 = 0) =1.9388669163312e-001 $\\
%            \hline
            $\Pr(X_2 = 0) = 1.000000000000000e-009$\\
            \hline
       \end{tabular}
\end{minipage}

\end{table}
The optimal DAF distribution fixes both $X_1$ and $X_2$ to $'1'$ and sets the probability of $X$ to be
$\Pr(X = 1) = 0.442419$, as expected for the case where the relays limit the achievable rate. For the EAF, the
useless relay $2$ is fixed to $0$, to facilitate transmission with the useful relay $1$. In accordance, we
obtain time sharing proportions of $q_1 = 0.156947$ and $q_2 \approx 0$ for relay $1$ and relay $2$ respectively.
We note that in this scenario, we actually have that even the single-relay TS-EAF outperforms the two-relay DAF.

%%%%%%%%%%%%%%%%%%%%%%%%%%%%%%%%%%%%%%%%%%%%%%%%%%%%%%%%%%%%%%%%%%%%%%%%%%%%%%%%%%%%%%%
%%%%%%%%%%%%%%%%%%%%%%%%%%%%%%%%%%%%%%%%%%%%%%%%%%%%%%%%%%%%%%%%%%%%%%%%%%%%%%%%%%%%%%%
%%%%%%%%%%%%%%%%%%%%%%%%%%%%%%%%%%%%%%%%%%%%%%%%%%%%%%%%%%%%%%%%%%%%%%%%%%%%%%%%%%%%%%%
%%%%%%%%%%%%%%%%%%%%%%%%%%%%%%%%%%%%%%%%%%%%%%%%%%%%%%%%%%%%%%%%%%%%%%%%%%%%%%%%%%%%%%%
%%%%%%%%%%%%%%%%%%%%%%%%%%%%%%%%%%%%%%%%%%%%%%%%%%%%%%%%%%%%%%%%%%%%%%%%%%%%%%%%%%%%%%%
%%%%%%%%%%%%%%%%%%%%%%%%%%%%%%%%%%%%%%%%%%%%%%%%%%%%%%%%%%%%%%%%%%%%%%%%%%%%%%%%%%%%%%%
%%%%%%%%%%%%%%%%%%%%%%%%%%%%%%%%%%%%%%%%%%%%%%%%%%%%%%%%%%%%%%%%%%%%%%%%%%%%%%%%%%%%%%%

\section{The Gaussian Relay Channel}
\label{sec:Gauss_relay}

In this section we investigate the application of estimate-and-forward with time-sharing
to the Gaussian relay channel. For this channel, the common practice it to use Gaussian codebooks and
Gaussian quantization at the relay. The rate in Gaussian scenarios where coded modulation is applied, is usually
analyzed by applying DAF at the relay. In this section we show that when considering coded modulation, one should select the
relay strategy according to the channel condition: Gaussian selection seems a good choice when the SNR at the relay
is low and DAF appears to be superior when the relay enjoys high SNR conditions. However, for
intermediate SNR there is much room
for optimizing the estimation mapping at the relay.

\begin{figure}[ht]
     \epsfxsize=0.6\textwidth \leavevmode\centering\epsffile{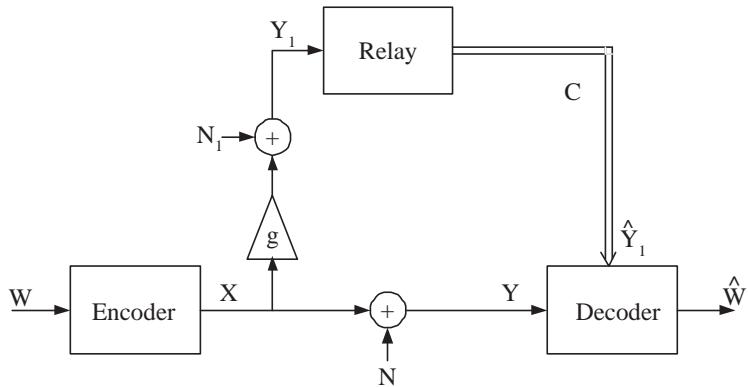}
    \caption{The Gaussian relay channel with a finite capacity noiseless relay link between the relay and the
    destination.}
    \label{fig:Gauss_relay}
\end{figure}

In the following we first recall the Gaussian relay channel with a Gaussian codebook, and
then we consider the Gaussian relay channel under BPSK modulation constraint. Since we focus on the mapping
at the relay we consider here the Gaussian relay channel with an orthogonal relay of finite
capacity $C$, also considered in
\cite{Goldsmith:2006}. This scenario is depicted in figure \ref{fig:Gauss_relay}.

Here $Y_1 = g \cdot X + N_1$ is the channel output at the relay, $Y = X + N$ is the channel output at the receiver, which decodes
the message based on $(Y^n, \hY_1^n)$. Let $\mW = \left\{1,2,...,2^{nR}\right\}$ denote the source message set, and let
the source have an average power constraint $P$:
\[
    \frac{1}{n}\sum_{i=1}^n x_i(w) \le P, \qquad \forall w \in \mW.
\]
The relay signal $\hY_1^n$ is transmitted to the destination through a finite-capacity noiseless link of
capacity $C$. For this scenario the expressions of \cite[theorem 6]{CoverG:79} specialize to
\begin{subequations}
\begin{eqnarray}
    \label{eqn:rate_Gauss}
    R & \le & I(X;Y,\hY_1)\\
    \label{eqn:constraint_Gauss}
    \mbox{subject to } C & \ge & I(\hY_1;Y_1|Y),
\end{eqnarray}
\end{subequations}
with the Markov chain $X,Y - Y_1 - \hY_1$.

We also consider in this section the DAF method whose information rate is given by (see \cite[theorem 1]{CoverG:79})
\[
    R_{DAF} = \min \left\{I(X;Y_1), I(X;Y)+C\right\},
\]
and the upper bound of \cite[theorem 3]{CoverG:79}:
\[
    R_{upper} = \min\left\{I(X;Y)+C, I(X;Y,Y_1)\right\}.
\]
We note that although these expressions were derived for the finite, discrete alphabets case, following the argument
in \cite[remark 30]{Kramer:2005}, they also hold for the Gaussian case.

\subsection{The Gaussian Relay Channel with Gaussian Codebooks}
When $X \sim \mN(0,P)$, i.i.d., then the channel outputs at the relay and the receiver are jointly Normal RVs:
\[
    \left(
        \begin{array}{c}
        y\\
        y_1
        \end{array} \right) \sim \mN\left( \left( \begin{array}{c}
                                                        0\\ 0 \end{array}
                                                        \right) , \left( \begin{array}{cc}
                                                                            P + \sigD & gP \\
                                                                                gP    & g^2P + \sigR \end{array} \right) \right).
\]
The compression is achieved by adding to $Y_1$ a zero mean independent Gaussian RV, $N_Q$:
\begin{equation}
    \label{eqn:def_qaussian_quant}
    \hY_1 = Y_1 + N_Q, \qquad N_Q \sim \mN(0, \sigQ).
\end{equation}
We refer to the assignment \eqref{eqn:def_qaussian_quant} as Gaussian-quantization estimate-and-forward (GQ-EAF).
Evaluating the expressions \eqref{eqn:rate_Gauss} and \eqref{eqn:constraint_Gauss}
with assignment \eqref{eqn:def_qaussian_quant} results in  (see also \cite{Goldsmith:2006}):
%\begin{eqnarray*}
%    I(X;Y)  & = & \log(1+P) +
%    I(Y_1;\hY_1|Y)  & = & \log\left(1 + \frac{1}{\sigQ} + \frac{gP}{\sigQ(P+1)}  \right)\\
%    I(X;\hY_1|Y)    & = & \log\left(1 + \frac{gP}{(1 + \sigQ)(P+1)}  \right)\\
%\end{eqnarray*}
\begin{subequations}
\begin{eqnarray}
    \label{eqn:rate_Gauss_evaluated}
    I(X;Y, \hY_1)   & = & \frac{1}{2}\log_2\left(1+P + \frac{gP}{1 + \sigQ}  \right)\\
        \label{eqn:constraint_Gauss_evaluated}
    I(Y_1;\hY_1|Y)  & = & \frac{1}{2}\log_2\left(1  + \frac{1+P+gP}{\sigQ(P+1)}  \right).
\end{eqnarray}
\end{subequations}
The feasibility condition \eqref{eqn:constraint_Gauss} yields
\[
    \sigQ\ge  \frac{1+P+gP}{(2^{2C}  - 1)(P+1)},
\]
and because maximizing the rate \eqref{eqn:rate_Gauss_evaluated} requires minimizing $\sigQ$, the resulting
GQ-EAF rate expression is
\[
    R  \le \frac{1}{2} \log_2\left(1+P + \frac{gP}{1 + \frac{1+P+gP}{(2^{2C}  - 1)(P+1)}}  \right).
\]
Now, when using Gaussian quantization at the relay
%(as it is the most efficient way to compress $Y_1$, the Gaussian channel output at the relay {\Huge reference} ),
it is
obvious that time sharing does not help: we need the minimum $\sigQ$ in order to maximize
the rate. This minimum is
obtained only when the entire capacity of the relay link is dedicated to the transmission of the (minimally)
quantized $Y_1$.
However, when we consider the Gaussian relay channel with coded modulation, the situation is quite different, as
we show in the remaining of this section.

\subsection{The Gaussian Relay Channel with Coded Modulation}
Consider the Gaussian relay channel where $X$ is an equiprobable BPSK signal of amplitude $\sqrt{P}$:
\begin{equation}
    \label{eqn:def_PX}
    \Pr(X = \sqrt{P}) = \Pr(X = -\sqrt{P}) = \frac{1}{2}.
\end{equation}
Under these conditions, the received symbols $(Y,Y_1)$ are no longer jointly Gaussian, but follow a Gaussian-mixture
distribution:
\begin{eqnarray*}
    f(y,y_1) & = & \Pr(X = \sqrt{P})f(y,y_1|x = \sqrt{P}) + \Pr(X = -\sqrt{P})f(y,y_1|x = -\sqrt{P}) \\
             & = & \frac{1}{2}\left(G_y(\sqrt{P},\sigD)G_{y_1}(g\sqrt{P} , \sigR) + G_y(-\sqrt{P},\sigD)G_{y_1}(-g\sqrt{P} , \sigR)\right),
\end{eqnarray*}
where
\begin{equation}
    \label{eqn:def_G}
    G_x(a,b) \triangleq \frac{1}{\sqrt{2 \pi b}} e ^{-\frac{(x-a)^2}{2 b} }.
\end{equation}
Contrary to the Gaussian codebook case, where it is hard to identify a mapping $p(\hy_1|y_1)$ that will be superior to
 Gaussian quantization (if indeed such a mapping exists), in this case it is a natural question to compare the
Gaussian mapping of \eqref{eqn:def_qaussian_quant}, which induces a Gaussian-mixture distribution for $\hY_1$
with other possible mappings. In the case of binary inputs it is natural to consider binary mappings
for $\hY_1$. We can predict that such mappings will do well at high SNR on the source-relay link,
when the probability of error for symbol-by-symbol detection at the relay is small, with a much smaller
complexity than Gaussian quantization. We start by considering
 two types of  hard-decision (HD) mappings:
\begin{enumerate}
    \item The first mapping is HD-EAF: The relay first makes a hard decision about every received $Y_1$ symbol,
    determining whether it
    is positive or negative, and then randomly decides if it is going to transmit this decision or transmit
    an erasure symbol $E$ instead. The probability of transmitting an erasure, $1 - P_{\ners}$, is used to adjust the conference
    rate such that the feasibility constraint is satisfied. Therefore, the conditional distribution $p(\hY_1|Y_1)$ is
    given by:
    \begin{subequations}
        \begin{eqnarray}
            \label{eqn:def_p_hy1_given_y1_HD_eq1}
            p(\hY_1|Y_1 > 0) & = & \left\{
                        \begin{array}{cl}
                            P_{\ners} &, 1\\
                            1 - P_{\ners} &, E
                        \end{array}
                    \right.\\
            \label{eqn:def_p_hy1_given_y1_HD_eq2}
            p(\hY_1|Y_1 \le 0) & = & \left\{
                        \begin{array}{cl}
                            P_{\ners} &, -1\\
                            1 - P_{\ners} &, E
                        \end{array}
                    \right..
        \end{eqnarray}
    \end{subequations}
    This choice is motivated by the time-sharing method considered
    in section \ref{sec:timeshare_single}: after making a hard decision on the received symbol's sign --- positive
    or negative, the relay applies TS to that decision so that the rate required to transmit the resulting random variable
    is less than $C$. This facilitates transmission to the destination through the conference link.
    Since the entropy of the sign decision is $1$, then when $C \ge 1$ we can transmit the sign decisions directly without using
    an erasure. Therefore,
    we expect that for values of $C$ in the range $C > 1$, this mapping
    will not exceed the rate obtained for $C=1$. The focus is, therefore, on values of $C$ that are less than $1$.
    The expressions for this assignment are given in appendix \ref{append:Gauss-deriv-HD-EAF}.

    \item The second method is deterministic hard-decision. In this approach, we select a threshold $T$ such that the
    range of $Y_1$ is partitioned into three regions: $Y_1 < -T, -T \le Y_1 \le T, Y_1 > T$. Then, according to the
    value of each received $Y_1$ symbol, the corresponding $\hY_1$ is deterministically determined:
    \begin{eqnarray}
        \hY_1 = \left\{
            \begin{array}{cl}
                1, & Y_1 > T\\
                E, & -T \le Y_1 \le T\\
                -1, & Y_1 < -T
            \end{array}
        \right..
    \end{eqnarray}
    The threshold $T$ is selected such that the achievable rate is maximized subject to satisfying
    the feasibility constraint. We refer to this method as deterministic HD (DHD). Therefore, this is
    another type of TS in which  the erasure probability is determined by the fraction of the time
    the relay input is between $-T$ to $T$.
    This method should be better than HD-EAF at high relay SNR since for HD-EAF, erasure is selected without
    any regard to the quality of the decision - both good sign decisions and bad sign decisions are
    erased with the same probability. However in DHD, the erased area is the area where the decisions have
    low quality in the first place and all high quality decisions are sent. However, at low relay SNR and
    small capacity for the relay-destination link, HD-EAF may perform better than DHD since the
    erased area  (i.e. the region between $-T$ to $+T$) for the DHD mapping has to be very large
    to allow 'squeezing' the estimate through the relay link,
    while HD-EAF may require less compression of the HD output.
    The expressions for evaluating the rate of the DHD assignment are given in appendix \ref{sec:expressions_DHD}.
\end{enumerate}

\begin{figure}[h]
    \centering
    \scalebox{0.7}{\includegraphics{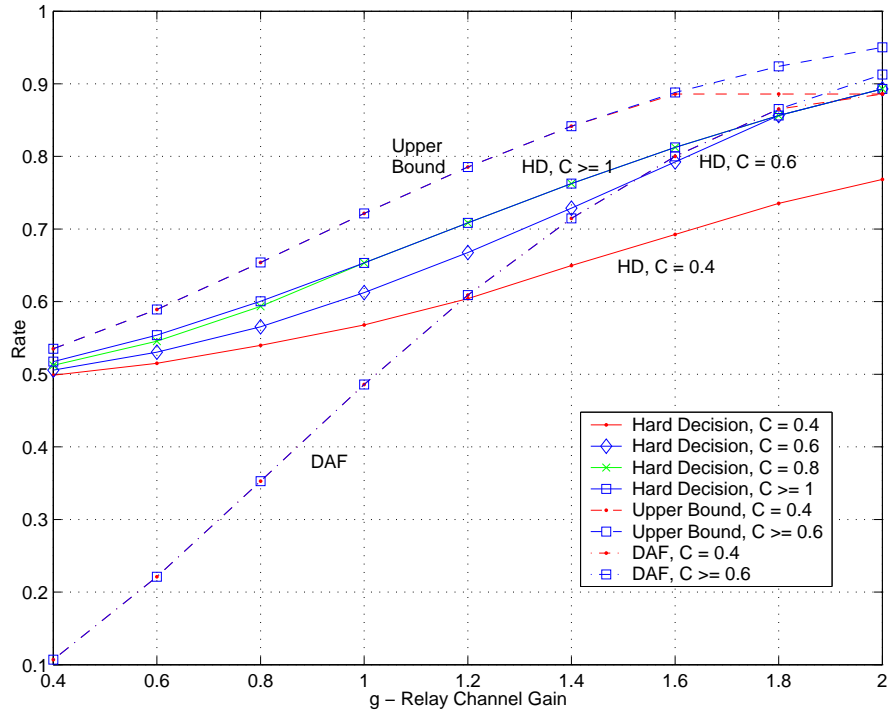}}
    \caption{Information rate with BPSK and hard decision EAF mapping at the relay vs. relay channel gain $g$,
        for different values of $C$.}
    \label{fig:hard-decision-vs-g}
    \vspace{-0.2cm}
\end{figure}

We now examine the performance of each technique using numerical evaluation:
first, we examine the achievable rates with HD-EAF. The expressions are evaluated for $\sigR = \sigD = 1$ and
$P = 1$. For every pair of values $(g,C)$ considered, the maximum $P_{\ners}$ was selected. Figure \ref{fig:hard-decision-vs-g}
depicts the achievable rate vs. $g$ for $ 0.4 \le C \le 2$, together with the upper bound and the decode-and-forward rate.
As can be observed from figure \ref{fig:hard-decision-vs-g}, the information rate of HD-EAF increases with $C$
until $C = 1$ and then remains constant.
It is also seen that for small values of $g$, HD-EAF is better than DAF. This region of $g$  increases with $C$,
and for $C \ge 1$ the crossover value of $g$
is approximately $1.71$. However, even for $g = 2$, DAF is only $2.5\%$ better than HD-EAF.

\begin{figure}[!h]
    \centering
    \scalebox{0.69}{\includegraphics{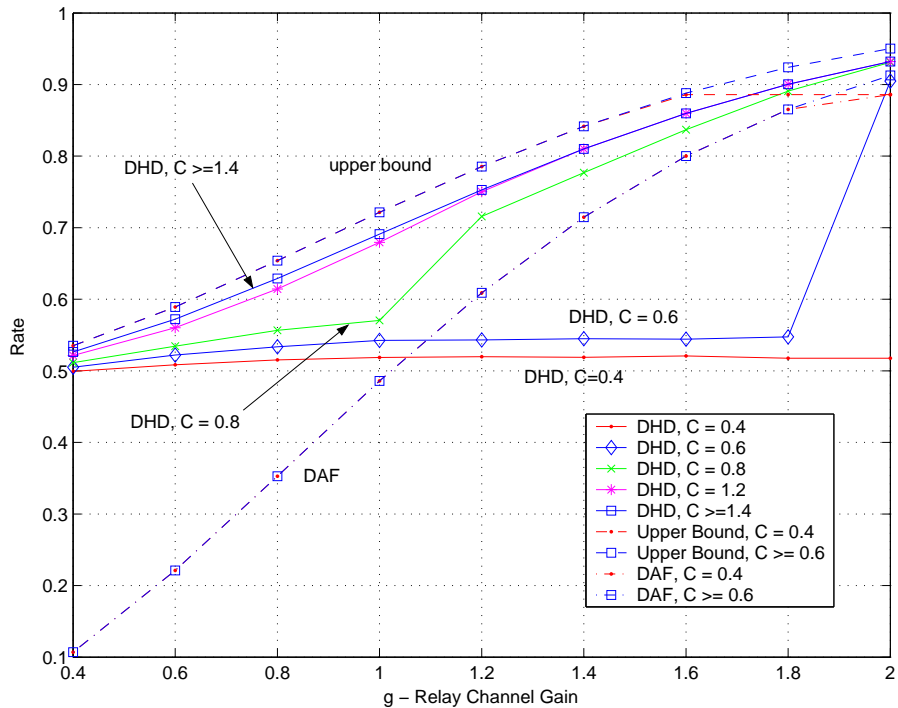}}
    \caption{Information rate with BPSK, for deterministic hard decision at the relay vs. relay channel gain $g$, for
    different values of $C$.}
    \label{fig:DHD-vs-g}
\end{figure}

Next, examine DHD: as can be seen from figure \ref{fig:DHD-vs-g}, for small values of $C$, DAF exceeds
the information rate of DHD for values of $g$ greater than $1$, but for $C \ge 0.8$, DHD is superior to
DAF, and in fact DAF approaches DHD from below. Another phenomena obvious from the
figure (esp. for $C = 0.8$), is the existence of a threshold: for low values of $C$ there is some $g$ at which the DHD rate
exhibits a jump.
\begin{figure}
    \centering
    \scalebox{0.69}{\includegraphics{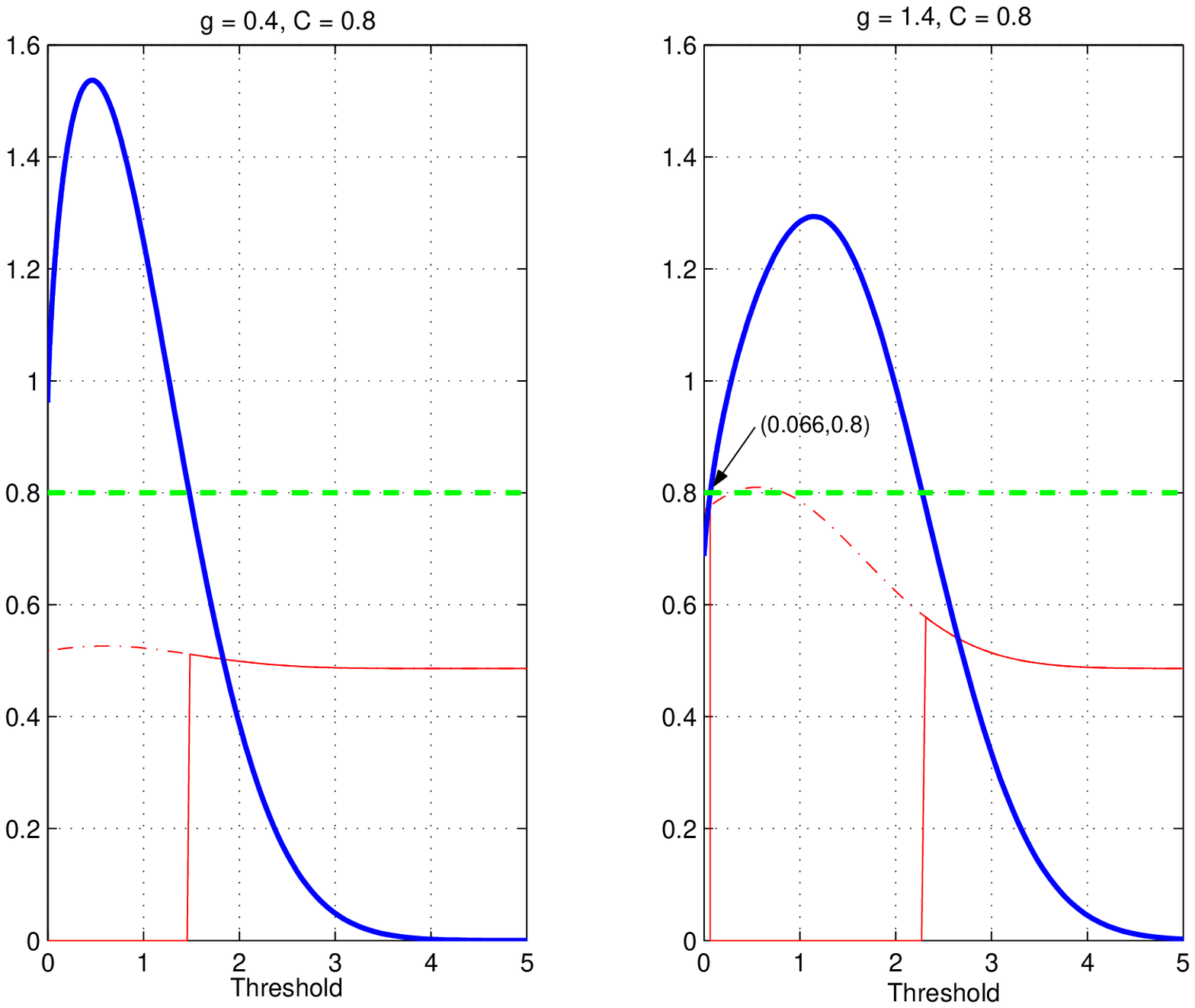}}
    \caption{$I(\hY_1;Y_1|Y)$ and $I(X;\hY_1,Y)$ vs. Threshold $T$ for $(g,C) = (0.4,0.8)$ (left) and
        $(g,C) = (1.4,0.8)$ (right). The bold solid line represents $I(\hY_1,Y_1|Y)$, the bold dashed line represents $C = 0.8$,
        $I(X;Y,\hY_1)$ is represented by the dash-dot line and the resulting information rate is depicted with the solid line. }
    \label{fig:DHD-Explanation}
\end{figure}
This can be explained by looking at figure \ref{fig:DHD-Explanation}, which depicts
the values of $I(X;\hY_1,Y)$ and $I(\hY_1;Y_1|Y)$ vs. the threshold $T$: the bold-solid graph of
$I(\hY_1;Y_1|Y)$ can intersect the bold-dashed horizontal line representing $C$ at two values of $T$. We also note that
for small $T$ the value of $I(X;\hY_1,Y)$ is generally greater than for large $T$. Now, the jump can be explained as follows: as
shown in appendix \ref{sec:HDH-Explanation}, for small $T$ and $g$, $I(\hY_1;Y_1|Y)$ is bounded from below.
Now, if this bound value is greater than $C$ then the intersection will occur only at a large value of $T$, hence
the small rate. When $g$ increases, the value of $I(\hY_1;Y_1|Y)$ for small $T$ decreases accordingly, until
at some $g$ it intersects $C$ for a small $T$ as well as for a large $T$, as indicated by the arrow in the
right-hand part
of figure \ref{fig:DHD-Explanation}. This allows us to obtain the
rates in the region of small $T$ which are in general higher than the rates for large $T$ and this
is the source of the jump in the achievable rate.

\FloatBarrier

\subsection{Time-Sharing Deterministic Hard-Decision (TS-DHD)}
It is clearly evident from the above numerical evaluation that none of the two mappings, HD-EAF and DHD, is universally better than
the other: when $g$ is small and $C$ is less than $1$, then HD-EAF performs better than DHD, since the erased region is too large,
and when $g$ increases, DHD performs better than HD-EAF since it erases only the low quality information. It is therefore natural to consider
a third mapping which combines both aspects of binary mapping at the relay, namely deterministically erasing low quality information and
then randomly gating the resulting discrete variable in order to allow its transmission over the conference link.
This hybrid mapping is given in the following equation:
    \begin{subequations}
    \label{eqn:def_TS-DHD}
        \begin{eqnarray}
            \label{eqn:def_TS-DHD_eq1}
            p(\hY_1|Y_1 > T) & = & \left\{
                        \begin{array}{cl}
                            P_{\ners} &, 1\\
                            1 - P_{\ners} &, E
                        \end{array}
                    \right.\\
            \label{eqn:def_TS-DHD_eq2}
            p(\hY_1 = E \;|\;|Y_1| \le T) & = & 1\\
            \label{eqn:def_TS-DHD_eq3}
            p(\hY_1|Y_1 < -T) & = & \left\{
                        \begin{array}{cl}
                            P_{\ners} &, -1\\
                            1 - P_{\ners} &, E
                        \end{array}
                    \right..
        \end{eqnarray}
    \end{subequations}
In this mapping, the region $|Y_1| \le T$ is always erased, and the complement region is erased with probability $P_{\ers} = 1- P_{\ners}$.
Of course, now both $T$ and $P_{\ers}$ have to be optimized. The expressions for TS-DHD can be found in appendix \ref{appndx:expressions_TS_DHD}.
Figure \ref{fig:compare_HD-EAF_DHD_TS-DHD} compares the performance of
DHD, HD-EAF and TS-DHD. As can be seen, the hybrid method enjoys the benefits of both types of mappings and is the superior method.
\begin{figure}[!h]
    \centering
    \scalebox{0.69}{\includegraphics{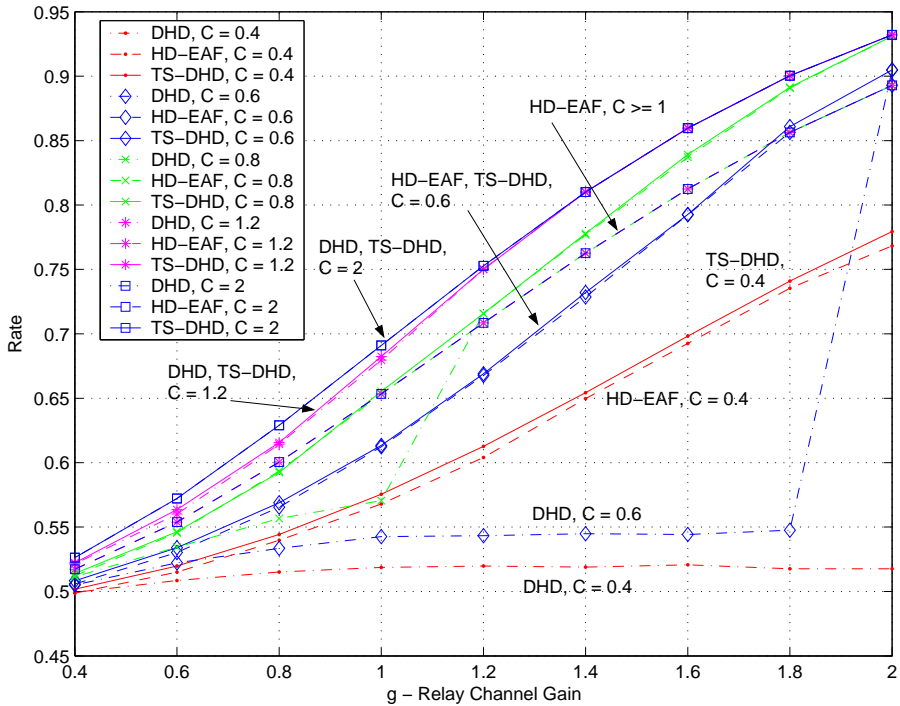}}
    \caption{Information rate with BPSK, for  HD-EAF, DHD and TS-DHD at the relay vs. relay channel gain $g$, for
    different values of $C$.}
    \label{fig:compare_HD-EAF_DHD_TS-DHD}
\end{figure}

\begin{figure}[!h]
    \centering
    \scalebox{0.67}{\includegraphics{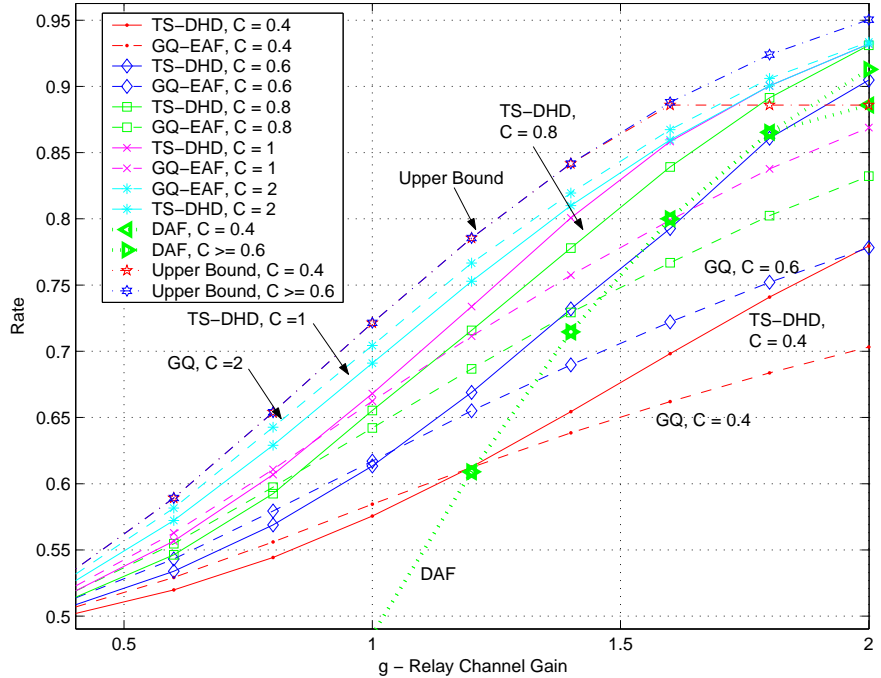}}
    \caption{Information rate with BPSK, for DAF,  TS-DHD and GQ-EAF at the relay vs. relay channel gain $g$, for
    different values of $C$.}
    \label{fig:compare-HD-EAF-GQ-EAF}
\end{figure}

Next, figure \ref{fig:compare-HD-EAF-GQ-EAF} compares the performance of TS-DHD, GQ-EAF, and DAF.
As can be seen from the figure, Gaussian quantization is not always the optimal choice: for $C = 0.6$ (the lines with
diamond-shaped markers) we have that
GQ-EAF is the best method for $g < 1.05$, for $1.05 < g < 1.55$ TS-DHD is the best method and for $g>1.55$
DAF achieves the highest rate.
For $C = 1$ (x-shaped markers) TS-DHD is superior to both GQ-EAF and DAF for $g > 0.9$ and  for $C = 2$, GQ-EAF is the superior method for all $g \le 2$.
This suggests that for the practical Gaussian relay scenario, where the modulation constraint is taken into account, there is
room to optimize the mapping at the relay since the choice of Gaussian quantization is not always optimal.

\begin{figure}[!h]
    \centering
    \scalebox{0.67}{\includegraphics{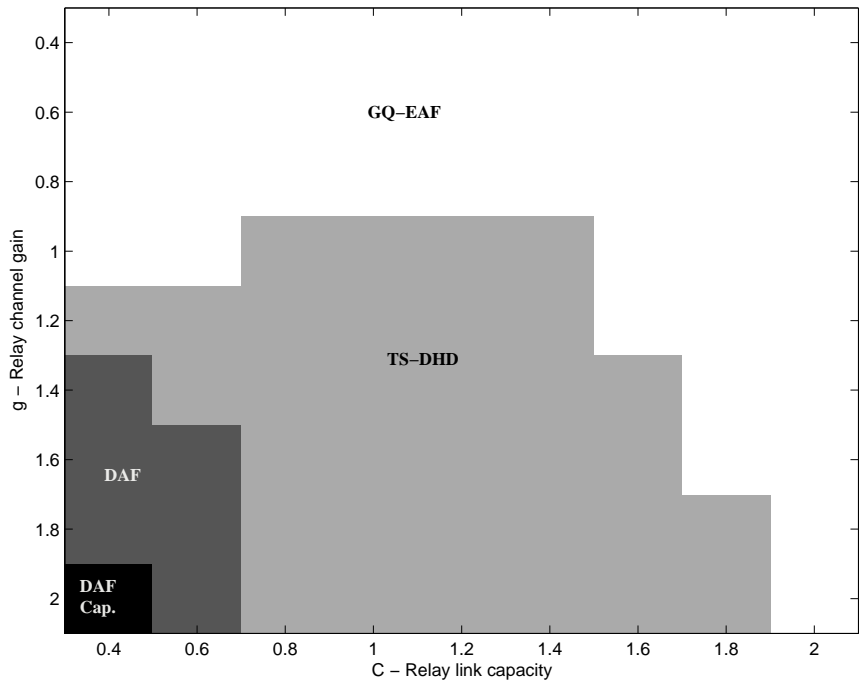}}
    \caption{The best cooperation strategy (out of DAF, TS-DHD and GQ-EAF)
        for the Gaussian relay channel with BPSK transmission.}
    \label{fig:DAF-EAF-Regions}
\end{figure}

Lastly, figure \ref{fig:DAF-EAF-Regions} depicts the regions in the g-C plane in which each of the methods considered here is superior,
in a similar manner to  \cite[figure 2]{Goldsmith:2006}\footnote{The block shapes are due to the step-size of $0.2$ in the values of $g$ and $C$ used
for evaluating the rates. In the final version we will present an evaluation over a finer grid (such an evaluation
requires several weeks to complete).}.
As can be observed from the figure, in the noisy region of small $g$ and also in the region of very large $C$,
GQ-EAF is superior, and in the
strong relay region of medium-to-high $g$ and medium-to-high $C$, TS-DHD is the superior method.
 DAF is superior small $C$ and high $g$.
%In the transition region where $C \in [0.6,1.8]$ and $g \in [1,2]$, the two hard decision methods, HD-EAF and DHD are
%superior.
In a sense, the TS-DHD method is a hybrid method between the DAF which makes a hard-decision on the
entire block and GQ-EAF which makes a soft decision every symbol, therefore it is superior in the transition region
between the region where DAF is distinctly better, and the region where GQ-EAF is distinctly superior.

\FloatBarrier

\subsection{When the SNR on the Direct Link Approaches $0$ ($\sigD \rightarrow \infty$)}
In this subsection we analyze the relaying strategies discussed in this section as the SNR on the direct link $ X - Y $
approaches zero. Because TS-DHD is a hybrid method combining
both DHD and HD-EAF, we analyze the behavior of the components rather than the hybrid, to gain more insight.
This analysis is particularly useful when trying to numerically evaluate the rates, since as the direct-link SNR goes to zero,
   the computer's numerical accuracy does not allow to numerically obtain the rates using the general expressions.
%  Therefore, in order to examine the behavior
%of the three EAF relay mappings considered in this section at low SNR on the direct link, it is required to derive analytical approximations
%to the rate expressions.

First we note that
when the SNR of the direct link $ X - Y $ approaches $0$ we have that $I(X;Y) \rightarrow 0$ as well.
To see this we write
\begin{eqnarray*}
    I(X;Y)  & = & h(Y) - h(Y|X)\\
            & = & h(Y) - h(X + N|X)\\
            & = & h(Y) - h(N),
\end{eqnarray*}
with $h(Y) = -\int_{-\infty}^{\infty}f(y) \log_2(f(y)) dy$, and from \eqref{eqn:f_Y_HC}
\begin{eqnarray*}
    f(Y)& = & \frac{1}{2}\left(G_y(\sqrt{P},\sigD) + G_y(-\sqrt{P},\sigD)\right)\\
        & = & \frac{1}{2}\left( \frac{1}{\sqrt{2 \pi \sigD}}e^{-\frac{(y-\sqrt{P})^2}{2\sigD}}
            +\frac{1}{\sqrt{2 \pi \sigD}} e^{-\frac{(y+\sqrt{P})^2}{2\sigD}}\right)\\
        & = & \frac{1}{\sqrt{2 \pi \sigD}}e^{-\frac{y^2}{2\sigD}}\left(\frac{1}{2} e^{\frac{y\sqrt{P}}{\sigD}}
            +\frac{1}{2} e^{-\frac{y\sqrt{P}}{\sigD}}\right)e^{-\frac{P}{2\sigD}}\\
        & = & \frac{1}{\sqrt{2 \pi \sigD}}e^{-\frac{y^2}{2\sigD}}
             \cosh\left(\frac{y\sqrt{P}}{\sigD}\right)e^{-\frac{P}{2\sigD}}\\
        &\stackrel{\sigD \rightarrow \infty}{\approx} & \frac{1}{\sqrt{2 \pi \sigD}}e^{-\frac{y^2}{2\sigD}}\\
        & \triangleq & G_y(0,\sigD),
\end{eqnarray*}
where the approximation is in the sense that for small $|y|$ we have $\cosh(|y|) \approx 1$ and for large $|y|$, $e^{-\frac{y^2}{2\sigD}}$
drives the entire expression to zero as $e^{-\frac{y^2}{2\sigD}}$,
for $\sigD \rightarrow \infty$.
This approximation reflects the intuitive notion that as the variance increases to infinity, the two-component, symmetric Gaussian
mixture resembles more and more a zero-mean Gaussian RV with the same variance.
Therefore, for low SNR, the output is very close to a zero-mean Normal
RV with variance $\sigD$, and $h(Y) \approx h(N)$,\footnote{For $\sigma = 20$ we have that $\int_{-\infty}^{\infty} |f_Y(y) - G_y(0,\sigD)|dy < 0.001$,
for $\sigma = 55$, $h(Y) - h(N) \approx 0.001$ and for $\sigma = 200$, $h(Y) - h(N) < 0.0001$.} hence
\[
    I(X;Y) \stackrel{\sigD \rightarrow \infty}{\longrightarrow} 0.
\]
Note that the upper bound and the decode-and-forward rate in this case are both equal to
\[
    R_{DAF} = R_{upper} = \min\left\{C,I(X;Y_1)\right\}.
\]

Now, let us evaluate the rate for HD-EAF as the SNR goes to zero. From \eqref{eqn:rate_Gauss}:
\[
    R \le I(X;Y,\hY_1) = I(X;\hY_1) + I(X;Y|\hY_1),
\]
and
\begin{eqnarray*}
    I(X;Y | \hY_1) & = & h(Y|\hY_1) - h(Y | X, \hY_1)\\
                   & = & \Pr(\hY_1 = 1) h(Y|\hY_1 = 1) + \Pr(\hY_1 = E) h(Y| \hY_1 = E) +
                        \Pr(\hY_1 = -1) h(Y|\hY_1 = -1) - h(N).
\end{eqnarray*}
Using  appendix \ref{append:Gauss-deriv}, equations \eqref{eqn:cond_entropy_hy1_is_1} -- \eqref{eqn:cond_f_y1_pos},
we have
\begin{eqnarray*}
    h(Y|\hY_1 = 1)  & = & -\int_{y = -\infty}^{\infty} f_{Y|\hY_1}(y|\hy_1 = 1) \log_2 \left(f_{Y|\hY_1}(y|\hy_1 = 1)\right) dy,\\
    f_{Y|\hY_1}(y|\hy_1 = 1)  & = & \frac{f_{Y,Y_1}(y,y_1>0)P_{\ners}}{\Pr(Y_1>0)P_{\ners}} = \frac{f_{Y,Y_1}(y,y_1>0)}{\Pr(Y_1>0)},\\
    f_{Y,Y_1}(y,y_1>0)      & = & \frac{1}{2}\left( f_{Y,Y_1|X}(y,y_1>0|x = \sqrt{P}) + f_{Y,Y_1|X}(y,y_1>0|x = -\sqrt{P}) \right)\\
                    & = & \frac{1}{2}\left( G_y(\sqrt{P},\sigD) \Pr(Y_1>0|X = \sqrt{P}) + G_y(-\sqrt{P},\sigD) \big(1 - \Pr(Y_1>0|X = \sqrt{P})\big)\right)\\
                    & = & \frac{1}{\sqrt{2 \pi \sigD}}e^{-\frac{y^2}{2\sigD}}\left(\frac{1}{2} e^{\frac{y\sqrt{P}}{\sigD}}\Pr(Y_1>0|X = \sqrt{P})
            +\frac{1}{2} e^{-\frac{y\sqrt{P}}{\sigD}}\big(1-\Pr(Y_1>0|X = \sqrt{P})\big)\right)e^{-\frac{P}{2\sigD}}\\
                    & = & \frac{1}{\sqrt{2 \pi \sigD}}e^{-\frac{y^2}{2\sigD}}
                    \left(\frac{\left(\frac{1}{2}-\delta\right) e^{\frac{y\sqrt{P}}{\sigD}}
            +\left(\frac{1}{2}+\delta\right) e^{-\frac{y\sqrt{P}}{\sigD}}}{2}\right)e^{-\frac{P}{2\sigD}}\\
                    & = & \frac{1}{\sqrt{2 \pi \sigD}}e^{-\frac{y^2}{2\sigD}}
                    \left(\frac{1}{2}\cosh\left(\frac{y\sqrt{P}}{\sigD}\right)
            -\delta \sinh\left(\frac{y\sqrt{P}}{\sigD}\right)\right)e^{-\frac{P}{2\sigD}}\\
                    & \stackrel{(a)}{\approx} & \frac{1}{2} G_y(0,\sigD),
\end{eqnarray*}
when $\sigD \rightarrow \infty$ and $\delta \in \left[-\frac{1}{2},\frac{1}{2}\right]$ is selected such that
$\Pr(Y_1>0|X = \sqrt{P}) = \frac{1}{2} - \delta$.
The approximation in (a) is because for small $|y|$, $\sinh\left(\frac{y\sqrt{P}}{\sigD}\right) \approx 0$ and
$\cosh\left(\frac{y\sqrt{P}}{\sigD}\right) \approx 1$, and for large $|y|$, both
$ e^{-\frac{y^2}{2\sigD}}\sinh\left(\frac{y\sqrt{P}}{\sigD}\right) \rightarrow 0$ and
$ e^{-\frac{y^2}{2\sigD}}\cosh\left(\frac{y\sqrt{P}}{\sigD}\right) \rightarrow 0$.
%Note that for the symmetric case we consider here $\delta = 0$.
Hence
\begin{eqnarray*}
    h(Y|\hY_1 = 1) & \approx & -\int_{y = -\infty}^{\infty} \frac{G_y(0,\sigD)}{2\Pr(Y_1>0)} \log_2 \left(\frac{G_y(0,\sigD)}{2\Pr(Y_1>0)}\right) dy\\
                & = & -\frac{1}{2\Pr(Y_1>0)}\int_{y = -\infty}^{\infty} G_y(0,\sigD)
                    \left[\log_2 \left(G_y(0,\sigD)\right) - \log_2 \left(2\Pr(Y_1>0)\right)\right] dy\\
                & = & \frac{1}{2\Pr(Y_1>0)} \left[h(N) + \log_2 \left(2\Pr(Y_1>0)\right)\right],
\end{eqnarray*}
and using $\Pr(Y_1 > 0) = \Pr(Y_1 \le 0) = \frac{1}{2}$ and $h(Y|\hY_1 = 1) = h(Y| \hY_1 = -1)$, we obtain
\begin{eqnarray*}
    h(Y|\hY_1) & \approx & \frac{1}{2}P_{\ners}h(N) + (1 - P_{\ners})h(N) + \frac{1}{2}P_{\ners} h(N) \\
                    & = &    h(N).
\end{eqnarray*}
Therefore, at low SNR, $Y$ and $\hY_1$ become independent.
Then, $I(X;Y | \hY_1)  =  h(Y|\hY_1)  - h(N) \approx 0$ and the information rate becomes (see
appendix \ref{appndx:appndxHD-EAF-highSNR})
\begin{eqnarray*}
    R \le I(X;\hY_1) & = & H(\hY_1) - H(\hY_1|X)\\
%                    & = & H\left(\frac{1}{2}P_{\ners}, 1 - P_{\ners} ,\frac{1}{2}P_{\ners}\right)
%                            - H\left(P_1 P_{\ners}, 1 - P_{\ners}, (1-P_1)P_{\ners}\right)\\
%                    & = & -P_{\ners} \log_2\left(\frac{1}{2}P_{\ners}\right) -(1 - P_{\ners})\log_2(1 - P_{\ners})+  P_1 P_{\ners} \log_2(P_1 P_{\ners})\\
%                    &   & \quad     +(1 - P_{\ners})\log_2(1 - P_{\ners})  +  (1-P_1)P_{\ners}\log_2((1-P_1)P_{\ners})\\
%                    & = & -P_{\ners} \log_2\left(P_{\ners}\right) +P_{\ners}  +  P_1 P_{\ners} \log_2(P_1) + P_1 P_{\ners} \log_2(P_{\ners})\\
%                    &   & \quad       +  (1-P_1)P_{\ners}\log_2(1-P_1) + (1-P_1)P_{\ners}\log_2(P_{\ners}) \\
%                    & = &  P_{\ners}(1  +  P_1  \log_2(P_1) +  (1-P_1)\log_2(1-P_1) ) \\
                    & = &  P_{\ners}(1  -H ( P_1  ,1-P_1 )),
\end{eqnarray*}
where $H(\cdot)$ is the discrete entropy for the specified discrete distribution and $P_1 = \Pr(Y_1 > 0 | X = \sqrt{P})$.
Now, consider the feasibility condition $C \ge I(Y_1;\hY_1|Y)$:
\begin{eqnarray*}
    I(Y_1;\hY_1|Y)  & = & H(\hY_1|Y) - H(\hY_1|Y_1,Y)\\
                    & \stackrel{(a)}{\approx} & H(\hY_1) - H(\hY_1|Y_1)\\
%                    & = & H\left(\frac{1}{2}P_{\ners}, 1 - P_{\ners} ,\frac{1}{2}P_{\ners}\right) -
%                        H(P_{\ners},1-P_{\ners})\\
%                    & = & - 2 \frac{1}{2}P_{\ners} \log_2\left(\frac{1}{2}P_{\ners}\right)
%                        - (1 - P_{\ners}) \log_2\left(1 - P_{\ners}\right) + P_{\ners} \log_2(P_{\ners})\\
%                    &   & \quad        + (1 - P_{\ners}) \log_2\left(1 - P_{\ners}\right)\\
                    & = &  P_{\ners},
\end{eqnarray*}
where (a) follows from the independence of $Y$ and $\hY_1$ at low SNR, see appendix \ref{appndx:appndxHD-EAF-highSNR}.
Therefore, for low SNR, we set $P_{\ners} = \min\left\{ C,1\right\}$
and the rate becomes
\[
    R \le \min\left\{ C,1\right\}( 1 - H ( P_1  ,1-P_1 )).
\]

For the GQ-EAF we first approximate $f(Y,\hY_1)$ at low SNR starting with \eqref{eqn:joint_y_hy1_gq_eaf}:
\begin{eqnarray*}
    f_{Y,\hY_1}(y,\hy_1) & = & \frac{1}{2}\left(G_y(\sqrt{P},\sigD)G_{\hy_1}(g\sqrt{P},\sigR+\sigQ) +
            G_y(-\sqrt{P},\sigD)G_{\hy_1}(-g\sqrt{P},\sigR+\sigQ) \right)\\
            & = & \frac{1}{\sqrt{2 \pi \sigD}} e^{-\frac{y^2}{2\sigD}}
                    \left(\frac{1}{2} G_{\hy_1}(g\sqrt{P},\sigR+\sigQ) e^{\frac{y\sqrt{P}}{\sigD}}+
                       \frac{1}{2} G_{\hy_1}(-g\sqrt{P},\sigR+\sigQ) e^{\frac{-y\sqrt{P}}{\sigD}} \right)e^{-\frac{P}{2\sigD}}\\
            & \approx & G_y(0,\sigD)f_{\hY_1}(\hy_1),
\end{eqnarray*}
as $e^{\pm \frac{y\sqrt{P}}{\sigD}} \approx 1$ in the region when $G_{\hy_1}$ is significant, for both $X = \sqrt{P}$
or $X = -\sqrt{P}$.
We conclude that as the direct SNR approaches 0, $Y$ and $\hY_1$ become independent.
Now, the rate is given by:
\begin{eqnarray}
    R & \le & I(X;Y,\hY_1) \nonumber \\
      &  =  & h(Y,\hY_1) - h(Y,\hY_1|X) \nonumber \\
      &  =  & h(Y) + h(\hY_1) - h(X+N, gX+N_1+N_Q|X) \nonumber \\
      &  =  & h(Y) + h(\hY_1) - h(N, N_1+N_Q|X)\nonumber \\
      &  =  & h(Y) - h(N|X) + h(\hY_1) - h(N_1 + N_Q|X)\nonumber \\
      &  =  & I(X;Y) + I(X;\hY_1)\nonumber \\
      & \approx & I(X;\hY_1)\nonumber \\
      \label{eqn:GQ-EAF-at-low-SNR}
      & = & h(\hY_1) - h(N_1+N_Q).
\end{eqnarray}
The  feasibility condition becomes:
\begin{eqnarray}
    C & \ge & I(\hY_1;Y_1|Y) \nonumber\\
        & = & h(\hY_1|Y) - h(\hY_1|Y,Y_1) \nonumber\\
    \label{eqn:cond_C_lowSNR_GQ}
        & \approx & h(\hY_1) - h(N_Q),
\end{eqnarray}
with
\[
    f_{\hY_1}(\hy_1) = \frac{1}{2}\left[G_{\hy_1}(g\sqrt{P},\sigR+\sigQ) + G_{\hy_1}(-g\sqrt{P},\sigR+\sigQ)\right].
\]

For DHD, as $\sigD \rightarrow \infty$ we have
\begin{eqnarray*}
    I(X;\hY_1;Y) & = & I(X;Y)  + I(X;\hY_1|Y) \\
        & \approx & I(X;\hY_1|Y)\\
        & = & H(\hY_1|Y) - H(\hY_1|Y,X)\\
        & \stackrel{(a)}{\approx} & H(\hY_1) - H(\hY_1|X)\\
        & = & I(X;\hY_1)
\end{eqnarray*}
where (a) follows from the independence of $Y$ and $Y_1$ as $\sigD \rightarrow \infty$ and the fact that
$\hY_1$ is a deterministic function of $Y_1$, combined with the fact that given $X$, $Y_1$ and $Y$ are independent.
The feasibility condition becomes
\[
    C  \ge  H(\hY_1|Y) \approx  H(\hY_1).
\]
Because $I(X;\hY_1)$ is not a monotone function of $T$ we have to optimize over $T$ to find the actual rate.

As can be seen from the expression for HD-EAF, when the SNR on the direct link decreases, the capacity of the
conference link acts as a scaling factor on the rate of the binary channel from the source to the relay.
\begin{figure}[!h]
    \centering
    \scalebox{0.7}{\includegraphics{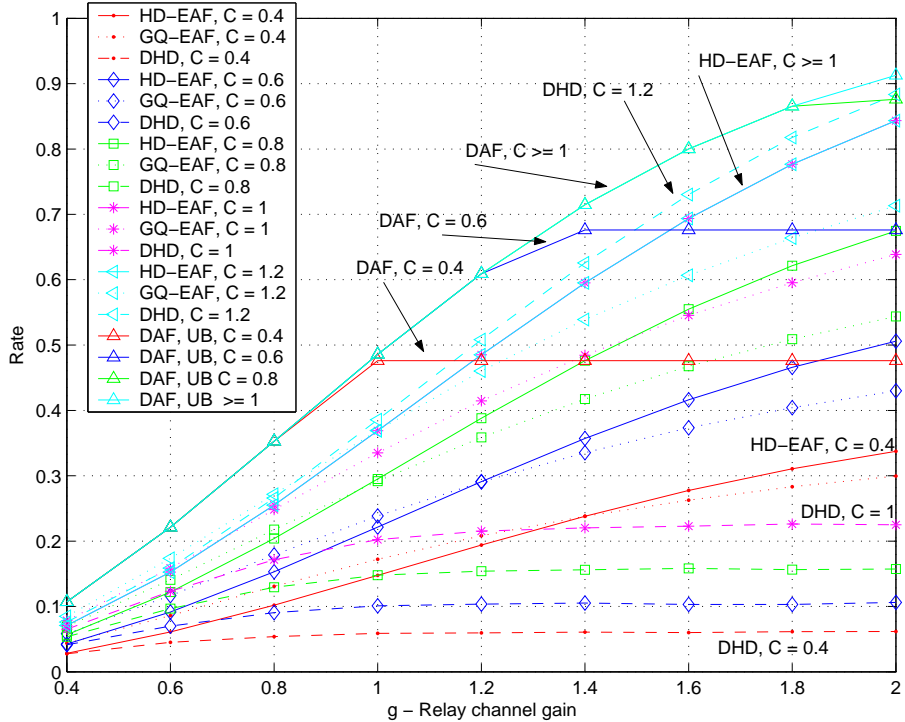}}
    \caption{Information rate with DAF, DHD, HD-EAF and GQ-EAF vs.
        relay channel gain $g$, for different values of $C$, at low SNR on the source-relay link.}
    \label{fig:low-direct-snr}
\end{figure}
In figure \ref{fig:low-direct-snr} we plotted the information rate for DHD, HD-EAF, GQ-EAF and DAF (which coincides with
the upper bound). Comparing the three EAF strategies we note that DHD, which at intermediate SNR on the source-relay channel performs well for $C \ge 0.8$,
has the worst performance at low SNR up to $C = 1.2$. At $C = 1.2$, DHD becomes the best technique out of the three.
For $C < 1.2$ and high SNR on the
source-relay channel, HD-EAF outperforms both DHD and GQ-EAF. For low SNR on the source-relay channel, GQ-EAF is again superior.

\subsection{Discussion}
We make the following observations:
\begin{itemize}
    \item As noted at the beginning of this section, for low SNR on the source-relay link,
    GQ-EAF outperforms TS-DHD. To see why, consider the
    distribution of $Y_1$:
    \begin{eqnarray*}
        f_{Y_1}(y_1) & = & G_{y_1}(0,\sigR) \cosh\left(\frac{g \sqrt{P} y_1}{\sigR}\right) e^{-\frac{g^2P}{2\sigR}}\\
        &  \stackrel{g \rightarrow 0}{\approx} & G_{y_1}(0,\sigR) \left(1 - \frac{g^2P}{2\sigR}\right),
    \end{eqnarray*}
    where the approximation is obtained using the first order Taylor expansion, and the fact that for large
    values of $Y_1$, $G_{y_1}(0,\sigR)$ dominates the expression. Therefore, as $g \rightarrow 0$, $Y_1$
    approaches a zero-mean Gaussian RV: $Y_1 \stackrel{\mathcal{D}}{\rightarrow} \mN(0,\sigR)$.
    As discussed in \cite[ch. 13.1]{cover-thomas:it-book},
    the closer the reconstruction variable is to the original variable, the better the quantization performance are expected to be. Therefore
    it should be natural to guess that GQ will perform better at low relay link SNR.

    \item At the other extreme, as $g \rightarrow \infty$, consider the DAF strategy:
    as $g \rightarrow \infty$, have that
    \begin{eqnarray*}
        h(Y_1) & = & -\int_{y_1 = -\infty}^{\infty}
            \frac{1}{2}\left[G_{y_1}(g\sqrt{P},\sigR) + G_{y_1}(-g\sqrt{P},\sigR)\right]\times\\
            &  & \qquad \qquad \qquad
            \log_2\left(\frac{1}{2}\left[G_{y_1}(g\sqrt{P},\sigR) + G_{y_1}(-g\sqrt{P},\sigR)\right]
            \right)dy_1\\
        & \stackrel{g \rightarrow \infty}{\approx}&  1 - \int_{y_1 = -\infty}^{\infty} \frac{1}{2}G_{y_1}(g\sqrt{P},\sigR)
            \log_2 G_{y_1}(g\sqrt{P},\sigR) dy_1 \\
        &   & \qquad \qquad \qquad - \int_{y_1 = -\infty}^{\infty} \frac{1}{2}G_{y_1}(-g\sqrt{P},\sigR)
            \log_2 G_{y_1}(-g\sqrt{P},\sigR) dy_1\\
        & = & 1 + h(N_1),
    \end{eqnarray*}
    and therefore,
    \[
        I(X;Y_1)  = h(Y_1) -  h(Y_1|X) \approx 1 + h(N_1) -  h(N_1) = 1 = H(X).
    \]
    Hence,
    \[
        R_{DAF}  = \min\left\{I(X;Y_1), I(X;Y)+C\right\} = \min \left\{1, I(X;Y)+C \right\},
    \]
    which is the maximal rate. Therefore, as $g \rightarrow \infty$ DAF provides the optimal rate.

    \item We can expect that at intermediate SNR, methods that balance between the soft-decision per symbol of GQ-EAF and
    the hard-decision on the entire codeword of DAF, will be superior to both.
    Furthermore, we believe that as the SNR decreases, increasing the
    cardinality of $\hY_1$ accordingly will improve the performance.
\end{itemize}

\section{Multi-Step Cooperative Broadcast Application}
\label{sec:application_multi_step}
% In relaying we first need to find a common knowledge that both the receiver and the
% relay share. The relay helps the receiver by refining this common knowledge. In the DAF method
% the common knowledge is the set of messages $\mW$. In our new relay method the common knowledge is the
% set $\stypm(Y_1|\xvec_2^m)$. Since this set can always be used as common knowledge for conferencing, we
% can apply the same idea used in theorem \ref{thm:main_thm} to generate common knowledge in multi-step conferencing for
% cooperative broadcast.

\begin{figure}[h]
    \centering
    \scalebox{0.6}{\includegraphics{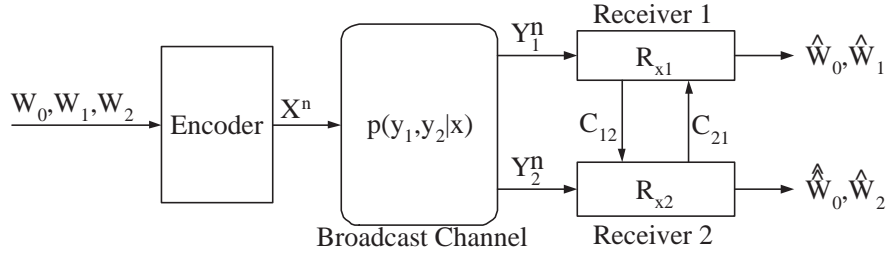}}
    \caption{The broadcast channel with cooperating receivers. The encoder sends three messages, a common message $W_0$,  a private message to $\Rgood$,
    $W_1$, and a private message to $\Rbad$, $W_2$. $\hat{W}_0$ and $\hat{\hat{W}}_0$ are the estimates of $W_0$ at
    $\Rgood$ and $\Rbad$ respectively.}
    \label{fig:three_msg_bc}
\end{figure}

In this section we consider the cooperative broadcast (BC) scenario. In this scenario, one transmitter communicates with two receivers. In its most
general form, the transmitter sends three independent messages: a common message intended for both receivers and two private messages,
one for each receiver, where all three messages are encoded into a single channel codeword $X^n$.
Each receiver gets a noisy version of the codeword, $Y_1^n$
at $\Rgood$ and $Y_2^n$ at $\Rbad$. After reception, the receivers exchange messages in a K-cycle conference over noiseless
conference links  of finite capacities $C_{12}$ and $C_{21}$.
Each conference message is based on the channel output at each receiver and the conference messages previously received
from the
other receiver, in a similar manner to the conference defined by Willems in~\cite{Willems:83} for the cooperative MAC.
After conferencing, each receiver decodes its message.
This scenario is depicted in figure
\ref{fig:three_msg_bc}. This setup was studied in \cite{DraperFK:03} for the
single common message case over the independent BC (i.e. $p(\yvec_1,\yvec_2|\xvec) = \prod_{i=1}^n p(y_{1,i}|x_i)p(y_{2,i}|x_i)$),
and in \cite{RonSer:2005} for the general setup with a single cycle of conferencing.

    \subsection{Definitions}

    We use the standard definition for the discrete memoryless general broadcast channel
    given in \cite{Cover:98}.
    We define a cooperative coding scheme as follows:
    \begin{definition}
        {\em A $\left(C_{12}, C_{21} \right)$-admissible K-cycle conference} consists of the following elements:
        \begin{enumerate}
            \item $K$ message sets from $\Rgood$ to $\Rbad$, denoted by
                $\mW_{12}^{(1)}$, $\mW_{12}^{(2)}$,...,$\mW_{12}^{(K)}$, and $K$ message sets from $\Rbad$ to $\Rgood$,
                denoted by $\mW_{21}^{(1)}$, $\mW_{21}^{(2)}$,...,$\mW_{21}^{(K)}$.
                Message
                set $\mW_{12}^{(k)}$ consists of $2^{nR_{12}^{(k)}}$ messages and message
                set $\mW_{21}^{(k)}$ consists of $2^{nR_{21}^{(k)}}$ messages.
            \item $K$ mapping functions, one for each conference step from $\Rgood$ to $\Rbad$:
                \[
                    h_{12}^{(k)}: \mY_1^n \times \mW_{21}^{(1)} \times \mW_{21}^{(2)} \times ... \times
                                \mW_{21}^{(k-1)} \mapsto \mW_{12}^{(k)},
                \]
                and $K$ mapping functions, one for each conference step from $\Rbad$ to $\Rgood$:
                \[
                    h_{21}^{(k)}: \mY_2^n \times \mW_{12}^{(1)} \times \mW_{12}^{(2)} \times ... \times
                        \mW_{12}^{(k)} \mapsto \mW_{21}^{(k)},
                \]
                where $k = 1,2,...,K$.
        \end{enumerate}
%        Let $R_{12}^{(k)} = \frac{1}{n}\log_2\left( ||\mW_{12}^{(k)}|| \right)$, and
%        $R_{21}^{(k)} = \frac{1}{n}\log_2\left( ||\mW_{21}^{(k)}|| \right)$. Then
    The conference rates satisfy:
        \[
            C_{12} = \sum_{k = 1}^K R_{12}^{(k)}, \qquad C_{21} = \sum_{k = 1}^K R_{21}^{(k)}.
        \]
    \end{definition}
    \begin{definition}
        {\em A $(2^{nR_0},2^{nR_1},2^{nR_2},n,C_{12},C_{21},K)$ code} for the general broadcast channel with
        a common message and two independent private messages, consists of three sets of source messages,
        $\mM_0 = \left\{1, 2,...,2^{nR_0}\right\}$, $\mM_1 = \left\{1, 2,...,2^{nR_1}\right\}$ and
                $\mM_2 = \left\{1, 2,...,2^{nR_2}\right\}$,
%        three conference message sets,
%        \begin{eqnarray*}
%            \mW_{21}^a &  = & \Big\{1,2,...,2^{n R_{21}^a} \Big\},\\
%            \mW_{12}   &  = & \Big\{1,2,...,2^{n R_{12}}   \Big\},\\
%            \mW_{21}^b &  = & \Big\{1,2,...,2^{n R_{21}^b} \Big\},
%        \end{eqnarray*}
        a mapping function at the transmitter,
        \[
           f: \mM_0 \times \mM_1 \times \mM_2 \mapsto \mX^n,
        \]
%        three relay functions,
%        \begin{eqnarray*}
%             &h_{21}^a:&  \mY_2^n  \mapsto \mW_{21}^a,\\
%             &h_{12}:  &  \mW_{21}^a \times \mY_1^n \mapsto \mW_{12},\\
%             &h_{21}^b:&  \mW_{12} \times \mY_2^n \mapsto \mW_{21}^b,
%        \end{eqnarray*}
%        with
%        \[
%            R_{21}^a  \le  C_{21}^a, \quad   R_{12}  \le  C_{12} \;\;\mbox{    and     }\;\;  R_{21}^b  \le  C_{21}^b ,
%        \]
%        where $C_{21}^a = \alpha C_{21}$ and $C_{21}^b = \left(1 - \alpha\right) C_{21}$;
        A $\left(C_{12}, C_{21} \right)$-admissible $K$-cycle conference,
        and two decoders,
        \begin{eqnarray*}
            & g_1: & \mW_{21}^{(1)} \times \mW_{21}^{(2)}\times ... \times \mW_{21}^{(K)} \times \mY_1^n \mapsto \mM_0 \times \mM_1, \\
            & g_2: & \mW_{12}^{(1)} \times \mW_{12}^{(2)}\times ... \times \mW_{12}^{(K)} \times \mY_2^n \mapsto \mM_0 \times \mM_2.
        \end{eqnarray*}
    \end{definition}
    \begin{definition}
        The {\em average probability of error} is defined as
        the average probability that at least one of the receivers does not decode its message pair correctly:
        \[
            \Pe = \Pr\left(g_1\left(W_{21}^{(1)}, W_{21}^{(2)}, ..., W_{21}^{(K)}, Y_1^n\right)\ne (M_0,M_1) \mbox{ or }
                g_2\left(W_{12}^{(1)}, W_{12}^{(2)}, ...,W_{12}^{(K)}, Y_2^n\right) \ne (M_0,M_2)\right),
        \]
        where we assume that each message is selected uniformly and independently over its respective message set.
    \end{definition}
%    \begin{definition}
%        Let $\typm(A)$ denote the set of $\delta$-weakly typical sequences of length $m$ generated
%        by the distribution $p_A(a)$ on $\mA$
%        %the i.i.d. distribution $\prod_{l=1}^m p(a_l)$, $a_l \in \mA$,
%        as defined in \cite[ch. 3]{cover-thomas:it-book}.
%        For $\avec_0^m \in \typm(A)$, define the set $\ttyp(A,\avec_0^m)$ to be the set of all
%        typical vectors $\avec^n \in \typ(A)$ such that their first $m$ elements
%        satisfy $\avec^m = \avec_0^m$.
%    \end{definition}

\subsection{The Cooperative Broadcast Channel with Two Independent and One Common Message}
\label{sec:multi-step-general-bc}
We first present the general result for the cooperative broadcast scenario with a $K$-cycle conference.
Denote with $\hYvec_1 = \left(\hY_1^{(1)}, \hY_1^{(2)},..., \hY_1^{(K)} \right)$ and
$\hYvec_2 = \left(\hY_2^{(1)}, \hY_2^{(2)},..., \hY_2^{(K)} \right)$. Let $R_1$ and $R_2$ be the private rates to $\Rgood$ and
$\Rbad$ respectively, and let $R_0$ denote the rate of the common information. Then, the following rate triplets are achievable:

\begin{theorem}
    \label{thm:multi-step-general-bc}
    \it
        Consider the general broadcast channel $\left(\mX, p(y_1,y_2|x), \mY_1 \times \mY_2\right)$ with cooperating
        receivers, having noiseless conference
    links of finite capacities $C_{12}$ and $C_{21}$ between them. Let the receivers hold a conference that
    consists of $K$ cycles. Then, any rate triplet $(R_0, R_1, R_2)$ satisfying
    \begin{subequations}
        \begin{eqnarray}
            R_0 & \le & \min\left\{I\left(W;Y_1,\hYvec_2\right), I\left(W;\hYvec_1,Y_2\right) \right\}\\
            R_1 & \le & I(U;Y_1, \hYvec_2|W)\\
            R_2 & \le & I(V;\hYvec_1, Y_2|W)\\
            R_1 + R_2 & \le & I(U;Y_1, \hYvec_2|W) + I(V;\hYvec_1, Y_2|W) - I(U;V|W),
        \end{eqnarray}
    \end{subequations}
    subject to,
    \begin{subequations}
        \begin{eqnarray}
        \label{eqn:c12_constr_multi_step}
            C_{12}  & \ge & I(Y_1; \hYvec_1, \hYvec_2|Y_2)\\
        \label{eqn:c21_constr_multi_step}
            C_{21}  & \ge & I(Y_2; \hYvec_2, \hYvec_1|Y_1),
        \end{eqnarray}
    \end{subequations}
    for some joint distribution
        \begin{eqnarray}
        \label{eqn:distributions}
         &  & p\left(w,u,v,x,y_1,y_2,\hy_1^{(1)}, \hy_1^{(2)},...,\hy_1^{(K)},\hy_2^{(1)}, \hy_2^{(2)},...,\hy_2^{(K)}\right) =\nonumber\\
         &  & \phantom{xxx} p(w,u,v,x) p(y_1,y_2|x) p\left(\hy_1^{(1)}|y_1\right) p\left(\hy_2^{(1)}|y_2,\hy_1^{(1)}\right)\cdot\cdot\cdot
                p\left(\hy_1^{(k)}|y_1,\hy_1^{(1)},\hy_1^{(2)},...,\hy_1^{(k-1)},\hy_2^{(1)},\hy_2^{(2)},...,\hy_2^{(k-1)}\right)\times\nonumber\\
         &  &  \phantom{xxx} p\left(\hy_2^{(k)}|y_2,\hy_1^{(1)},\hy_1^{(2)},...,\hy_1^{(k)},\hy_2^{(1)},\hy_2^{(2)},...,\hy_2^{(k-1)}\right)
                \cdot\cdot\cdot p\left(\hy_1^{(K)}|y_1,\hy_1^{(1)},\hy_1^{(2)},...,\hy_1^{(K-1)},\hy_2^{(1)},\hy_2^{(2)},...,\hy_2^{(K-1)}\right)\nonumber\\
         &  &  \phantom{xxx} \times p\left(\hy_2^{(K)}|y_2,\hy_1^{(1)},\hy_1^{(2)},...,\hy_1^{(K)},\hy_2^{(1)},\hy_2^{(2)},...,\hy_2^{(K-1)}\right),
    \end{eqnarray}
    is achievable.
    The cardinality of the $k$'th auxiliary random variables are bounded by:
    \begin{eqnarray*}
        ||\mhY_1^{(k)}|| & \le & ||\mY_1|| \times \prod_{l=1}^{k-1} ||\mhY_1^{(l)}|| \times \prod_{l=1}^{k-1} ||\mhY_2^{(l)}|| + 1, \qquad
            \qquad k = 1,2,...,K\\
        ||\mhY_2^{(k)}|| & \le & ||\mY_2|| \times \prod_{l=1}^{k} ||\mhY_1^{(l)}|| \times \prod_{l=1}^{k-1} ||\mhY_2^{(l)}|| + 1,
            \qquad \qquad k = 1,2,...,K.
    \end{eqnarray*}
\end{theorem}

\begin{proof}
    \subsubsection{Overview of Strategy}
    The coding strategy is based on combining the BC code construction of \cite{ElGamalM:81}, after incorporating the common message into the
    construction, with the $K$-cycle conference of
    \cite{Kaspi:85}. The transmitter constructs a broadcast code to split the rate between the three message sets. This
    is done independently of the relaying scheme.
    Each receiver generates its conference messages according to the construction of \cite{Kaspi:85}.
    After $K$ cycles of conferencing
    each receiver decodes its information based on its channel output and the conference messages received from the other receiver.

    \subsubsection{Code Construction at The Transmitter}
    \begin{itemize}
    \item
    Fix all the distributions in \eqref{eqn:distributions}. Fix $\eps > 0$ and let $n > 1$. Let $\delta > 0$ be a positive number whose
    value is determined in the following steps.
    Let $R(W) = \min\Big\{I\left(W;Y_1,\hYvec_2\right), I\left(W;\hYvec_1,Y_2\right) \Big\}$. Let $S_{[W]\delta}^{(n)}$ denote the
    set of all $\wvec \in \mW^n$ sequences such that $\wvec \in \stypd(W)$ and $\stypd(U,V|\wvec)$ is non-empty, as defined in
    \cite[corollary 5.11]{YeungBook}. From \cite[corollary 5.11]{YeungBook} we  have that
    $||S_{[W]\delta}^{(n)}|| \ge 2^{n(H(W)-\phi)}$, where $\phi \rightarrow 0$ as $\delta \rightarrow 0$ and $n \rightarrow \infty$.

    \item Pick $2^{n(R(W) - \eps)}$ sequences from $S_{[W]\delta}^{(n)}$ in a uniform and independent manner according to
    \[
        \Pr(\wvec) = \left\{
                \begin{array}{cl}
                    \frac{1}{||S_{[W]\delta}^{(n)}||} & ,\wvec \in S_{[W]\delta}^{(n)}\\
                    0   & ,\mbox{otherwise}.
                \end{array}
            \right.
    \]
    Label these sequences with $l \in \mM_0 \triangleq \left\{1,2,...,2^{n(R(W)-\eps)}\right\}$.

    \item For each sequence $\wvec(l)$, $l \in \mM_0$, consider the set $\stypdp(U|\wvec(l))$ ,$\delta' = \delta\max\left\{||\mU||, ||\mV||\right\}$.
    Since the sequences $\wvec \in \mW^n$ are selected such that $\stypd(U,V|\wvec(l))$ is non-empty and since
    $(\uvec, \vvec) \in \stypd(U,V|\wvec(l))$ implies  $\uvec \in \stypdp(U|\wvec(l))$, then also $\stypdp(U|\wvec(l))$ in non-empty, and by
    \cite[theorem 5.9]{YeungBook}, $||\stypdp(U|\wvec(l))|| \ge 2^{n(H(U|W) - \psi)}$,
    $\psi \rightarrow 0$ as $\delta' \rightarrow 0$ and $n \rightarrow \infty$.

    \item For each $l \in \mM_0$ pick $2^{n(I(U;Y_1,\hYvec_2|W)-\eps)}$ sequences in a uniform and independent manner from $\stypdp(U|\wvec(l))$ according
    to
    \[
        \Pr(\uvec|l) =  \left\{
            \begin{array}{cl}
                \frac{1}{||\stypdp(U|\wvec(l))||} & ,\uvec \in \stypdp(U|\wvec(l))\\
                0 & , \mbox{otherwise}.
            \end{array}
        \right.
    \]
    Label these sequences with $\uvec(i|l)$, $i \in \mZ_1 \triangleq \left\{1,2,...,2^{n(I(U;Y_1,\hYvec_2|W)-\eps)}\right\}$. Similarly,
    pick $2^{n(I(V;\hYvec_1, Y_2|W)-\eps)}$ sequences in a uniform and independent manner from $\stypdp(V|\wvec(l))$ according
    to
    \[
        \Pr(\vvec|l) =  \left\{
            \begin{array}{cl}
                \frac{1}{||\stypdp(V|\wvec(l))||} & ,\vvec \in \stypdp(V|\wvec(l))\\
                0 & , \mbox{otherwise}.
            \end{array}
        \right.
    \]
    Label these sequences with $\vvec(j|l)$, $j \in \mZ_2 \triangleq \left\{1,2,...,2^{n(I(V;\hYvec_1, Y_2|W)-\eps)}\right\}$.
    $\delta$ is selected such that $||S_{[W]\delta}^{(n)}|| \ge 2^{n(R(W)-\eps)}$, and $\forall l \in \mM_0$ we have
    that $||\stypdp(U|\wvec(l))|| \ge 2^{n(I(U;Y_1,\hYvec_2|W)-\eps)}$ and
    $||\stypdp(V|\wvec(l))|| \ge 2^{n(I(V;\hYvec_1, Y_2|W)-\eps)}$.

    \item Partition the set $\mZ_1$ into $2^{nR_1}$ subsets $B_{w_1}$,  $w_1 \in \mM_1 = \left\{1,2,...,2^{nR_1}\right\}$, let \\
    $B_{w_1} = \Big[(w_1 - 1)2^{n(I(U;Y_1,\hYvec_2|W)- R_1 - \eps)} + 1, w_1 2^{n(I(U;Y_1,\hYvec_2|W)- R_1 - \eps)} \Big]$. Similarly partition
    the set $\mZ_2$ into $2^{nR_2}$ subsets $C_{w_2}$,  $w_2 \in \mM_2 = \left\{1,2,...,2^{nR_2}\right\}$, let \\
    $C_{w_2} = \left[(w_2 - 1)2^{n(I(V;\hYvec_1,Y_2|W)- R_2 - \eps)} + 1, w_2 2^{n(I(V;\hYvec_1,Y_2|W)- R_2 - \eps)} \right]$.

    \item For each triplet $(l,w_1,w_2)$ consider the set
    \[
        \mD(w_1,w_2|l) \triangleq \left\{(m_1,m_2): m_1 \in B_{w_1}, m_2 \in C_{w_2}, \left(\uvec(m_1|l), \vvec(m_2|l)\right)
            \in \stypdp(U,V|\wvec(l)) \right\}.
    \]
    By \cite[lemma on pg. 121]{ElGamalM:81}, we have that taking $n$ large enough we can make
    $\Pr\left(||\mD(w_1,w_2|l)||  = 0\right) \le \eps$ for any arbitrary $\eps > 0$, as long as
    \begin{subequations}
        \begin{eqnarray}
        \label{eqn:R1_cond_lemma}
            R_1  & \le & I(U;Y_1,\hYvec_2|W)\\
        \label{eqn:R2_cond_lemma}
            R_2 & \le & I(V;\hYvec_1,Y_2|W)\\
        \label{eqn:R1_and_R2_cond_lemma}
            R_1 + R_2 & \le & I(U;Y_1,\hYvec_2|W) + I(V;\hYvec_1,Y_2|W) - I(U;V|W).
        \end{eqnarray}
    \end{subequations}
    Note that the individual rate constraints are required to guarantee that the sets $B_{w_1}$ and $C_{w_2}$ are non-empty.

    \item For each $l \in \mM_0$, we pick a unique pair of $(m_1(w_1,w_2,l), m_2(w_1,w_2,l)) \in \mD(w_1,w_2|l)$,
    $(w_1,w_2) \in \mM_1 \times \mM_2$.  The transmitter generates the codeword $\xvec(l,w_1,w_2)$ according to\\
    $p(\xvec(l,w_1,w_2)) = \prod_{i=1}^n p(x_i|u_i(m_1(w_1,w_2,l)),v_i(m_2(w_1,w_2,l)),w_i(l))$.
    When transmitting the triplet $(l,w_1,w_2)$ the transmitter outputs $\xvec(l,w_1,w_2)$.
    \end{itemize}

        \subsubsection{Codebook Generation at the Receivers}
        \begin{itemize}
            \item For the first conference step from $\Rgood$ to $\Rbad$, $\Rgood$ generates a codebook
                with $2^{nR_{12}'^{(1)}}$ codewords indexed by
                $z_{12}^{(1)}\in \mZ_{12}^{(1)} = \left\{1,2,...,2^{nR_{12}'^{(1)}}\right\}$ according to the distribution
                $p\left(\hy_1^{(1)}\right)$:
                $p\left(\hyvec_1^{(1)}(z_{12}^{(1)})\right) = \prod_{i=1}^n p\left(\hy_{1,i}^{(1)}(z_{12}^{(1)})\right)$.
                $\Rgood$ uniformly and independently partitions the message set
                $\mZ_{12}^{(1)}$ into $2^{nR_{12}^{(1)}}$ subsets indexed by
                $w_{12}^{(1)} \in \mW_{12}^{(1)} = \left\{1,2,...,2^{nR_{12}^{(1)}}\right\}$. Denote these subsets
                with $\mS_{12,w_{12}^{(1)}}^{(1)}$.
            \item For the first conference step from $\Rbad$ to $\Rgood$, $\Rbad$ generates a codebook with
                $2^{nR_{21}'^{(1)}}$ codewords indexed by $z_{21}^{(1)} \in \mZ_{21}^{(1)} = \left\{1,2,..., 2^{nR_{21}'^{(1)}}\right\}$
                for each codeword $\hyvec_1^{(1)}(z_{12}^{(1)})$, $z_{12}^{(1)} \in \mZ_{12}^{(1)}$, in an i.i.d.
                manner according to
                $p\left(\hyvec_2^{(1)}(z_{21}^{(1)}|z_{12}^{(1)})\right)=  \prod_{i=1}^n p\left(\hy_{2,i}^{(1)}(z_{21}^{(1)}|z_{12}^{(1)})\Big|\hy^{(1)}_{1,i}(z_{12}^{(1)})\right)$.
%                $z_{21}^{(1)} \in \mZ_{21}^{(1)}$.
                $\Rbad$ uniformly and independently partitions the message set $\mZ_{21}^{(1)}$
                 into $2^{nR_{21}^{(1)}}$ subsets indexed by
                $w_{21}^{(1)} \in \mW_{21}^{(1)} = \left\{1,2,...,2^{nR_{21}^{(1)}}\right\}$. Denote these subsets
                with $\mS_{21,w_{21}^{(1)}}^{(1)}$.
            \item For the $k$'th conference step from $\Rgood$ to $\Rbad$, $\Rgood$ considers each combination of
                $z_{12}^{(1)},z_{12}^{(2)},...,z_{12}^{(k-1)}$,
                $z_{21}^{(1)},z_{21}^{(2)},...,z_{21}^{(k-1)}$. For each combination, $\Rgood$ generates a codebook with $2^{nR_{12}'^{(k)}}$
                messages indexed by $z_{12}^{(k)} \in \mZ_{12}^{(k)} = \left\{1,2,...,2^{nR_{12}'^{(k)}}\right\}$,
                according to the distribution
                 $p\left(\hy_1^{(k)}|\hy_1^{(1)},\hy_1^{(2)},...,\hy_1^{(k-1)},\hy_2^{(1)},\hy_2^{(2)},...,\hy_2^{(k-1)}\right)$.
                $\Rgood$ uniformly and independently partitions the message set
                $\mZ_{12}^{(k)}$ into $2^{nR_{12}^{(k)}}$ subsets indexed by
                $w_{12}^{(k)} \in \mW_{12}^{(k)} = \left\{1,2,...,2^{nR_{12}^{(k)}}\right\}$. Denote these subsets
                with $\mS_{12,w_{12}^{(k)}}^{(k)}$.
            \item The codebook for the $k$'th conference step from $\Rbad$ to $\Rgood$ is generated in a parallel manner for each combination
                of $z_{12}^{(1)},z_{12}^{(2)},...,z_{12}^{(k)}$, $z_{21}^{(1)},z_{21}^{(2)},...,z_{21}^{(k-1)}$.

        \end{itemize}

    \subsubsection{Decoding and Encoding at $\Rgood$ at the $k$'th Conference Cycle ($k \le K$) for Transmission
        Block $i$}
    \label{sec:DecEncMultiStepRgood}
        $\Rgood$ needs first to decode the message $z_{21}^{(k-1)}$ sent from $\Rbad$ at the $(k-1)$'th cycle.
        To that end, $\Rgood$ uses $w_{21}^{(k-1)}$, the index received from $\Rbad$ at the $(k-1)$'th conference
        step. In
        decoding $z_{21}^{(k-1)}$ we assume that all the previous $z_{21}^{(1)},z_{21}^{(2)},...,z_{21}^{(k-2)}$
        were correctly decoded at $\Rgood$. We denote the $\hyvec_2^{(k)}$ sequences corresponding to
        $z_{21}^{(1)},z_{21}^{(2)},...,z_{21}^{(k-2)}$ by\\
         $\hyvec_2(1), \hyvec_2(2), ...,\hyvec_2(k-2)$, and
        similarly define $\hyvec_1(1), \hyvec_1(2) ,..., \hyvec_1(k-1)$.
        \begin{itemize}
            \item $\Rgood$ first generates the set $\mL_1(k-1)$ defined by:
                \begin{eqnarray*}
                    &  & \mL_1(k-1) = \bigg\{z_{21}^{(k-1)} \in  \mZ_{21}^{(k-1)} :
                        \Big(\hyvec_2^{(k-1)}(z_{21}^{(k-1)}|z_{12}^{(1)},z_{12}^{(2)},...,z_{12}^{(k-1)},z_{21}^{(1)},z_{21}^{(2)},...,z_{21}^{(k-2)}),\\
                    &  & \phantom{xxxxxxxxxxxxx} \hyvec_1(1),\hyvec_1(2),...,\hyvec_1(k-1),\hyvec_2(1),\hyvec_2(2),...,\hyvec_2(k-2),\yvec_1(i)\Big)\in \styp\bigg\}.
                \end{eqnarray*}
            \item $\Rgood$ then looks for a unique $z_{21}^{(k-1)} \in \mZ_{21}^{(k-1)}$ such that
                    $z_{21}^{(k-1)} \in \mL_1(k-1) \bigcap \mS_{21,w_{21}^{(k-1)}}^{(k-1)}$. If there is none or
                    there is more than one, an error is declared.
            \item From an argument similar to \cite{Kaspi:85}, the probability of error can be made arbitrarily small
                by taking $n$ large enough as long as
                \[
                    R_{21}'^{(k-1)} < I\left(\hY_2^{(k-1)};Y_1\big| \hY_1^{(1)},\hY_1^{(2)},...,\hY_1^{(k-1)},
                        \hY_2^{(1)},\hY_2^{(2)},...,\hY_2^{(k-2)}\right) + R_{21}^{(k-1)} - \eps.
                \]
                Here, $k > 1$, since for the first conference message from $\Rgood$ to $\Rbad$ no
                decoding takes place.
        \end{itemize}
        In generating the $k$'th conference message to $\Rbad$, it is assumed that all the previous $k-1$ messages from
        $\Rbad$ were decoded correctly.
        \begin{itemize}
            \item $\Rgood$ looks for a message $z_{12}^{(k)} \in \mZ_{12}^{(k)}$ such that
                \begin{eqnarray*}
                    & & \Big(\hyvec_1^{(k)}(z_{12}^{(k)}|z_{12}^{(1)},z_{12}^{(2)},...,z_{12}^{(k-1)},z_{21}^{(1)},z_{21}^{(2)},...,z_{21}^{(k-1)}),\\
                    & & \phantom{xxxx}    \hyvec_1(1),\hyvec_1(2),...,\hyvec_1(k-1),\hyvec_2(1),\hyvec_2(2),...,\hyvec_2(k-1),\yvec_1(i) \Big) \in \styp.
                \end{eqnarray*}
                From the argument in \cite{Kaspi:85}, the probability that such a sequence exists can be made arbitrarily close to $1$
                by taking $n$ large enough as long as
                \[
                    R_{12}'^{(k)} > I\left(\hY_1^{(k)};Y_1\Big|\hY_1^{(1)},\hY_1^{(2)},...,\hY_1^{(k-1)},\hY_2^{(1)},\hY_2^{(2)},...,\hY_2^{(k-1)} \right) + \eps.
                \]
            \item $\Rgood$ looks for the partition of $\mZ_{12}^{(k)}$ into which $z_{12}^{(k)}$ belongs. Denote the index of this
                partition with $w_{12}^{(k)}$.
            \item $\Rgood$ transmits $w_{12}^{(k)}$ to $\Rbad$ through the conference link.
        \end{itemize}

    \subsubsection{Decoding and Encoding at $\Rbad$ at the $k$'th Conference Step ($k \le K$) for Transmission
        Block $i$}
    \label{sec:DecEncMultiStepRbad}
        Using similar arguments to section \ref{sec:DecEncMultiStepRgood}, we obtain the following
        rate constraints:
        \begin{itemize}
            \item Decoding $z_{12}^{(k)}$ at $\Rbad$ can be done with an arbitrarily small
                probability of error by taking $n$ large enough as long as
                \[
                    R_{12}'^{(k)} < I\left(\hY_1^{(k)};Y_2\big| \hY_1^{(1)},\hY_1^{(2)},...,\hY_1^{(k-1)},
                        \hY_2^{(1)},\hY_2^{(2)},...,\hY_2^{(k-1)}\right) + R_{12}^{(k)} - \eps.
                \]
            \item Encoding $z_{21}^{(k)}$ can be done with an arbitrarily small probability of error
                by taking $n$ large enough as long as
                \[
                    R_{21}'^{(k)} > I\left(\hY_2^{(k)};Y_2\Big|\hY_1^{(1)},\hY_1^{(2)},...,\hY_1^{(k)},
                                \hY_2^{(1)},\hY_2^{(2)},...,\hY_2^{(k-1)} \right) + \eps.
                \]
        \end{itemize}

    \subsubsection{Combining All Conference Rate Bounds}
    \label{sec:combining_bounds_general}
        First consider the bounds on $R_{12}'^{(k)}$, $k = 1,2,...,K$:
        \begin{eqnarray*}
        &  &    I\left(\hY_1^{(k)};Y_1\Big|\hY_1^{(1)},\hY_1^{(2)},...,\hY_1^{(k-1)},\hY_2^{(1)},\hY_2^{(2)},...,\hY_2^{(k-1)} \right) + \eps
                < R_{12}'^{(k)} < \\
        &  &  \phantom{xxxxx}I\left(\hY_1^{(k)};Y_2\big| \hY_1^{(1)},\hY_1^{(2)},...,\hY_1^{(k-1)},
                        \hY_2^{(1)},\hY_2^{(2)},...,\hY_2^{(k-1)}\right) + R_{12}^{(k)} - \eps.
        \end{eqnarray*}
        This can be satisfied only if
        \begin{eqnarray*}
        &  &  I\left(\hY_1^{(k)};Y_2\big| \hY_1^{(1)},\hY_1^{(2)},...,\hY_1^{(k-1)},
                        \hY_2^{(1)},\hY_2^{(2)},...,\hY_2^{(k-1)}\right) + R_{12}^{(k)} - \eps >  \\
        &  &  \phantom{xxxxx}  I\left(\hY_1^{(k)};Y_1\Big|\hY_1^{(1)},\hY_1^{(2)},...,\hY_1^{(k-1)},\hY_2^{(1)},\hY_2^{(2)},...,\hY_2^{(k-1)} \right) + \eps\\
        & \Rightarrow  &   R_{12}^{(k)}  >  H\left(\hY_1^{(k)} \big| Y_2, \hY_1^{(1)},\hY_1^{(2)},...,\hY_1^{(k-1)},
                        \hY_2^{(1)},\hY_2^{(2)},...,\hY_2^{(k-1)}\right) \\
        &  &  \phantom{xxxxx}  -H\left(\hY_1^{(k)}\Big|Y_1,\hY_1^{(1)},\hY_1^{(2)},...,\hY_1^{(k-1)},\hY_2^{(1)},\hY_2^{(2)},...,\hY_2^{(k-1)} \right) + 2\eps\\
        &   &  \phantom{xxx} =  I\left(\hY_1^{(k)} ;Y_1\big| Y_2, \hY_1^{(1)},\hY_1^{(2)},...,\hY_1^{(k-1)},
                        \hY_2^{(1)},\hY_2^{(2)},...,\hY_2^{(k-1)}\right) +2\eps.
        \end{eqnarray*}
        Hence
        \begin{eqnarray}
            C_{12} & = & \sum_{k = 1}^K R_{12}^{(k)} \nonumber\\
                   & \ge & \sum_{k = 1}^K \bigg(I\left(\hY_1^{(k)} ;Y_1\big| Y_2, \hY_1^{(1)},\hY_1^{(2)},...,\hY_1^{(k-1)},
                        \hY_2^{(1)},\hY_2^{(2)},...,\hY_2^{(k-1)}\right) +2\eps\bigg)\nonumber\\
                   & = & \sum_{k = 1}^{K} \bigg[I\left(\hY_1^{(k)} ;Y_1\big| Y_2, \hY_1^{(1)},\hY_1^{(2)},...,\hY_1^{(k-1)},
                        \hY_2^{(1)},\hY_2^{(2)},...,\hY_2^{(k-1)}\right)\nonumber\\
                   &   & \phantom{xxxxxx} + I\left(\hY_2^{(k)} ;Y_1\big| Y_2, \hY_1^{(1)},\hY_1^{(2)},...,\hY_1^{(k)},
                        \hY_2^{(1)},\hY_2^{(2)},...,\hY_2^{(k-1)}\right)\bigg] + 2K\eps \nonumber\\
%                   &   & \phantom{xxxxxxxxxx} +I\left(\hY_1^{(K)} ;Y_1\big| Y_2, \hY_1^{(1)},\hY_1^{(2)},...,\hY_1^{(K-1)},
%                        \hY_2^{(1)},\hY_2^{(2)},...,\hY_2^{(K-1)}\right)  \nonumber\\
%                   &   & \phantom{xxxxxxxxxxxxxx} + I\left(\hY_2^{(K)} ;Y_1\big| Y_2, \hY_1^{(1)},\hY_1^{(2)},...,\hY_1^{(K)},
%                        \hY_2^{(1)},\hY_2^{(2)},...,\hY_2^{(K-1)}\right) + 2K\eps \nonumber\\
                   & = &  \sum_{k = 1}^{K} I\left(\hY_1^{(k)},\hY_2^{(k)} ;Y_1\big| Y_2, \hY_1^{(1)},\hY_1^{(2)},...,\hY_1^{(k-1)},
                        \hY_2^{(1)},\hY_2^{(2)},...,\hY_2^{(k-1)}\right) + 2K\eps \nonumber\\
%                   &   & \phantom{xxxxxxx} +I\left(\hY_1^{(K)}, \hY_2^{(K)} ;Y_1\big| Y_2, \hY_1^{(1)},\hY_1^{(2)},...,\hY_1^{(K-1)},
%                        \hY_2^{(1)},\hY_2^{(2)},...,\hY_2^{(K-1)}\right) + 2K\eps\nonumber\\
                   \label{eqn:constr_c12_general_bc}
                   & = &   I\left( \hY_1^{(1)},\hY_1^{(2)},...,\hY_1^{(K)},
                        \hY_2^{(1)},\hY_2^{(2)},...,\hY_2^{(K)};Y_1\big| Y_2\right) +2K\eps,
        \end{eqnarray}
        and similarly
        \begin{equation}
        \label{eqn:constr_c21_general_bc}
            C_{21} \ge I\left( \hY_1^{(1)},\hY_1^{(2)},...,\hY_1^{(K)},
                        \hY_2^{(1)},\hY_2^{(2)},...,\hY_2^{(K)};Y_2\big| Y_1\right) +2K\eps.
        \end{equation}
         This provides the rate constraints on the conference auxiliary variables of \eqref{eqn:c12_constr_multi_step} and
         \eqref{eqn:c21_constr_multi_step}.

         \subsubsection{Decoding at $\Rgood$}
         $\Rgood$ uses $\yvec_1(i)$ and $\hyvec_2^{(1)},\hyvec_2^{(2)},...,\hyvec_2^{(K)}$ received from $\Rbad$, to decode $(l_i,w_{1,i})$ as follows:
         \begin{itemize}
            \item $\Rgood$ looks for a unique message $l \in \mM_0$ such
            \[
                \big(\wvec(l),\yvec_1(i),\hyvec_2^{(1)},\hyvec_2^{(2)},...,\hyvec_2^{(K)}\big) \in \styp.
            \]
            From the point-to-point channel capacity theorem (see \cite{ElGamalM:81}), this can be done with an arbitrarily
            small probability of error by taking $n$ large enough as long as
            \begin{equation}
            \label{eqn:constr_r0_decode_Rgood}
                R_0 \le I(W;Y_1,\hYvec_2).
            \end{equation}
            Denote the decoded message $\hat{l}_i$. Now $\Rgood$ decodes $w_{1,i}$ by looking for a unique $k \in \mZ_1$
            such that
            \[
                \big(\uvec(k|\hat{l}_i),\wvec(\hat{l}_i),\yvec_1(i),\hyvec_2^{(1)},\hyvec_2^{(2)},...,\hyvec_2^{(K)}\big) \in \styp.
            \]
            If a unique such $k$ exists, then denote the decoded index with $\hat{k}=k$. Now $\Rgood$ looks for the partition of $\mZ_1$ into which $\hat{k}$ belongs and sets $\hw_{1,i}$
            to be the index of that partition: $\hat{k} \in B_{\hw_{1,i}}$.
            Similarly to  the proof in \cite[ch 14.6.2]{cover-thomas:it-book}, assuming successful decoding
            of $l_i$, the probability of error can be made arbitrarily small by taking $n$ large enough as long as
            \[
                \frac{1}{n}\log_2||\mZ_1|| \le I(U;Y_1,\hYvec_2|W),
            \]
            which is satisfied by construction.

         \end{itemize}

         \subsubsection{Decoding at $\Rbad$}
         Repeating similar steps for decoding at $\Rbad$ we get that decoding $l_i$ can be done with an arbitrarily
         small probability of error by taking $n$ large enough as long as
         \begin{equation}
         \label{eqn:constr_r0_decode_Rbad}
            R_0 \le I(W; \hYvec_1,Y_2),
         \end{equation}
         and assuming successful decoding of $l_i$, decoding $w_{2,i}$ with an arbitrarily small probability of error
         requires that
         \[
                \frac{1}{n}\log_2||\mZ_2|| \le I(V;\hYvec_1,Y_2|W),
         \]
         which again is satisfied by construction.

         Finally, collecting \eqref{eqn:R1_cond_lemma}, \eqref{eqn:R2_cond_lemma},
         \eqref{eqn:R1_and_R2_cond_lemma}, \eqref{eqn:constr_r0_decode_Rgood} and \eqref{eqn:constr_r0_decode_Rbad} give
         the achievable rate constraints of theorem \ref{thm:multi-step-general-bc}, and \eqref{eqn:constr_c12_general_bc}
         and \eqref{eqn:constr_c21_general_bc} give the conference rate constraints of the theorem.
\end{proof}

\subsection{The Cooperative Broadcast Channel with a Single Common Message}
\label{sec:multi-step-single-common-message}

In the single common message cooperative broadcast scenario,
a single transmitter sends a message to two receivers encoded in a single channel codeword
$X^n$. % where the superscript $n$ denotes the length of a vector.
\begin{figure}[ht]
     \epsfxsize=0.6\textwidth \leavevmode\centering\epsffile{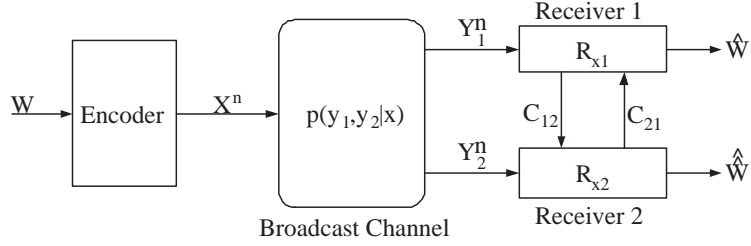}
    \caption{The broadcast channel with cooperating receivers, for the single common message case.
        $\hat{W}$ and $\hat{\hat{W}}$ are the estimates of $W$ at $\Rgood$ and $\Rbad$ respectively.}
    \label{fig:broadcast-cooperation-common}
\end{figure}
This scenario is depicted in figure \ref{fig:broadcast-cooperation-common}.
%The conference messages are
%functions of $Y_1^n$ (at $\Rgood$), $Y_2^n$ (at $\Rbad$),
%and the previous conference messages received from the
%other decoder, as defined by Willems in~\cite{Willems:83}.
After conferencing, each receiver decodes the message.
For this setup we have the following upper bound:
\begin{proposition}
         \label{prop:common_upper}
         {\it (\cite[theorem 6]{RonISIT05:05})}
         {\it
         Consider the general broadcast channel $(\mX, p(y_1,y_2|x), \mY_1 \times \mY_2)$ with cooperating
         receivers having noiseless conference links of finite capacities $C_{12}$ and $C_{21}$ between them.
         Then, for sending a common message to both receivers, any rate $R$ must satisfy
         \[
            R \! \le \!\sup_{p_X(x)} \! \min \! \Big\{I(X;Y_1) + C_{21}, I(X;Y_2) + C_{12}, I(X;Y_1,Y_2) \Big\}.
         \]}
\end{proposition}
In \cite{RonISIT05:05} we also derived the following achievable rate for
%the general broadcast channel with a single common message:
this scenario:
\begin{proposition}
        \label{prop:achive_common_one_step}
         {\it (\cite[theorem 5]{RonISIT05:05})} {\it
         Assume the broadcast channel setup of proposition \ref{prop:common_upper}.
         Then, for sending a common message to both receivers, any rate $R$ satisfying
         \begin{subequations}
             \begin{eqnarray}
                 R & \le & \sup_{p_X(x)}\Big[ \max\Big\{ R_{12}(p_X(x)), R_{21}(p_X(x)) \Big\} \Big], \nonumber\\
            \label{eqn:PrevResult1}
                 R_{12}(p_X(x)) & \triangleq & \min \Big( I(X;Y_1) + C_{21} ,  \max\big\{I(X;Y_2),
                        I(X;Y_2) - H(Y_1|Y_2,X) + \min\big(C_{12},H(Y_1|Y_2)\big)\big\} \Big),\phantom{xx}\\
             \label{eqn:PrevResult2}
                 R_{21}(p_X(x)) & \triangleq & \min \Big( I(X;Y_2) + C_{12} ,
                        \max \big\{I(X;Y_1), I(X;Y_1)  - H(Y_2|Y_1,X) + \min\big(C_{21},H(Y_2|Y_1)\big)\big\} \Big),\phantom{xx}
            \end{eqnarray}
         \end{subequations}
%         with the appropriate $C_{12} > H(Y_1|Y_2,X)$  or  $C_{21} > H(Y_2|Y_1,X)$ (the one used for the first cooperation step),
         is achievable.}
\end{proposition}

Note that this rate expression
depends only on the parameters of the problem and is, therefore, computable. In proposition
\ref{prop:achive_common_one_step}  the achievable rate  increases linearly with the cooperation
capacity. The downside of this method is that it %cannot be applied to any given conference capacity.
produces a rate increase over the non-cooperative rate only for conference links capacities that exceed
some minimum values.
Specializing the three independent messages result to the single common message case we obtain the
following achievable rate with a $K$-cycle conference for the general BC with a single common message:

\begin{corollary}
    \label{corr:single-coomon-message-with-multi-step}
    \it
    Consider the general broadcast channel with cooperating receivers, having noiseless conference
    links of finite capacities $C_{12}$ and $C_{21}$ between them. Let the receivers hold a conference that
    consists of $K$ cycles. Then, any rate $R$ satisfying
    \begin{equation}
        R = \max \left\{R_{12}, R_{21} \right\},
    \end{equation}
    is achievable.

    Here $R_{12}$ is defined as follows:
    \begin{equation}
        R_{12} = \sup_{p_X(x), \alpha \in [0,1]} \min \left\{ R_1, R_2 \right\},
    \end{equation}
    with
    \begin{subequations}
    \begin{eqnarray}
        \label{eqn:R_1}
        R_1 & = & I\left(X;Y_1,\hY_2^{(1)},\hY_2^{(2)},...,\hY_2^{(K-1)}\right) + \alpha C_{21},\\
        \label{eqn:R_2}
        R_2 & = & I\left(X;Y_2,\hY_1^{(1)},\hY_1^{(2)},...,\hY_1^{(K)}\right),
    \end{eqnarray}
    \end{subequations}
    subject to
    \begin{subequations}
    \begin{eqnarray}
        C_{12} & \ge & I\left(Y_1; \hY_1^{(1)},\hY_1^{(2)},...,\hY_1^{(K)},\hY_2^{(1)},\hY_2^{(2)},...,\hY_2^{(K-1)}\Big|Y_2\right),\\
        (1-\alpha)C_{21} & \ge & I\left(Y_2; \hY_1^{(1)},\hY_1^{(2)},...,\hY_1^{(K)},\hY_2^{(1)},\hY_2^{(2)},...,\hY_2^{(K-1)}\Big|Y_1\right),
    \end{eqnarray}
    \end{subequations}
    for the joint distribution
    \begin{eqnarray*}
         &  & p\left(x,y_1,y_2,\hy_1^{(1)}, \hy_1^{(2)},...,\hy_1^{(K)},\hy_2^{(1)}, \hy_2^{(2)},...,\hy_2^{(K-1)}\right) =\\
         &  & \phantom{xxx} p(x) p(y_1,y_2|x) p\left(\hy_1^{(1)}|y_1\right) p\left(\hy_2^{(1)}|y_2,\hy_1^{(1)}\right)\cdot\cdot\cdot
                p\left(\hy_1^{(k)}|y_1,\hy_1^{(1)},\hy_1^{(2)},...,\hy_1^{(k-1)},\hy_2^{(1)},\hy_2^{(2)},...,\hy_2^{(k-1)}\right)\times\\
         &  &  \phantom{xxx} p\left(\hy_2^{(k)}|y_2,\hy_1^{(1)},\hy_1^{(2)},...,\hy_1^{(k)},\hy_2^{(1)},\hy_2^{(2)},...,\hy_2^{(k-1)}\right)
                \cdot\cdot\cdot p\left(\hy_2^{(K-1)}|y_2,\hy_1^{(1)},\hy_1^{(2)},...,\hy_1^{(K-1)},\hy_2^{(1)},\hy_2^{(2)},...,\hy_2^{(K-2)}\right)\\
         &  &  \phantom{xxx} \times p\left(\hy_1^{(K)}|y_1,\hy_1^{(1)},\hy_1^{(2)},...,\hy_1^{(K-1)},\hy_2^{(1)},\hy_2^{(2)},...,\hy_2^{(K-1)}\right).
    \end{eqnarray*}
    The cardinality of the $k$'th auxiliary random variables are bounded by:
    \begin{eqnarray*}
        ||\mhY_1^{(k)}|| & \le & ||\mY_1|| \times \prod_{l=1}^{k-1} ||\mhY_1^{(l)}|| \times \prod_{l=1}^{k-1} ||\mhY_2^{(l)}|| + 1, \qquad
            \qquad k = 1,2,...,K\\
        ||\mhY_2^{(k)}|| & \le & ||\mY_2|| \times \prod_{l=1}^{k} ||\mhY_1^{(l)}|| \times \prod_{l=1}^{k-1} ||\mhY_2^{(l)}|| + 1,
            \qquad \qquad k = 1,2,...,K-1.
    \end{eqnarray*}
    $R_{21}$ is defined in a parallel manner to $R_{12}$, with $\Rbad$ performing the first conference step, and the appropriate change
    in the probability chain.

\end{corollary}

\bigskip
The proof of corollary \ref{corr:single-coomon-message-with-multi-step} is provided in appendix \ref{appndx:prof_corollary_single_common}.
\smallskip

We note that \cite[theorem 2]{DraperFK:03} presents a similar result for this scenario, under the constraint that the memoryless
broadcast channel can be decomposed as $p(\yvec_1,\yvec_2|\xvec) = \prod_{i=1}^n p(y_{1,i}|x_i)p(y_{2,i}|x_i)$, and
considering the sum-rate of the conference. Here we show that the same achievable rate expressions hold
for the general memoryless broadcast channel.
A recent result appears in \cite{Shlomo_BZ}, where
the single common message case for a Gaussian BC is considered.
In the multi-cycle conference considered in this section, we let the auxiliary RVs follow a more
general chain than that of \cite{Shlomo_BZ} --- which results in a larger achievable rate.

\subsection{A Single-Cycle Conference with TS-EAF}
Consider the case where only a single cycle of conferencing between the receivers is allowed.
 Specializing corollary \ref{corr:single-coomon-message-with-multi-step} to a single cycle case
 we obtain
\begin{subequations}
 \begin{eqnarray}
    \label{eqn:two-step_TSEAF-R_1}
    R_1 & = & I(X;Y_1) + C_{21}\\
    \label{eqn:two-step_TSEAF-R_2}
    R_2 & = & I(X;Y_2, \hY_1^{(1)})\\
    \label{eqn:two-step_TSEAF-C12}
    C_{12} & \ge & I(Y_1; \hY_1^{(1)}|Y_2),
 \end{eqnarray}
\end{subequations}
and the TS-EAF assignment is
\[
    p(\hy_1^{(1)}|y_1) = \left\{
        \begin{array}{cl}
            q_1, & \hy_1^{(1)} = y_1\\
            1-q_1, & \hy_1^{(1)} = \Omega \notin \mY_1.
        \end{array}
    \right.
\]
Applying the TS-EAF assignment to \eqref{eqn:two-step_TSEAF-C12} and \eqref{eqn:two-step_TSEAF-R_2} we obtain
\begin{eqnarray*}
    C_{12} & \ge & I(Y_1; \hY_1^{(1)}|Y_2)\\
        & = & H(Y_1|Y_2) - H(Y_1|Y_2, \hY_1^{(1)})\\
        & = & H(Y_1|Y_2) - q_1 H(Y_1|Y_2, Y_1) - (1-q_1)H(Y_1|Y_2)\\
        & = & q_1 H(Y_1|Y_2)\\
    R_2 & = & I(X;Y_2, \hY_1^{(1)}) \\
        & = & I(X;Y_2) + H(X|Y_2) - H(X|Y_2,\hY_1^{(1)})\\
        & = & I(X;Y_2) + H(X|Y_2) - (1-q_1) H(X|Y_2) - q_1 H(X|Y_2,Y_1)\\
        & = & I(X;Y_2) + q_1 I(X;Y_1|Y_2).
\end{eqnarray*}
Maximizing $R_2$ requires maximizing $q_1 \in [0,1]$. Therefore setting $q_1 = \left[\frac{C_{12}}{H(Y_1|Y_2)}\right]^*$, we
obtain $R_2 = I(X;Y_2) + \left[\frac{C_{12}}{H(Y_1|Y_2)}\right]^* I(X;Y_1|Y_2)$. Combining with $R_1$ we have
that the rate when $\Rbad$ decodes first is given by
\[
    R_{12} = \min \left\{I(X;Y_1) + C_{21}, I(X;Y_2) + \left[\frac{C_{12}}{H(Y_1|Y_2)}\right]^* I(X;Y_1|Y_2)\right\},
\]
and by symmetric argument we can obtain $R_{21}$. We conclude that the rate for the single-cycle conference with TS-EAF is given by
\begin{eqnarray*}
    R & = &  \sup_{p(x)} \min\left\{ R_{12}, R_{21} \right\},\\
    R_{12} & = & \min \left\{I(X;Y_1) + C_{21}, I(X;Y_2) + \left[\frac{C_{12}}{H(Y_1|Y_2)}\right]^* I(X;Y_1|Y_2)\right\}\\
    R_{21} & = & \min \left\{I(X;Y_1) + \left[\frac{C_{21}}{H(Y_2|Y_1)}\right]^* I(X;Y_2|Y_1), I(X;Y_2) + C_{12}\right\}.
\end{eqnarray*}
We note that this rate is always better than the point-to-point rate and also better than the joint-decoding rate of
proposition \ref{prop:achive_common_one_step} (whenever cooperation can provide a rate increase).
However, as in proposition \ref{prop:achive_common_one_step}, at least one receiver has to satisfy the Slepian-Wolf
condition for the full cooperation rate to be
achieved. We also note that using TS-EAF with more than two steps does not improve upon this result.

\begin{figure}[htb]
                         \centering
                         \scalebox{0.60}{\includegraphics{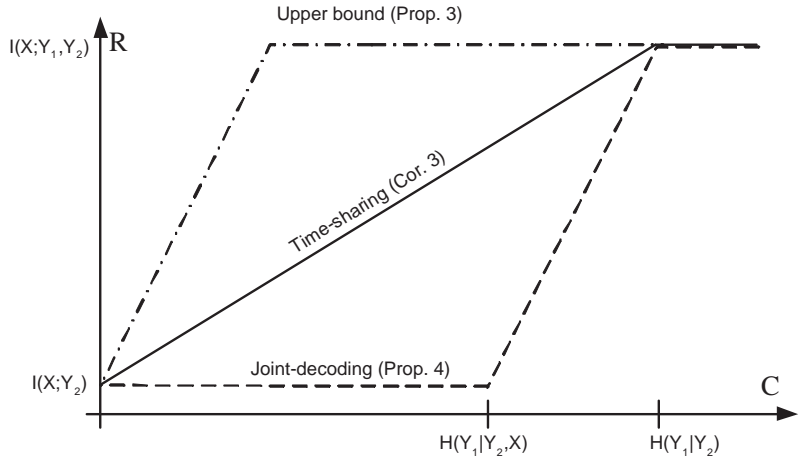}}
                         \caption{\small The achievable rate $R$ vs. conference capacity $C$,
                              for proposition \ref{prop:common_upper} (dashed-dot),
                              proposition \ref{prop:achive_common_one_step} (dashed) and corollary \ref{corr:single-coomon-message-with-multi-step} (solid),
                              for the symmetric broadcast channel.}
                         \label{fig:compare:ft_and_ts}
                  \end{figure}

Finally, we demonstrate the results of proposition \ref{prop:achive_common_one_step} and corollary \ref{corr:single-coomon-message-with-multi-step} through
a symmetric BC example: consider the symmetric broadcast channel where $\mY_1 = \mY_2 = \mY$ and
              \[
                  p_{Y_1|Y_2,X}(a|b,x) = p_{Y_2|Y_1,X}(a|b,x),
              \]
              for any $a,b \in \mY \times \mY$ and $x \in \mX$. Let $C_{21} = C_{12} = C$.
              For this scenario we have that $R_{12} = R_{21}$, in corollary \ref{corr:single-coomon-message-with-multi-step} and
              also $R_{12}(p_X(x)) = R_{21}(p_X(x))$ in proposition \ref{prop:achive_common_one_step}. The resulting rate is depicted in
              figure \ref{fig:compare:ft_and_ts} for a fixed probability $p(x)$.
              We can see that for this case, time-sharing exceeds joint-decoding for all values of $C$. Both methods meet the upper
              bound at $C = H(Y_1|Y_2)$. We note that this is a corrected version of the figure in \cite{ron:ISIT06}.

\section{Conclusions}
\label{sec:conclu}

In this  paper we considered the EAF technique using time-sharing on the auxiliary RVs. We first showed that incorporating
joint-decoding at the destination into the EAF technique results in a special case of the classic EAF of
\cite[theorem 6]{CoverG:79}. We then used the time-sharing assignment of the auxiliary RVs to obtain an
easily computable achievable rate for the multiple-relay case, which can be compared against the DAF-based results, to select the highest rate
for any given scenario.
Next, we showed that for the Gaussian relay channel with coded modulation, the Gaussian auxiliary RV
assignment is not always optimal, and a TS-EAF implementing a per-symbol hard decision may sometimes perform better.
Finally, we considered a third application of TS-EAF to the cooperative broadcast scenario with a multi-cycle
conference. We first derived an achievable rate for the general channel, and then we specialized it to the single-cycle
conference for which we obtained an explicit achievable rate. This rate is superior to the explicit expression that
can be obtained with joint-decoding.

\section{Acknowledgements}
% The authors with to thank Gerhard Karmer and Shlomo Shamai for discussions regarding this work.
In the final version.

\useRomanappendicesfalse
\appendices
\setcounter{equation}{0}
\renewcommand{\theequation}{\thesection.\arabic{equation}}

\section{Expressions for Section \ref{sec:Gauss_relay}}
\label{append:Gauss-deriv}

% We first recall the definition in eqaution \eqref{eqn:def_G} for $G_x(a,b)$:
% \[
%     G_x(a,b) = \frac{1}{\sqrt{2 \pi b}} e ^{-\frac{(x-a)^2}{2 b} }.
% \]
\subsection{Hard-Decision Estimate-and-Forward}
\label{append:Gauss-deriv-HD-EAF}
We evaluate $I(X;\hY_1,Y)$, with $p(\hY_1|Y_1)$ given by \eqref{eqn:def_p_hy1_given_y1_HD_eq1}
and \eqref{eqn:def_p_hy1_given_y1_HD_eq2} using:
\[
    I(X;\hY_1,Y) = I(X;\hY_1) + I(X;Y|\hY_1).
\]
\begin{enumerate}
    \item Evaluating $I(X;\hY_1)$:  Note that both $X$ and $\hY_1$ are discrete RVs,
        therefore $I(X;\hY_1)$ can be evaluated using the
        discrete entropies. The conditional distribution of $\hY_1$ given $X$ is given by:
        \begin{equation}
        \label{eqn:def_p_hy1_given_x}
            p(\hY_1|X=\sqrt{P}) = \left\{
                                        \begin{array}{cr}
                                        P_1 \cdot P_{\ners}, &              1\\
                                        1-P_{\ners}, &                      E\\
                                        (1-P_1) P_{\ners},&                 -1
                                        \end{array}
                                    \right.
        \end{equation}
        where
        \[
            P_1 = \Pr(Y_1 > 0 | X = \sqrt{P}).
        \]
        $p(\hY_1|X=-\sqrt{P})$ can be obtained from $p(\hY_1|X=\sqrt{P})$ by switching $1$ and $-1$ in \eqref{eqn:def_p_hy1_given_x}.

    \item Evaluating $I(X;Y|\hY_1)$: write first
        \[
            I(X;Y|\hY_1) = h(Y| \hY_1) - h(Y | \hY_1, X),
        \]
        and we note that
        \[
            h(Y | \hY_1, X) = h(X + N | \hY_1 , X) = h(N | \hY_1 , X) = h(N) = \frac{1}{2} \log_2 (2 \pi e \sigD).
        \]
        Using the chain rule we write
        \[
            h(Y| \hY_1) = p(\hY_1 = 1) h(Y| \hY_1 = 1) + p(\hY_1 = E) h(Y | \hY_1 = E) + p(\hY_1 = -1) h(Y | \hY_1 = -1),
        \]
        $p(\hY_1)$ can be obtained by combining \eqref{eqn:def_PX} and \eqref{eqn:def_p_hy1_given_x} which results in
        \begin{equation}
        \label{eqn:appndx_p_hy1}
            p(\hY_1) = \left\{
                        \begin{array}{cr}
                            \frac{1}{2}P_{\ners}, & 1\\
                            1 - P_{\ners} , & E\\
                            \frac{1}{2}P_{\ners}, & -1
                        \end{array}
                    \right.,
        \end{equation}
        and we note that $h(Y | \hY_1 = E) = h(Y)$, since erasure is equivalent to no prior information.
        Finally we note that by definition
        \begin{eqnarray}
            h(Y) & = &  -\int_{y = -\infty}^{\infty} f(y) \log_2(f(y)) dy, \nonumber\\
            f(Y) & = & \Pr(X = \sqrt{P}) f(Y | X = \sqrt{P}) + \Pr(X = -\sqrt{P}) f(Y | X = -\sqrt{P})\nonumber\\
            \label{eqn:f_Y_HC}
                 & = & \frac{1}{2}\left(G_y(\sqrt{P},\sigD) + G_y(-\sqrt{P},\sigD) \right),
        \end{eqnarray}
        where
        \begin{equation}
            \label{eqn:def_G}
                G_x(a,b) = \frac{1}{\sqrt{2 \pi b}} e ^{-\frac{(x-a)^2}{2 b} }.
        \end{equation}
        Next, we have
        \begin{eqnarray}
                        \label{eqn:cond_entropy_hy1_is_1}
            h(Y|\hY_1 = 1) & = & -\int_{y = -\infty}^{\infty} f(y|\hy_1 = 1) \log_2(f(y|\hy_1 = 1)) dy\\
            f(Y|\hY_1 = 1) & = & \frac{f(Y,\hY_1 = 1)}{\Pr(\hY_1 = 1)}\nonumber\\
                        & = & \frac{f(Y,Y_1 > 0)P_{\ners}}{\Pr(Y_1 > 0)P_{\ners}}\nonumber\\
            \label{eqn:cond_f_hy1_is_1}
                        & = & \frac{f(Y,Y_1 > 0)}{\Pr(Y_1 > 0)},\\
            f(Y,Y_1 > 0) & = & \Pr(X = \sqrt{P}) f(Y,Y_1 > 0 | X = \sqrt{P}) + \Pr(X = -\sqrt{P}) f(Y,Y_1 > 0 | X = -\sqrt{P})\nonumber\\
            \label{eqn:cond_f_y1_pos}
                        & = & \frac{1}{2}\left( f(Y,Y_1 > 0 | X = \sqrt{P}) + f(Y,Y_1 > 0 | X = -\sqrt{P})\right).
        \end{eqnarray}
        Using
        \[
                    f_{Y,Y_1}(y,y_1 | x )  = \mN\left(
                    \left( \begin{array}{c}
                            x\\ g \cdot x
                        \end{array}\right),
                        \left( \begin{array}{cc}
                            \sigD & 0\\ 0 & \sigR
                            \end{array}
                            \right)
                                    \right)  = G_y(x,\sigD)G_{y_1}(g\cdot x,\sigR),
        \]
        we obtain
        \[
                  f(Y,Y_1 > 0 | X )  =  \int_{y_1 = 0}^{\infty} f(y,y_1 | x ) dy_1 = G_y(x, \sigD)
                        \int_{y_1 = 0}^{\infty} G_{y_1}(g \cdot x, \sigR) dy_1.
        \]
\end{enumerate}

Next we need to evaluate $I(\hY_1;Y_1|Y) = h(Y_1|Y) - h(Y_1|Y, \hY_1)$:
\begin{enumerate}
        \item $h(Y_1|Y) = h(Y,Y_1) - h(Y)$. Here
        \begin{eqnarray*}
            h(Y,Y_1) & = & -\int_{y = -\infty}^{\infty}\int_{y_1 = -\infty}^{\infty} f(y,y_1) \log_2(f(y,y_1)) dy \;dy_1,\\
            f(Y,Y_1) & = & \frac{1}{2}\left(f(Y,Y_1|X = \sqrt{P}) + f(Y,Y_1|X = -\sqrt{P})\right),\\
            f(Y,Y_1|X ) & = & G_y(x,\sigD)G_{y_1}(g \cdot x ,\sigR).
        \end{eqnarray*}

        \item By the definition of conditional entropy we have
         \[
            h(Y_1|Y, \hY_1) = p(\hY_1 = 1) h(Y_1| Y, \hY_1 = 1) + p(\hY_1 = E) h(Y_1 |Y, \hY_1 = E) + p(\hY_1 = -1) h(Y_1 | Y, \hY_1 = -1),
        \]
        where $h(Y_1 |Y, \hY_1 = E) = h(Y_1 |Y )$,
        and for $\hY_1 = 1$, for example, we have
        \[
            h(Y_1 |Y, \hY_1 = 1) = -\int_{y = -\infty}^{\infty} \int_{y_1 = -\infty}^{\infty}
                f(y,y_1|\hy_1 = 1)\log_2(f(y_1|y, \hy_1 = 1)) dy \; dy_1.
        \]
        Finally, we need to derive the distributions $f(y,y_1|\hy_1 = 1)$ and $f(y_1|y, \hy_1 = 1)$.
        Begin with
        \begin{eqnarray*}
            &   & f_{Y,Y_1|\hY_1}(y,y_1|\hy_1 = 1) = \frac{f_{Y,Y_1,\hY_1}(y,y_1,\hy_1 = 1)}{\Pr(\hy_1 = 1)}\\
            &   & \phantom{XXXXXX}\qquad  = \frac{f_{Y,Y_1,\hY_1}(y,y_1,y_1 > 0)P_{\ners}}{\Pr(y_1 > 0)P_{\ners}} = f(y,y_1|y_1 > 0) = \left\{
                                                            \begin{array}{cl}
                                                                \frac{f_{Y,Y_1}(y,y_1)}{\Pr(Y_1>0)} , & y_1 > 0\\
                                                                0                           , & y_1 \le 0
                                                            \end{array}
                                                        \right.
        \end{eqnarray*}
        and due to the symmetry, $\Pr(Y_1 > 0)  = \Pr(Y_1 \le 0) = \frac{1}{2}$.
        We also have
        \begin{eqnarray*}
        f(Y_1|Y , \hY_1 = 1) & =  & \frac{f(Y_1,Y|\hY_1 = 1) }{f(Y|\hY_1 = 1)}  =  \frac{f(Y_1,Y|Y_1 >0 ) }{f(Y|Y_1 > 0)} = \frac{\frac{f(Y_1,Y) }{\Pr(Y_1>0)} }{\frac{f(Y,Y_1 > 0)}{\Pr(Y_1>0)}} = \frac{f(Y_1,Y)}{f(Y,Y_1 > 0)}, \quad Y_1 > 0\\
        f(Y_1|Y , \hY_1 = 1) & = & 0, \quad Y_1 \le 0.
        \end{eqnarray*}
\end{enumerate}

\subsection{Evaluation of the Rate with DHD}
\label{sec:expressions_DHD}
We evaluate the achievable rate using $I(X;Y,\hY_1) = I(X;\hY_1) + I(X;Y|\hY_1)$.
The distribution of $\hY_1$ is given by:
\begin{eqnarray*}
    \Pr(\hY_1 = 1)  =  \Pr(Y_1 > T) & = & \frac{1}{2}\left(\Pr(Y_1 > T | X = \sqrt{P}) + \Pr(Y_1 > T | X = -\sqrt{P}) \right)\\
             & = & \frac{1}{2} \left(\int_{y_1 > T}G_{y_1}(g\sqrt{P},\sigR)dy_1 + \int_{y_1 > T}G_{y_1}(-g\sqrt{P},\sigR)dy_1\right)\\
    \Pr(\hY_1 = E)  =  \Pr(|Y_1| \le T) & = & \frac{1}{2}\left(\Pr(|Y_1| \le T | X = \sqrt{P})
                + \Pr(|Y_1| \le T | X = -\sqrt{P}) \right)\\
             & = & \frac{1}{2} \left(\int_{y_1 =-T}^T G_{y_1}(g\sqrt{P},\sigR)dy_1
                + \int_{y_1 =-T}^T G_{y_1}(-g\sqrt{P},\sigR)dy_1\right),
\end{eqnarray*}
and by symmetry, $\Pr(\hY_1 = 1) = \Pr(\hY_1 = -1)$ and $H(\hY_1|X = \sqrt{P}) = H(\hY_1|X = -\sqrt{P})$.
Therefore, we need the conditional distribution $p(\hY_1|X = \sqrt{P})$:
\begin{eqnarray*}
    \Pr(\hY_1 = 1 | X = \sqrt{P}) & = & \Pr(Y_1 > T| X = \sqrt{P}) = \int_{y_1 > T}G_{y_1}(g\sqrt{P},\sigR)dy_1\\
    \Pr(\hY_1 = -1 | X = \sqrt{P}) & = & \Pr(Y_1 < -T| X = \sqrt{P}) = \int_{y_1 <-T}G_{y_1}(g\sqrt{P},\sigR)dy_1\\
    \Pr(\hY_1 = E | X = \sqrt{P}) & = & 1 - \Pr(\hY_1 = 1 | X = \sqrt{P}) - \Pr(\hY_1 = -1 | X = \sqrt{P}).
\end{eqnarray*}
This allows us to evaluate $I(X;\hY_1) = H(\hY_1) - H(\hY_1|X)$. For evaluating $I(X;Y|\hY_1)$ note that
\[
    h(Y|\hY_1,X) = h(X + N | \hY_1,X) = h(N |\hY_1,X) = h(N) = \frac{1}{2}\log_2(2 \pi e \sigD),
\]
and we need only to evaluate $h(Y|\hY_1)$: by definition
\[
    h(Y|\hY_1) = \Pr(\hY_1 = 1)h(Y|\hY_1 = 1) + \Pr(\hY_1 = E)h(Y|\hY_1=E) + \Pr(\hY_1 = -1)h(Y|\hY_1 = -1),
\]
and note that $h(Y|\hY_1=E) = h(Y)$. Finally,
\begin{eqnarray*}
    h(Y|\hY_1 = 1) & = &   -\int_{y=-\infty}^{\infty} f(y|\hy_1 = 1) \log_2 (f(y | \hy_1 = 1)) dy\\
    f_{Y|\hY_1}(y|\hy_1 = 1) & = & f(y|y_1 >T) = \frac{f(y,y_1 >T)}{\Pr(Y_1 > T)}\\
    f_{Y,Y_1}(y,y_1 >T) & = & \frac{1}{2}\left(f(y,y_1 >T | X = \sqrt{P}) + f(y,y_1 >T| X = -\sqrt{P})\right)\\
         & = & \frac{1}{2}\left(G_y(\sqrt{P},\sigD)\Pr(Y_1>T|X = \sqrt{P}) + G_y(-\sqrt{P},\sigD)\Pr(Y_1>T|X = -\sqrt{P})\right).
\end{eqnarray*}

Evaluating $I(\hY_1;Y_1|Y)$ we have:
\begin{eqnarray*}
    I(\hY_1;Y_1|Y) & = & H(\hY_1|Y) - H(\hY_1|Y,Y_1) \\
     & \stackrel{(a)}{=} & H(\hY_1|Y)\\
     & = & H(\hY_1) + h(Y|\hY_1) - h(Y),
\end{eqnarray*}
where (a) is due to the deterministic mapping from $Y_1$ to $\hY_1$, and $h(Y)$ can be evaluated using
\eqref{eqn:f_Y_HC}.

\subsubsection{DHD when $T \rightarrow 0$}
\label{sec:HDH-Explanation}
As $T \rightarrow 0$ we have that $\Pr(\hY_1 = E) \rightarrow 0$ and $\hY_1$ converges in distribution to a
Bernoulli RV with probability $\frac{1}{2}$. Therefore
\begin{eqnarray*}
    f(Y,\hY_1 = 1) & = & \frac{1}{2}\left(G_y(\sqrt{P},\sigD)\Pr(Y_1>T|X = \sqrt{P}) + G_y(-\sqrt{P},\sigD)\Pr(Y_1>T|X = -\sqrt{P})\right)\\
        & \stackrel{T \rightarrow 0}{\approx} & \frac{1}{2}\left(G_y(\sqrt{P},\sigD)\Pr(Y_1>0|X = \sqrt{P})
                + G_y(-\sqrt{P},\sigD)\Pr(Y_1>0|X = -\sqrt{P})\right)\\
        &  =  & \frac{1}{2}\left(G_y(\sqrt{P},\sigD)P_+
                + G_y(-\sqrt{P},\sigD)(1 - P_+)\right),
\end{eqnarray*}
where $P_+ = \Pr(Y_1>0|X = \sqrt{P})$. Now, letting $g \rightarrow 0$ we have that $P_+ \rightarrow \frac{1}{2}$ and
therefore
\begin{eqnarray*}
    f(Y|\hY_1 = 1) & \stackrel{g \rightarrow 0, T \rightarrow 0}{\longrightarrow } & f(Y)\\
    \Rightarrow h(Y|\hY_1 = 1) & \stackrel{g \rightarrow 0, T \rightarrow 0}{\longrightarrow }& h(Y).
\end{eqnarray*}
We conclude that as $g \rightarrow 0, T \rightarrow 0$, then $h(Y|\hY_1) \rightarrow h(Y)$ and therefore the
$I(Y_1;\hY_1|Y)$ becomes
\[
    I(Y_1;\hY_1|Y)  =  H(\hY_1) + h(Y|\hY_1) - h(Y)  \stackrel{g \rightarrow 0, T \rightarrow 0}{\longrightarrow } 1
\]
Using the continuity of $I(Y_1;\hY_1|Y)$ we conclude that for small values of $g$, as $T$ decreases then
$I(Y_1;\hY_1|Y)$ is bounded from below. This implies that for small $g$ and small $C$ the feasibility
is obtained only for large $T$, which in turn implies low rate.

\subsection{Evaluating the Information Rate with TS-DHD}
\label{appndx:expressions_TS_DHD}
\subsubsection{Evaluating $I(X;Y,\hY_1)$}
We first write
\[
    I(X;Y,\hY_1) = I(X;\hY_1) + I(X;Y|\hY_1).
\]
Evaluating $I(X;\hY_1) = H(\hY_1) - H(\hY_1|X)$ requires the marginal of $\hY_1$.
Using the mapping defined in \eqref{eqn:def_TS-DHD}we find the marginal distribution of $\hY_1$:
\[
    \Pr(\hY_1) = \left\{
        \begin{array}{cl}
            1,  & (1-P_{\ers})\Pr(Y_1>T)\\
            E,  & \Pr(|Y_1| \le T) + P_{\ers} \Pr(|Y_1|>T)\\
            -1, & (1-P_{\ers})\Pr(Y_1 < -T)
        \end{array}
    \right.,
\]
where
\begin{eqnarray*}
    \Pr(Y_1 > T) = \Pr(Y_1 < -T) & = & \int_{y_1 = T}^{\infty} \frac{1}{2}\left[G_{y_1}(\sqrt{P},\sigR)+ G_{y_1}(-\sqrt{P},\sigR) \right]d y_1\\
    \Pr(|Y_1| < T) & = & \int_{y_1 = -T}^{T} \frac{1}{2}\left[G_{y_1}(\sqrt{P},\sigR)+ G_{y_1}(-\sqrt{P},\sigR) \right]d y_1.
\end{eqnarray*}
Also, due to symmetry we have that $H(\hY_1|X = \sqrt{P}) = H(\hY_1|X = -\sqrt{P})$, and therefore we need only to find the conditional
$\Pr(\hY_1|X = \sqrt{P})$:
\[
    \Pr(\hY_1|X = \sqrt{P}) = \left\{
        \begin{array}{cl}
            1,  & (1-P_{\ers})\Pr(Y_1>T|X = \sqrt{P})\\
            E,  & \Pr(|Y_1| \le T|X = \sqrt{P}) + P_{\ers} \Pr(|Y_1|>T|X = \sqrt{P})\\
            -1, & (1-P_{\ers})\Pr(Y_1 < -T|X = \sqrt{P})
        \end{array}
    \right.,
\]
and we note that $f_{Y_1|X} (y_1 | x = \sqrt{P}) = G_{y_1}(\sqrt{P},\sigR)$.

Next, we need to evaluate $I(X;Y|\hY_1) = h(Y|\hY_1) - h(Y | \hY_1,X)$. We first note that
\[
    h(Y| \hY_1,X) = h(X + N|X,\hY_1) = h(N|X , \hY_1) = h(N) = \frac{1}{2}\log_2(2 \pi e \sigR).
\]
Lastly, we have
\[
    h(Y|\hY_1) = \Pr(\hY_1 = 1) h(Y|\hY_1 = 1) + \Pr(\hY_1 = E) h(Y|\hY_1 = E) + \Pr(\hY_1 = -1)h(Y|\hY_1 = -1).
\]
We note that $h(Y|\hY_1 = E) = h(Y)$ and that $h(Y|\hY_1 = 1)$ and $h(Y|\hY_1 = -1)$ are calculated exactly as in
appendix \ref{sec:expressions_DHD} for the DHD case.

\subsubsection{Evaluating $I(\hY_1;Y_1|Y)$}
Begin by writing
\begin{eqnarray*}
    I(\hY_1;Y_1|Y) & = &  h(\hY_1|Y_1) - h(\hY_1|Y_1,Y) \\
        & = & h(Y|\hY_1) + H(\hY_1) - h(Y) - h(\hY_1|Y_1)
\end{eqnarray*}
where we used the fact that given $Y_1$, $\hY_1$ is independent of $Y$. All the terms in the above expressions have been calculated
in the previous subsection, except $h(\hY_1|Y_1)$:
\begin{eqnarray*}
    h(\hY_1|Y_1) & = & \Pr(\hY_1 > T) h(\hY_1|Y_1 > T) + \Pr(|Y_1| \le T) h(\hY_1||Y_1| \le T) + \Pr(Y_1 < -T) h(\hY_1|Y_1 < -T)\\
     & = & \Pr(\hY_1 > T) H(P_{\ers},1 - P_{\ers}) +  \Pr(\hY_1 < -T)H(P_{\ers},1 - P_{\ers}) \\
     & = & (1 - P(|Y_1| \le T)H(P_{\ers},1 - P_{\ers}).
\end{eqnarray*}

\subsection{Gaussian-Quantization Estimate-and-Forward}
Here the relay uses the assignment of equation \eqref{eqn:def_qaussian_quant}:
\[
    \hY_1 = Y_1 + N_Q, \qquad N_Q \sim \mN(0, \sigQ).
\]
We first evaluate
\begin{eqnarray*}
    I(X;Y,\hY_1) = h(Y,\hY_1) - h(Y,\hY_1|X):
\end{eqnarray*}
\begin{enumerate}
    \item
        \begin{eqnarray}
            h(Y,\hY_1) & = & - \int_{y = -\infty}^{\infty} \int_{\hy_1 = -\infty}^{\infty}
                f_{Y,\hY_1}(y, \hy_1) \log_2(f_{Y,\hY_1}(y,\hy_1)) dy \; d\hy_1\nonumber\\
        \label{eqn:joint_y_hy1_gq_eaf}
            f_{Y,\hY_1}(y,\hy_1) & = & \frac{1}{2}\left(G_y(\sqrt{P},\sigD)G_{\hy_1}(g\sqrt{P},\sigR + \sigQ)
                        +G_y(-\sqrt{P},\sigD)G_{\hy_1}(-g\sqrt{P},\sigR + \sigQ)\right).
        \end{eqnarray}

    \item We also have
    \begin{eqnarray*}
        h(Y,\hY_1|X) & = & h(X + N, gX + N_1 + N_Q|X)\\
                    & = & h( N,  N_1 + N_Q|X)\\
                    & = & h(N) + h(N_1 + N_Q)\\
                    & = & \frac{1}{2}\log_2\left((2\pi e)^2\sigD (\sigR + \sigQ)\right).
    \end{eqnarray*}
\end{enumerate}
Lastly we need to evaluate
\[
    I(\hY_1;Y_1|Y) = h(\hY_1|Y) - h(\hY_1 | Y_1,Y) = h(\hY_1,Y) - h(Y) - h(\hY_1 | Y_1,Y),
\]
where
\[
    h(\hY_1| Y_1, Y) = h(Y_1 + N_Q | Y_1,Y) = h(N_Q|Y_1,Y) = h(N_Q) = \frac{1}{2} \log_2(2 \pi e \sigQ).
\]

\subsection{Approximation of HD-EAF for $\sigD \rightarrow \infty$}
     \label{appndx:appndxHD-EAF-highSNR}
     Using \eqref{eqn:def_p_hy1_given_x} and \eqref{eqn:appndx_p_hy1} we can write
        \begin{eqnarray*}
            R \le I(X;\hY_1) & = & H(\hY_1) - H(\hY_1|X) \nonumber\\
                    & = & H\left(\frac{1}{2}P_{\ners}, 1 - P_{\ners} ,\frac{1}{2}P_{\ners}\right)
                            - H\left(P_1 P_{\ners}, 1 - P_{\ners}, (1-P_1)P_{\ners}\right) \nonumber\\
                    & = & -P_{\ners} \log_2\left(\frac{1}{2}P_{\ners}\right) -(1 - P_{\ners})\log_2(1 - P_{\ners})+  P_1 P_{\ners} \log_2(P_1 P_{\ners})\nonumber\\
                    &   & \quad     +(1 - P_{\ners})\log_2(1 - P_{\ners})  +  (1-P_1)P_{\ners}\log_2((1-P_1)P_{\ners})\nonumber\\
                    & = & -P_{\ners} \log_2\left(P_{\ners}\right) +P_{\ners}  +  P_1 P_{\ners} \log_2(P_1) + P_1 P_{\ners} \log_2(P_{\ners})\nonumber\\
                    &   & \quad       +  (1-P_1)P_{\ners}\log_2(1-P_1) + (1-P_1)P_{\ners}\log_2(P_{\ners}) \nonumber\\
                    & = &  P_{\ners}(1  +  P_1  \log_2(P_1) +  (1-P_1)\log_2(1-P_1) ) \nonumber\\
                    & = &  P_{\ners}(1  -H ( P_1  ,1-P_1 )).
            \end{eqnarray*}
        \begin{eqnarray*}
            I(Y_1;\hY_1|Y)  & = & h(\hY_1|Y) - h(\hY_1|Y_1,Y)\\
                    & \stackrel{(a)}{\approx} & H(\hY_1) - H(\hY_1|Y_1)\\
                    & = & H\left(\frac{1}{2}P_{\ners}, 1 - P_{\ners} ,\frac{1}{2}P_{\ners}\right) -
                        H(P_{\ners},1-P_{\ners})\\
                    & = & - 2 \frac{1}{2}P_{\ners} \log_2\left(\frac{1}{2}P_{\ners}\right)
                        - (1 - P_{\ners}) \log_2\left(1 - P_{\ners}\right) + P_{\ners} \log_2(P_{\ners})\\
                    &   & \quad        + (1 - P_{\ners}) \log_2\left(1 - P_{\ners}\right)\\
                    & = &  P_{\ners},
        \end{eqnarray*}
where in (a) we used the fact that $\hY_1$ and $Y$ are independent as $\sigD \rightarrow \infty$, and that given
$Y_1$, $\hY_1$ is independent of $Y$.

\setcounter{equation}{0}
\section{Proof of Corollary \ref{corr:single-coomon-message-with-multi-step}}
\label{appndx:prof_corollary_single_common}
%\begin{proof}

%    In the proof we combine channel coding and the multi-step
%    conference proposed by Kaspi in \cite{Kaspi:85}.
%
%    Fix $n$, $\alpha \in [0,1]$, $p(x)$, and for $k = 1,2,...,K$, fix $p\left(\hy_1^{(k)}|y_1,\hy_1^{(1)},\hy_1^{(2)},...,\hy_1^{(k-1)},\hy_2^{(1)},\hy_2^{(2)},...,\hy_2^{(k-1)}\right)$
%    and\\
%     $p\left(\hy_2^{(k)}|y_2,\hy_1^{(1)},\hy_1^{(2)},...,\hy_1^{(k)},\hy_2^{(1)},\hy_2^{(2)},...,\hy_2^{(k-1)}\right)$.
%
    In the following we highlight only the modifications from the general broadcast result due to the application of
    DAF to the last
    conference step from $\Rgood$ to $\Rbad$, and the fact that we transmit a single message.

    \subsubsection{Codebook Generation and Encoding at the Transmitter}
        The transmitter generates $2^{nR}$ codewords $\xvec$ in an i.i.d. manner according to
        $p(\xvec(w)) = \prod_{i=1}^n p(x_i(w))$, $w \in \mW = \left\{1,2,...,2^{nR}\right\}$. For transmission
        of the message $w_i$ at time $i$ the transmitter outputs $\xvec(w_i)$.

    \subsubsection{Codebook Generation at the $\Rgood$}

             The $K$ conference steps from $\Rgood$ to $\Rbad$ are carried out exactly as in section \ref{sec:DecEncMultiStepRgood}.
             The first $K-1$ steps from $\Rbad$ to $\Rgood$ are carried out as in section \ref{sec:DecEncMultiStepRbad}.
             The $K$'th conference step from $\Rbad$ to $\Rgood$, is different from that of theorem \ref{thm:multi-step-general-bc},
             as after the $K$'th step from $\Rgood$ to $\Rbad$, $\Rbad$ may decode the message
             since $\Rbad$ received all the $K$ conference messages from $\Rgood$. Then, $\Rbad$ uses decode-and-forward for
                its $K$'th conference transmission to $\Rgood$. Therefore, $\Rbad$ simply partitions $\mW$ into $2^{n \alpha C_{21}}$
                subsets in a uniform and independent manner.
%        \end{itemize}

    \subsubsection{Encoding and Decoding at the $K$'th Conference Step from $\Rbad$ to $\Rgood$}
        \begin{itemize}
%            \item Encoding at $\Rgood$ at the $K$'th conference step is done as described in section
%                \ref{sec:DecEncMultiStepRgood}.
            \item Before the $K$'th conference step, $\Rbad$ decodes its message using his channel input and all the
            $K$ conference messages received from $\Rgood$. This can be done with an arbitrarily small probability of error as long as \eqref{eqn:R_2} is satisfied.
            \item  Having decoded its message, $\Rbad$ uses the decode-and-forward strategy to select the
                $K$'th conference message to $\Rgood$. The conference capacity allocated to this step is
                $R_{21}^{(K)} = \alpha C_{21}$.
            \item Having received the $K$'th conference message from $\Rbad$, $\Rgood$ can now
                decode its message using the information received at the first $K-1$ steps,
                and combining it with the information from the last step using the decode-and-forward
                decoding rule. This gives rise to \eqref{eqn:R_1}.
        \end{itemize}

    \subsubsection{Combining All the Conference Rate Bounds}
        The bounds on $R_{12}'^{(k)}$, $k = 1,2,...,K$ can be obtained as in section \ref{sec:combining_bounds_general}:
        \begin{eqnarray*}
            C_{12} & = & \sum_{k = 1}^K R_{12}^{(k)}\\
                   & \ge &   I\left( \hY_1^{(1)},\hY_1^{(2)},...,\hY_1^{(K)},
                        \hY_2^{(1)},\hY_2^{(2)},...,\hY_2^{(K-1)};Y_1\big| Y_2\right) +2K\eps,
        \end{eqnarray*}
        and similarly
        \[
            (1-\alpha)C_{21} \ge I\left( \hY_1^{(1)},\hY_1^{(2)},...,\hY_1^{(K)},
                        \hY_2^{(1)},\hY_2^{(2)},...,\hY_2^{(K-1)};Y_2\big| Y_1\right) +2K\eps,
        \]
        where $(1-\alpha)C_{21}$ is the total capacity allocated to the first $K-1$ conference steps from $\Rbad$ to $\Rgood$.
         This provides the rate constraints on the conference auxiliary variables.


\begin{thebibliography}{10}

\bibitem{Meulen:71}
E. C. van der Meulen.
\newblock {``Three-Terminal Communication Channels"}.
\newblock {\em Adv. Appl. Probab.},vol. 3, pp. 120--154, 1971.

\bibitem{CoverG:79}
T.~M. Cover and A.~A. {El Gamal}.
\newblock {``Capacity Theorems for the Relay Channel"}.
\newblock {\em IEEE Trans. Inform. Theory}, IT-25(5):572--584, 1979.

\bibitem{GuptaKumar:2003}
P.~Gupta and P.~R.~Kumar.
\newblock{``Towards an Information Theory of Large Networks: An Achievable Rate Region"}.
\newblock{\em IEEE Trans. Inform. Theory}, 49(8):1877--1894, 2003.

\bibitem{XieKumar:2004}
L.~-L.~Xie and P.~R.~Kumar.
\newblock{``A Network Information Theory for Wireless Communication: Scaling Laws and Optimal Operation"}.
\newblock{\em IEEE Trans. Inform. Theory}, 50(5):748--767, 2004.

\bibitem{XieKumar:2005}
L.~-L.~Xie and P.~R.~Kumar.
\newblock{``An Achievable Rate for the Multiple-Level Relay Channel"}.
\newblock{\em IEEE Trans. Inform. Theory}, 51(4):1348--1358, 2005.

\bibitem{Kramer:2003}
G.~Kramer, M.~Gastpar, and P.~Gupta.
\newblock{``Capacity Theorems for Wireless Relay Channels"}.
\newblock {\em Proc. 41st Allerton Conf.  Communications, Control, and Computing}, pp. 1074--1083, Monticello, IL, 2003.

\bibitem{Madsen:2005}
B.~Wang, J.~Zhang and A.~Host-Madsen.
\newblock{``On the Capacity of MIMO Relay Channels"}.
\newblock{\em IEEE Trans. Inform. Theory}, 51(1):29--43, 2005.

\bibitem{Kramer:2005}
G.~Kramer, M.~Gastpar, and P.~Gupta.
\newblock{``Cooperative Strategies and Capacity Theorems for Relay Networks"}.
\newblock{\em IEEE Trans. Inform. Theory}, 51(9):3037--3063 , 2005.

\bibitem{Gastpar:2002}
M. Gastpar, G. Kramer and P. Gupta.
\newblock{``The Multiple-Relay Channel: Coding and Antenna-Clustering Capacity"}.
\newblock{\em Proc. IEEE Int. Symp. Inform. Theory (ISIT)}, Lausanne, Switzerland, 2002, pg. 136.

%\bibitem{SchienGallager:2000}
%B.~Schein and R.~Gallager.
%\newblock {``The Gaussian Parallel Relay Network"}.
%\newblock {\em Proc. IEEE Int. Symp. Inform. Theory (ISIT)}, Sorrento, Italy, 2000, pg. 22.

\bibitem{ElGamalH:2006}
L.~Lifeng, L.~Ke and H.~El-Gamal.
\newblock{``The Three-Node Wireless Network: Achievable Rates and Cooperation Strategies"}.
\newblock{\em IEEE Trans. Inform. Theory},  52(3):805--828,  2006.

\bibitem{Goldsmith:2006}
C. T. K. Ng, I. Maric, A. J. Goldsmith, S. Shamai and R. D. Yates.
\newblock{``Iterative and One-Shot Conferencing in Relay Channels"}.
\newblock {\em Proc. IEEE Inform. Theory Workshop (ITW)}, Punta del Este, Uruguay, 2006.

%\bibitem{Motani:2005}
%H.~F.~Chong, M.~Motani and  H.~K.~Garg.
%\newblock{``New Coding Strategies for the Relay Channel"}.
%\newblock{\em Proc. IEEE Int. Symp. Inform. Theory (ISIT)}, Adelaide,  Australia, 2005, pp. 1086--1090.

\bibitem{DraperFK:03}
S.~C. Draper, B.~J. Frey, and F.~R. Kschischang.
\newblock {``Interactive Decoding of a Broadcast Message"}.
\newblock {\em Proc. 41st Allerton Conf.}, % on Communication, Control and Computing},
 Urbana, IL, 2003.

\bibitem{RonSer:2005}
R. Dabora and S.~D. Servetto,
\newblock {``Broadcast Channels with Cooperating Decoders"}.
\newblock {{\em  IEEE Trans. Inform. Theory}}, to appear.

\bibitem{LiagV:2005}
Y. Liang and V. V. Veeravalli.
\newblock{``Cooperative Broadcast Relay Channels"}.
\newblock{Submitted to the {\em IEEE Trans. Inform. Theory}}, July 2005.

\bibitem{ElGamal:06}
A. El-Gamal, M. Mohseni and S. Zahedi,
\newblock{``Bounds on Capacity and Minimum Energy-per-Bit for AWGN Relay Channels"}.
\newblock{\em IEEE Trans. Inform. Theory}, IT-52(4):1545--1561, 2006.

\bibitem{HostMadsen:05}
A. Host-Madsen, and J. Zhang.
\newblock{``Capacity Bounds and Power Allocation for Wireless Relay Channels"}.
\newblock{\em IEEE Trans. Inform. Theory}, IT-51(6):2020--2040, 2006.

\bibitem{Laneman:2000}
J. N. Laneman and G. W. Wornell.
\newblock{``Energy-Efficient Antenna Sharing and Relaying for Wireless Networks"}.
\newblock{\em Proc. IEEE Wireless Communications and Networking Conference (WCNC)} 2000, vol. 1,  pp. 7--12.

\bibitem{Bao:2005}
X. Bao and J. Li.
\newblock{``Decode-Amplify-Forward (DAF): A New Class of Forwarding Strategy for Wireless Relay Channels"}.
\newblock{\em Proc. 6th IEEE Workshop on Signal Proc. Adv. in Wireless Comm. (SPAWC) }, New York, 2005,  pp. 816--820.

\bibitem{Kramer:Asi05}
G. Kramer.
\newblock{``Distributed and Layered Codes for Relaying"}.
\newblock{\em Proc. 39th Asilomar Conf. on Signals, Systems and Computers}, 2005, pp. 1752--1756.

\bibitem{Stankovic:05}
L. Zhixin, V. Stankovic and X. Zixiang.
\newblock{``Wyner-Ziv Coding for the Half-Duplex Relay Channel"}.
\newblock{\em Proc. IEEE Int. Conf. on Acoustics, Speech, and Signal Processing (ICASSP)}, Philadelphia, 2005,
vol. 5, pp. 1113--1116.

\bibitem{Marton:79}
K.~Marton.
\newblock {``A Coding Theorem for the Discrete Memoryless Broadcast Channel"}.
\newblock {\em IEEE Trans. Inform. Theory}, IT-25(3):306--311, 1979.

\bibitem{Motani:06}
M. Motani, H.-F. Chong and H. K. Garg.
\newblock{``Backward Decoding Strategies for the Relay Channel"}.
\newblock{\em MSRI Workshop: Mathematics of Relaying and Cooperation in Communication Networks}, Berkeley, 2006.

\bibitem{YeungBook}
R.~W. Yeung.
\newblock{\em A First Course in Information Theory}.
\newblock Springer, 2002.

\bibitem{cover-thomas:it-book}
T.~M. Cover and J.~Thomas.
\newblock {\em {Elements of Information Theory}}.
\newblock John Wiley and Sons Inc., 1991.

\bibitem{WZ:1976}
A.~Wyner and J.~Ziv.
\newblock{``The Rate-Distortion Function for Source Coding with Side Information at the Decoder"}.
\newblock {\em IEEE Trans. Inform. Theory}, 22(1):1--10, 1976.

\bibitem{Willems:83}
F.~M.~J. Willems.
\newblock {``The Discrete Memoryless Multiple Access Channel with Partially
  Cooperating Encoders"}.
\newblock {\em IEEE Trans. Inform. Theory}, 29(3):441--445, 1983.

\bibitem{RonISIT05:05}
R.~Dabora and S.~D. Servetto.
\newblock{``On the Rates for the General Broadcast Channel with Partially Cooperating Receivers"}.
\newblock{\em Proc. IEEE Int. Symp. Inform. Theory (ISIT)}, Adelaide,  Australia, 2005, pp. 2174--2178.

\bibitem{Cover:98}
T.~M. Cover.
\newblock {``Comments on Broadcast Channels"}.
\newblock {\em IEEE Trans. Inform. Theory}, 44(6):2524--2530, 1998.

\bibitem{ElGamalM:81}
A.~A. {El Gamal} and E.~C. van~der Meulen.
\newblock {``A Proof of Marton's Coding Theorem for the Discrete Memoryless
  Broadcast Channel"}.
\newblock {\em IEEE Trans. Inform. Theory}, IT-27(1):120--122, 1981.

\bibitem{Kaspi:85}
A.~H. Kaspi.
\newblock{``Two-Way Source Coding with a Fidelity Criterion"}.
\newblock {\em IEEE Trans. Inform. Theory}, IT-31(6):735--740, 1985.

\bibitem{Shlomo_BZ}
A. Steiner, A. Sanderovich and S. Shamai.
\newblock{``Broadcast Cooperation Strategies for Two Colocated Users"}.
\newblock{Submitted to the {\em IEEE Trans. Inform. Theory}}, August 2007.

\bibitem{ron:ISIT06}
R.~Dabora and S.~D. Servetto.
\newblock{``A Multi-Step Conference for Cooperative Broadcast"}.
\newblock{\em Proc. IEEE Int. Symp. Inform. Theory (ISIT)}, Seattle, WA, July 2006.

\end{thebibliography}
\end{document}